\documentclass[final,1p,times]{elsarticle}
\usepackage{amssymb}
\usepackage{hyperref}
\usepackage{upgreek}
\usepackage{xspace}
\usepackage{color}

\newcommand{\dd}		{\mathrm{d}}

\newcommand{\mub}{\ensuremath{\mu_{B}}\xspace}

\journal{Nuclear Physics A}

\begin{document}

\begin{frontmatter}



\title{Loosely-bound objects produced in nuclear collisions at the LHC}


\author[label1,label2,label3]{Peter Braun-Munzinger}
\author[label4]{Benjamin D\"onigus\corref{cor1}}

\address[label1]{Research Division and ExtreMe Matter Institute EMMI\\ GSI Helmholtzzentrum f\"ur Schwerionenforschung, Darmstadt, Germany}
\address[label2]{Physikalisches Institut, Ruprecht-Karls-Universit\"{a}t Heidelberg, Heidelberg, Germany}
\address[label3]{Institute of Particle Physics and Key Laboratory of Quark and Lepton Physics (MOE)\\
Central China Normal University, Wuhan, China}
\address[label4]{Institut f\"{u}r Kernphysik, Johann Wolfgang Goethe-Universit\"{a}t Frankfurt, Frankfurt, Germany}

\cortext[cor1]{Corresponding author.}
\ead{b.doenigus@gsi.de}

\begin{abstract}
Loosely-bound objects such as light nuclei are copiously produced in proton-proton and nuclear collisions at the Large Hadron Collider (LHC), despite the fact that typical energy scales in such collisions exceed the binding energy of the objects by orders of magnitude. In this review we summarise the experimental observations, put them into context of previous studies at lower energies, and discuss the underlying physics. Most of the data discussed here were taken by the ALICE Collaboration during LHC Run1, which started in 2009 and ended in 2013. Specifically we focus on the production of (anti-)nuclei and (anti-)hypernuclei. Also included are searches for exotic objects like the H-dibaryon, a  possible $uuddss$ hexaquark state, or also a possible bound state of a $\Lambda$ hyperon and a neutron. Furthermore, the study of hyperon-nucleon and hyperon-hyperon interactions through measurements of correlations are briefly discussed, especially in connection with the possible existence of loosely-bound states composed of these baryons. In addition, some results in the strange and charmed hadron sector are presented, to show the capabilities for future measurements on loosely-bound objects in this direction. Finally, perspectives are given for measurements in the currently ongoing Run2 period of the LHC and in the future LHC Run3.  

\end{abstract}

\begin{keyword}


LHC \sep Heavy-ion collisions \sep (Anti-)(Hyper-)Nuclei \sep Antimatter \sep Exotica \sep Production mechanisms 
\end{keyword}

\end{frontmatter}

%\tableofcontents
%\listoffigures

%\listoftables

\section{Introduction}
\label{introduction}

The data collected at the Large Hadron Collider (LHC)~\cite{1748-0221-3-08-S08001} at different energies and for different collision systems, i.e. pp, p--Pb and Pb--Pb led to a large number of interesting observations regarding the production of composite objects such as light nuclei and hyper-nuclei. These are mainly obtained by the ALICE Collaboration~\cite{alice} using the particle identification capabilities of the ALICE detector.  Results on total and differential production cross sections were obtained for different collision systems and a number of observables, such as 
transverse momentum ($p_{\mathrm{T}}$) spectra or integrated production yields d$N$/d$y$. 

The high collision energies reached  at the LHC lead to a significant increase of the production probabilities for all particles compared to the measurements at the Relativistic Heavy-Ion Collider (RHIC) at the Brookhaven National Lab (BNL). The current top energy for Run2 (2015-2018) at the LHC is 13 TeV for pp collisions, 8 TeV per nucleon-pair in p--Pb collisions and 5.02 TeV per nucleon-pair in Pb--Pb collisions.   

\begin{figure}[!htb]
\begin{center}
\includegraphics[width=0.8\textwidth]{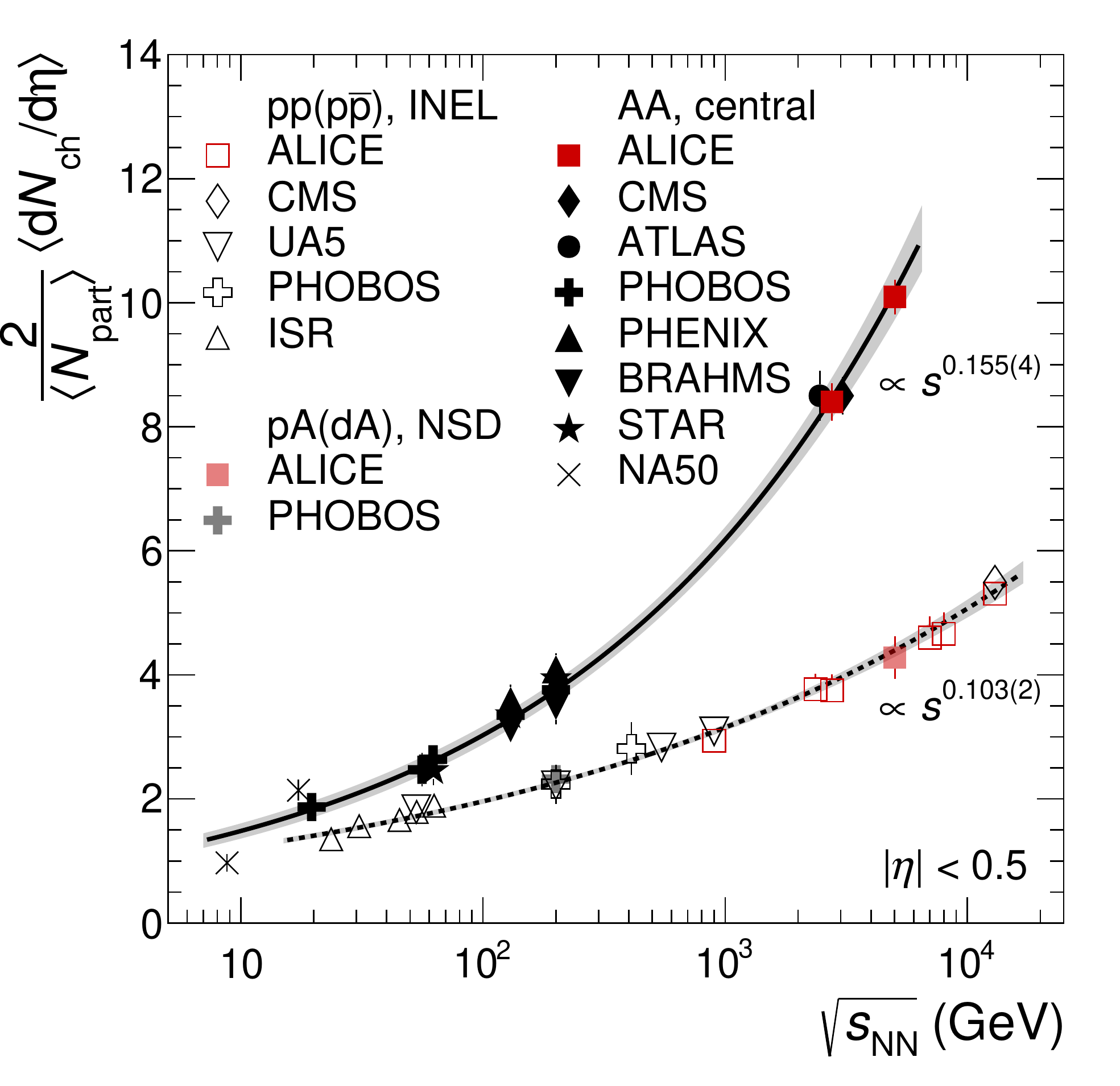}
\caption{\label{multiplicity} Charged particle multiplicity ($\langle \mathrm{d}N_{\mathrm{ch}}/\mathrm{d}\eta\rangle$) for high-energy hadron and nuclear collisions as a function of collision energy, shown as $\frac{2}{\langle N_{part} \rangle}\langle \mathrm{d}N_{\mathrm{ch}}/\mathrm{d}\eta\rangle$, where $\langle N_{part} \rangle$ is the mean number of participating nucleons. Measurements for central Pb--Pb and Au--Au collisions are shown together with inelastic pp and p$\overline{\mathrm{p}}$ collisions as a function of $\sqrt{s}$, together with those from non-single diffractive p--A and d--A collisions. The energy dependence of A--A, pp (p$\overline{\mathrm{p}}$) collisions is well described by power-law functions with different exponents. For details see~\cite{multiplicity_5TeV}.}
\end{center}
\end{figure}

In every central Pb--Pb collision at the LHC more than 21000 charged particles ($\approx$ 32000 particles including neutrals) are produced at $\sqrt{s_\mathrm{NN}} = $5.02 TeV~\cite{multiplicity_5TeV} and more than 17000 charged particles ($\approx$ 26000 particles including neutrals) are produced at $\sqrt{s_\mathrm{NN}} = $2.76 TeV~\cite{Adam:2016ddh}. This corresponds to an increase by about 25\%  while the energy is roughly doubled. The development of the mean charged particle multiplicity $\langle \mathrm{d}N_{\mathrm{ch}}/\mathrm{d}\eta\rangle$  with collision energy is shown in Figure~\ref{multiplicity}. The increase of produced particles for different colliding systems can be described by power-law functions, with $\langle \mathrm{d}N_{\mathrm{ch}}/\mathrm{d}\eta\rangle \propto s^{0.155}_{\mathrm{NN}}$ for central A--A and  $ \propto s^{0.103}$ for pp (p$\overline{\mathrm{p}}$), p--Pb and d--Au collisions. 

The ALICE Collaboration has published a large set of results for the production of hadrons composed of $u$, $d$ and $s$ quarks (e.g.~\cite{pKpi,pKpi_centrality,lambda_k0s,Abelev:2014uua,Abelev:2014laa}). These are of particular interest to understand the dynamics and production mechanisms at work to form mesons and baryons. Also the results on hadrons containing $c$ quarks are important to mention, especially since the charm quark is expected to be produced in initial hard collisions of partons.  All these particles are understood in the quark model as compact 'bags' with $q \bar{q}$ (mesons) or $qqq$ (baryons) configurations. In particular, their radii are less than 1 fm. It is by now well accepted that the yields of these hadrons in relativistic nuclear collisions can be well described in a thermal approach with temperature and baryo-chemical potential as the main parameters governing their production\cite{BraunMunzinger:2003zd,Becattini:2005xt,Braun-Munzinger:2015hba,Becattini:2016xct,Andronic:2017pug}. At LHC energy the value of the temperature parameter is around 156 MeV and the chemical potential vanishes, implying equal production yields for particles and anti-particles. We will below give a brief summary of this approach.

Light nuclei and hypernuclei with baryon number $B \leq 4$, in contrast, are objects composed of nucleons and hyperons and, hence, are generally not described in terms of quarks. Their radii are significantly larger than those of hadrons. Some of these, the deuteron and the hypertriton, are particularly loosely bound. These loosely-bound objects are systems which are stable against strong decays but with binding energies $E_B$ tiny compared to their masses and even much smaller than the typical nuclear potential binding them. The size of such systems scales then approximately $\propto \frac{1}{\sqrt{E_B}}$, independent of the nuclear potential. A case in point is the deuteron with $E_B=2.23$~MeV and a rms radius of $\sqrt{\langle r^2\rangle} \approx 2.1$~fm~\cite{nuclei:datatable}. A more dramatic case is the hypertriton in which a $\Lambda$ hyperon is bound to a deuteron by only 130~keV, leading to a size of approximately 10~fm. In contrast, the nuclei $^3$He and $^4$He  are more strongly bound and reasonably compact objects. As will be shown below, all these particles  are copiously produced at LHC energies. Their detailed production mechanism is, however,  not fully understood. Surprisingly, as we will demonstrate below, their yields in Pb--Pb collisions  can be described with the same thermal approach as for standard hadrons. Provocatively, their thermal production temperature $T$ coincides with those for all other hadrons, although for their binding energies the relation $E_B  \ll T \approx 156$~MeV. Here and in the following we use natural units with constants $\hbar=k_\mathrm{B}=c=1$ except in the figures.  

Light nuclei were actually first observed in high energy proton-nucleus collisions at the CERN PS accelerator~\cite{Cocconi:1960zz}. Already then these findings of unexpected large yields were considered a major surprise. 
The first application of thermal concepts to explain deuteron as well as $^3$He and t production in proton-induced collisions at high energy
is due to Hagedorn, who used a pre-cursor of his statistical approach~\cite{hagedorn60}  to describe the production of deuterons and other light nuclei in  collisions of 25 GeV protons with various nuclear targets. However, the results were at variance with the momentum dependence of measured deuteron yields. Better agreement was achieved with a 'coalescence' approach discussed below. The subject was taken up again only 25 years later with the beginning of the relativistic nuclear collisions program at the Brookhaven AGS and CERN SPS, see~\cite{Nagamiya:1992sg} for a review.

Using thermal concepts to describe the production of composite objects such as light nuclei was taken up again about 25 years ago~\cite{pbm1,pbm,thermalModel}. There it was noticed that treating such objects as point particles and using only mass and quantum leads to an excellent description of the available data and to predictions for data at much higher energies. With the new data from the RHIC and LHC accelerators  the empirical evidence is now becoming detailed enough to fully explore the consequences of this approach. We will focus mostly on results obtained at LHC energies but, where appropriate, also take data at lower energy into account. 

Because of the fragility of loosely bound particles and the high particle density in the hot fireballs at the hadronisation stage, one would expect that these objects cannot be produced at all, or if they would be produced should dissolve immediately after being formed. Consequently,
alternative approaches based on coalescence models have been developed to shed light on their production mechanism. These approaches have also met with some success but a number of issues and open questions remain. In this review we will contrast the thermal production picture with results using various coalescence models. 

The production of these loosely-bound states in abundances not reached before allows also  the spectroscopy of their ground-states and to study their branching ratios in (weak) decays. This holds for the known hyper-nuclei but also for other exotic states predicted by QCD or QCD inspired models and still to be discovered.  

In the following we provide first a short overview of the ALICE setup, highlighting the parts which play an important role later on. This is followed by remarks on the experimental capabilities in the strange and charmed hadron sector.
In the next part we describe the models which are used to understand and interpret the results mentioned above, namely the hydrodynamic picture utilized to explain the evolution of the created fireball and the models to describe the production of particles and in particular the loosely-bound states we are focusing on.  The results for the production of nuclei, hyper-nuclei and exotica are described and discussed in the following section.  We then give an outlook to results expected in LHC Run2 and Run3 (starting 2021), before we conclude and discuss the presented results.  

\section{ALICE setup}
\label{alice}

\begin{figure}[!ht]
\begin{center}
\includegraphics[width=0.95\textwidth]{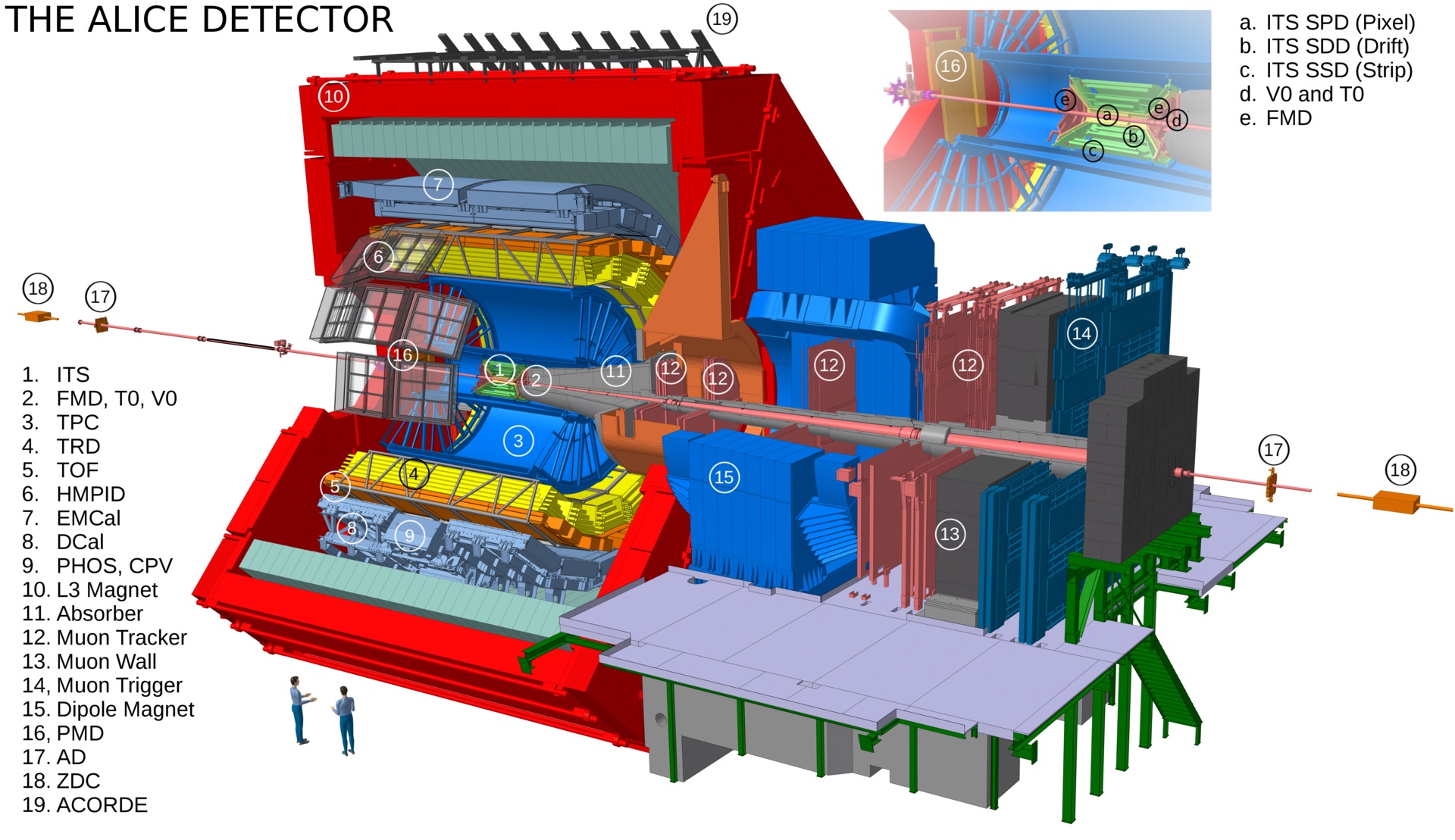}
\caption{\label{alice_setup} Artistic view of the ALICE detector setup. The central barrel is hosted inside the (red) L3 magnet, providing a solenoidal field of 0.5 T. This part of the setup is clearly separated from the muon spectrometer outside of the solenoid. The insert shows an enlargement of the collision area with the ITS silicon detectors inside the TPC. See text for the discussion of the main detectors and~\cite{alice, alice_performance} for more details.}
\end{center}
\end{figure}

The ALICE detector layout is optimised to study different signatures and observables of the quark-gluon plasma, the state expected from QCD thermodynamics when hadrons melt to a deconfined form of matter consisting of quarks and gluons. For this, the layout (see Figure~\ref{alice_setup}) consists of different detector types for vertexing, tracking and particle identification. It is split into a forward muon spectrometer, where hadrons are stopped by an absorber mixture of concrete and steel, and a central barrel part, which is used to study the production of hadrons and electrons at mid-rapidity (full coverage up to  pseudo-rapidity $|\eta| < 0.9$).  In addition to these different detectors, a set of zero-degree calorimeters is installed at 116 m distance from the interaction point.

The central barrel is making use of the solenoidal magnetic field of up to 0.5 T, provided by the L3 magnet,  to bend the tracks of charged particles,  thus allowing the measurement of the momentum $p$ and rigidity $p/z$, where $z$ denotes the charge number of the particles. The region of the interaction point is surrounded first by a thin Be beam pipe, followed by six layers of three different silicon detector types, mainly used for vertexing and tracking. The first two layers are made of silicon pixel detectors (SPD), followed by two layers of silicon strip detectors (SSD) and finally two layers of silicon drift detectors (SDD) allowing also particle identification using the specific energy loss (d$E$/d$x$) in the detector. The complete set of silicon detectors is called inner tracking system (ITS).

Tracks originating from the interaction point and traversing the beam pipe and the silicon detectors reach the time projection chamber (TPC), the main tracking device in the central barrel also used for particle identification using the specific energy loss~\cite{alice_tpc}. Particles with intermediate $p_{\mathrm{T}}$ reach the transition radiation detector (TRD) and also the time-of-flight detector (TOF). The TRD is designed to separate electrons from pions, through the measurement of transition radiation,  collected additionally to the ionisation in the Xe/CO$_2$ gas mixture. Further, The TRD can be used for triggering on (di-)electrons and  on highly ionising particles ($z>1$). The time-of-flight measurement is also used for particle identification, whereas the start-time of the flight-time measurement is generated by the T0 detector and the length is largely determined by the track reconstructed using the TPC.         

\begin{figure}[!htb]
\begin{center}
\includegraphics[width=0.7\textwidth]{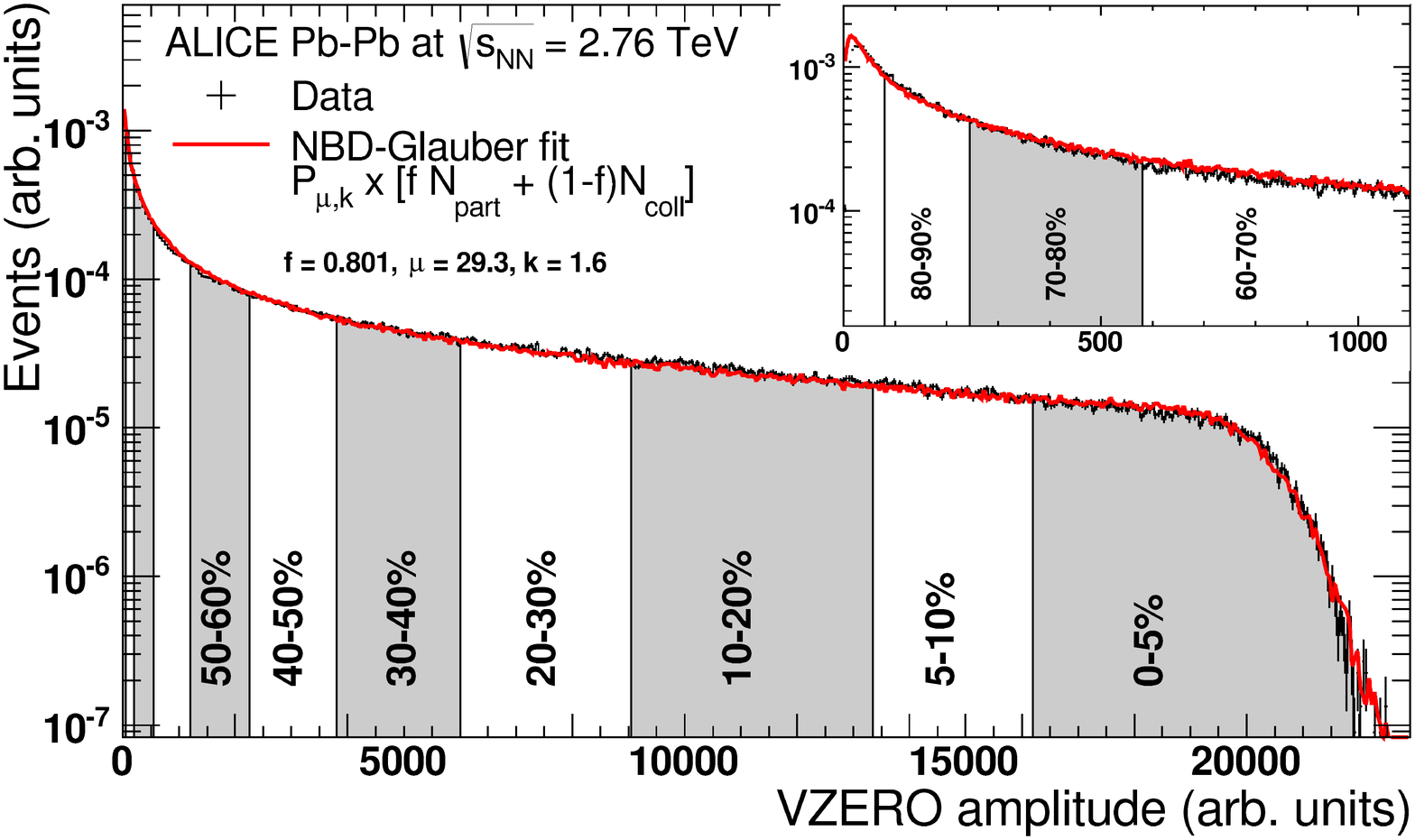}
\caption{\label{centrality_v0} Measured VZERO detector amplitude compared to a Glauber model, used to define the centrality classes. The figure is taken from~\cite{vzero1}, where also more details are given on the Glauber fit.}
\end{center}
\end{figure}

A mainly in heavy-ion collisions used quantity is the so-called centrality. The centrality is a measure for a given fraction of the multiplicity distribution. Usually it is defined as centrality class by selecting a percentile of the measured multiplicity distribution (e.g. 0-5\%). This is done comparing with a geometrical model, named after Roy J. Glauber, the Glauber model~\cite{Miller:2007ri}. By a Monte Carlo approach for this model, the number of participant nucleons ($N_{\mathrm{part}}$) and the number of inelastic nucleon-nucleon collisions ($N_{\mathrm{coll}}$) can be estimated, using the randomly selected impact parameters $b$ and the distribution of the nucleons of two nuclei
according to the relevant nuclear density distribution. When such a Monte Carlo simulation is compared to the measured distribution in a detector, the corresponding centrality classes can be defined. This is shown in Figure~\ref{centrality_v0} using the VZERO detector (in Figure~\ref{alice_setup} named V0), two scintillator hodoscope arrays, hosted also in the central barrel~\cite{alice_centrality,vzero1,vzero2}.      

In addition to the aforementioned detectors the ALICE setup consists of more specialised detectors as calorimeters, further forward detectors or detectors to trigger on cosmic rays reaching the ALICE detectors and used for their calibration and alignment.\\
A full description of the detector layout can be found in~\cite{alice,alice_performance}.

Because of these very nice capabilities, especially the excellent particle identification over a large range of rigidity, the ALICE detector is well suited to study stable, weakly and strongly decaying particles. In other words the ALICE setup at the LHC is a unique tool for the research discussed in this review.
   
\section{ALICE performance for strangeness and charm physics}
\label{charm_strangeness}
\begin{figure}[!htb]
\begin{center}
\includegraphics[width=0.9\textwidth]{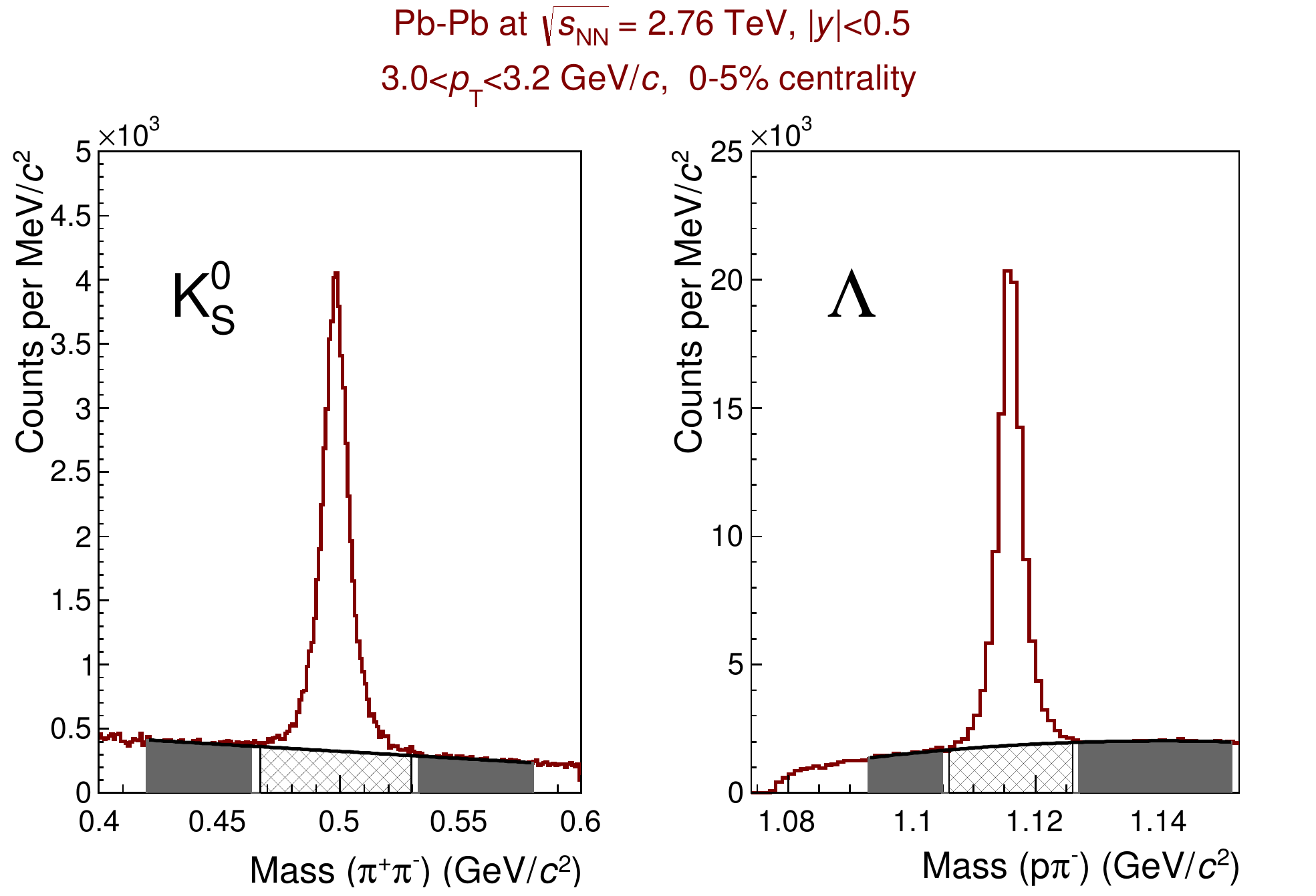}
\caption{\label{k0_lambda} Examples of reconstructed invariant mass spectra of K$^0_s$ and $\Lambda$ particles for a particular $p_\mathrm{T}$ bin and in the most central event class (0-5\%). Figure taken from~\cite{alice_performance}.}
\end{center}
\end{figure}

The excellent vertexing capabilities of the ALICE detector setup allows nicely to separate the primary vertex from secondary vertices stemming from weak decays. The resolution of the ITS is of about 60 $\mu m$, which leads to the clear and clean secondary vertex finding for the weak decays of hadrons containing strangeness, as K$^{0}_\mathrm{s}$ and $\Lambda$, whose mean decay length is of the order of several centimeters. Their signal is then reconstructed by a invariant mass analysis which lead to highly significant signals ($\sigma > 10$) as depicted in Figure~\ref{k0_lambda}, when either a track pair of $\pi^++\pi^-$ or a p$+\pi^-$ pair is found such that both originate from the same secondary vertex.

\begin{figure}[!htb]
\begin{center}
\includegraphics[width=0.49\textwidth]{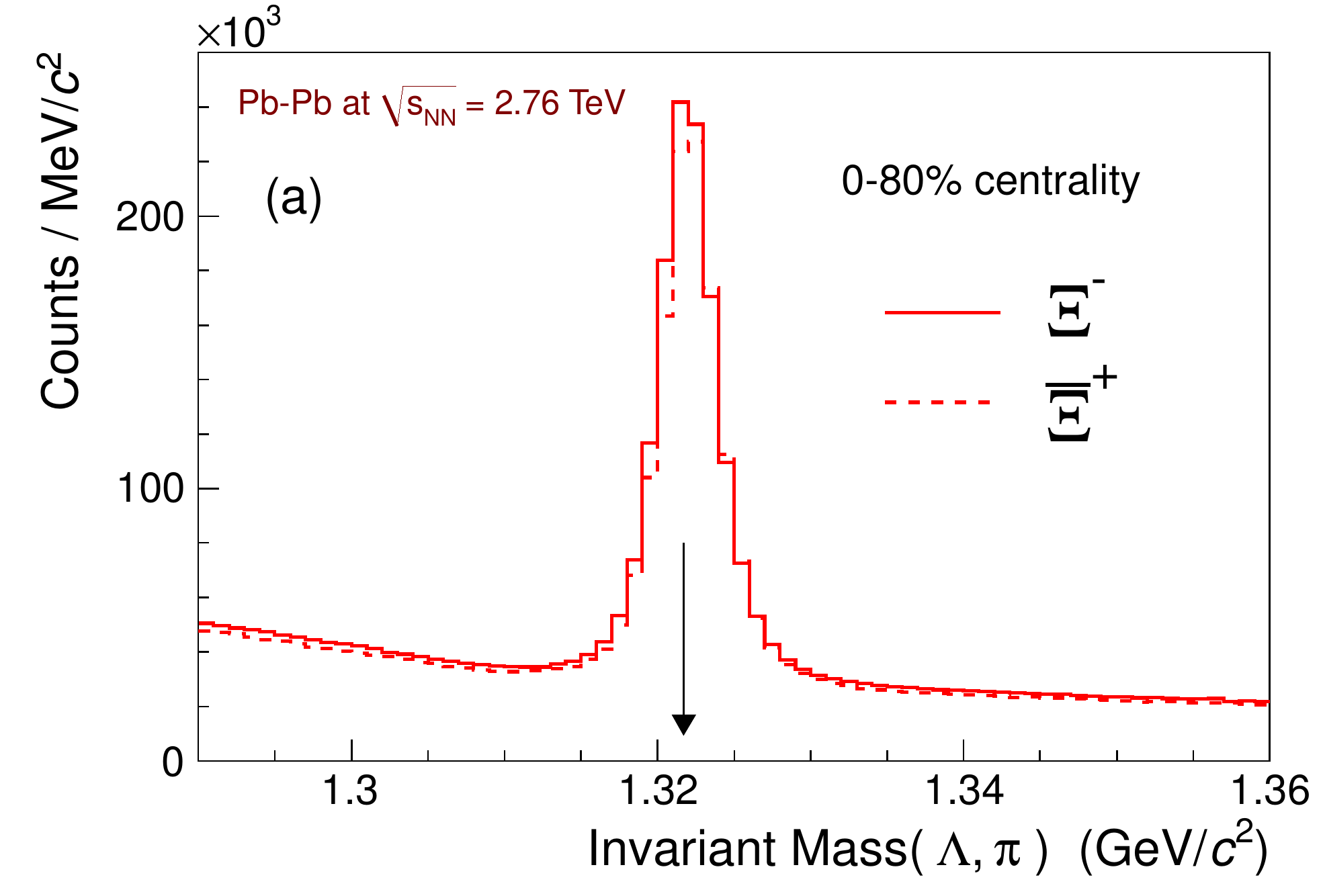}
\includegraphics[width=0.49\textwidth]{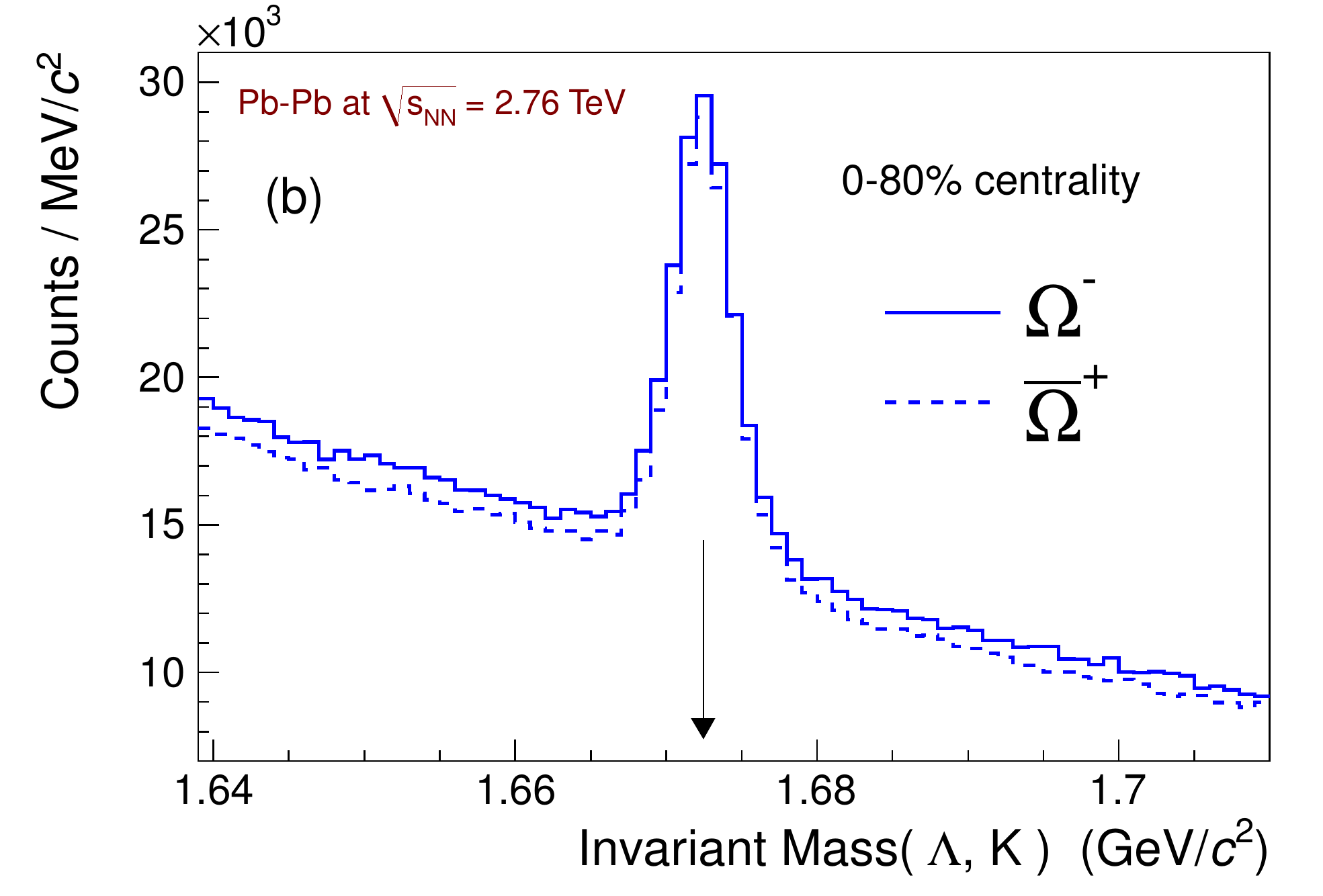}
\caption{\label{xi_omega} Examples of reconstructed invariant mass of $\Xi^{-}$ and $\Omega^{-}$ spectra in Pb--Pb collisions at $\sqrt{s_\mathrm{NN}}$ in 0-80\% centrality. The arrows indicate the expected positions in mass, averaged by the Particle Data Group~\cite{Agashe:2014kda}. Taken from~\cite{ABELEV:2013zaa}}
\end{center}
\end{figure}

The $\Lambda$ candidates can be further combined with a 'bachelor' $\pi$ or K meson. If the reconstructed $\Lambda$ flight-line comes close to a charged track which seems to be displaced from the primary vertex a cascade candidate can be reconstructed. An example of the invariant mass of $\Lambda + \pi^-$ and $\Lambda + \mathrm{K}^{-}$ is shown in Figure~\ref{xi_omega}. The signals correspond to the decays of $\Xi^-$ and $\Omega^-$.

\begin{figure}[!htb]
\begin{center}
\includegraphics[width=0.9\textwidth]{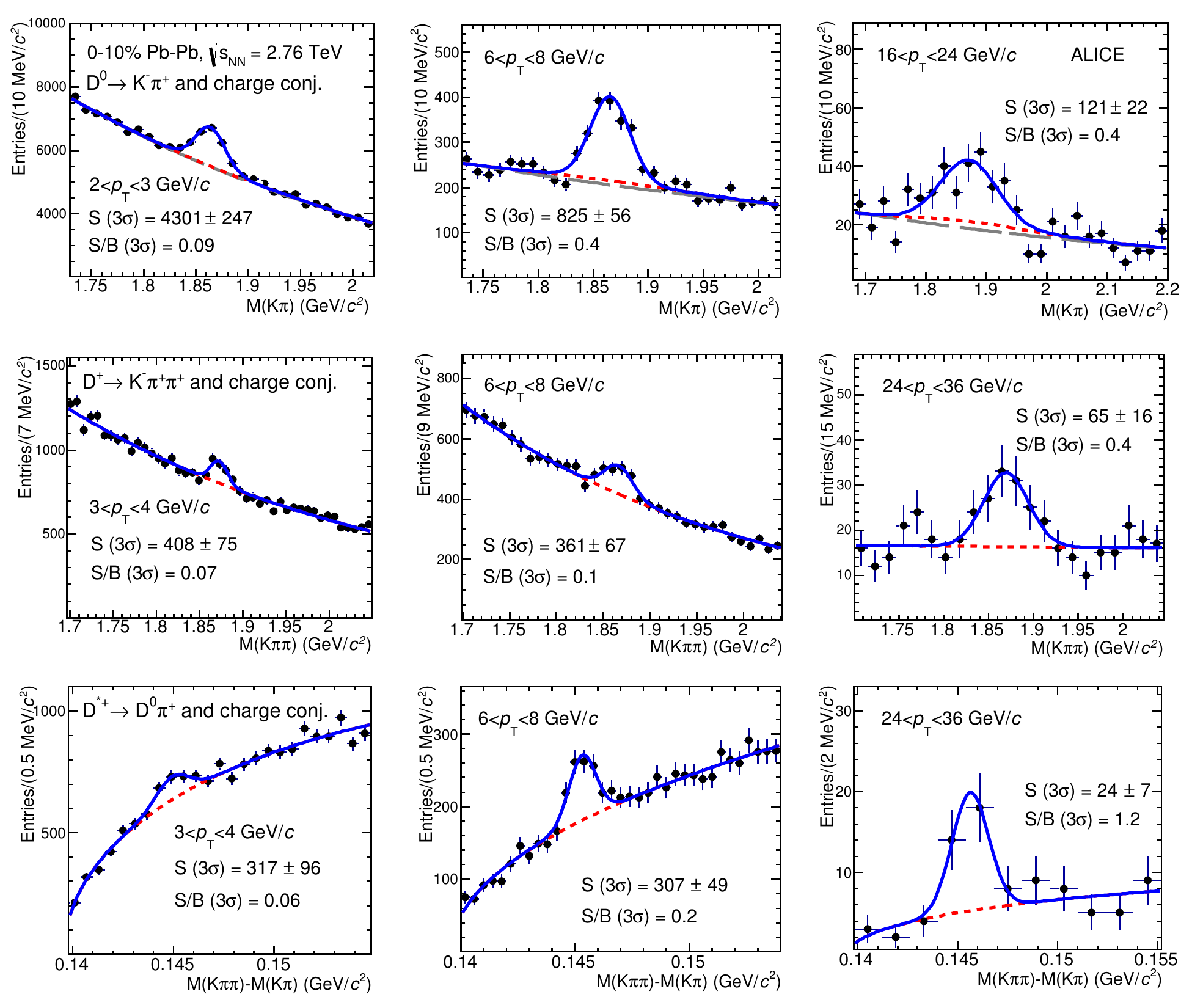}
\caption{\label{d_mesons} Reconstructed invariant mass spectra of D$^0$, D$^+$ and D$^{*+}$ (going from top to bottom) mesons, for three different transverse momentum intervals and for the 0-10\% central events. Taken from~\cite{Adam:2015sza}.}
\end{center}
\end{figure}

Furthermore, the weak decays of charmed mesons are in the order of several hundred micrometers and thus their secondary vertices are also separable from the primary vertex. Combining two or three tracks with displaced vertices and applying strong selection criteria, such as the cosine of the pointing angle to be close to unity, lead to invariant mass plots as visualized in Figure~\ref{d_mesons}.
   
These examples show the excellent performance and the capabilities in the strangeness and charm physics sector of the ALICE setup and with these the possibilities in studies connected to the topics discussed in this topical review.  

\section{Models}
\label{models}

\subsection{The evolution of the fireball}\label{hydro}

Relativistic heavy-ion collisions offer a unique way to study the matter created in these collisions as state of deconfined quarks and gluons usually called Quark-Gluon Plasma, or simply QGP. This QGP exists only for a very short time (about 10-15 fm/$c$ at the LHC) until it is cooled down below the pseudo-critical temperature $T_c$, while expanding, and the quarks and gluons start to hadronise. In this created hadron gas the particles may still interact and scatter, although the expansion during the hadronisation stage leads to low enough densities that inelastic collisions are rare. At this point the production yields of the particles are frozen in: the corresponding temperature is usually referred to as chemical freeze-out temperature $T_{chem}$, whereas the particles stop to scatter and stream freely below the kinetic freeze-out temperature $T_{kin}$. One of the important results from particle production studies  in relativistic nuclear collisions is \cite{BraunMunzinger:2003zz,Andronic:2005yp,Becattini:2005xt,Andronic:2017pug} that, for center-of-mass energies studied at RHIC and the LHC, the value of $T_{chem}$ closely agrees with $T_c$: chemical freeze-out takes place near the QCD phase boundary.

The latter two characteristic temperatures of the fireball evolution can be extracted from model analyses which will be described in a dedicated section (\ref{models}). The pseudo-critical temperature $T_c$ (pseudo-critical temperature because the transition in ultra-relativistic heavy-ion collisions at very high energy is found to be a cross-over transition) is studied using lattice QCD where different thermodynamic quantities are investigated. From these studies a $T_c$ of 154 $\pm$ 9\,MeV and a critical energy-density of $\epsilon_c = 0.34 \pm 0.13$\,GeV/fm$^3$ can be extracted~\cite{PhysRevD.90.094503,Borsanyi2010}. Thus, above the temperature $T_c$ a deconfined state of matter is created in heavy-ion collisions where temperatures of the fireball reach more than 300 MeV at the LHC as is, e.g., indicated by measurements of effective temperatures from spectra of direct photons~\cite{Adam2016235}.

\begin{figure}[!htb]
\begin{center}
\includegraphics[width=0.8\textwidth]{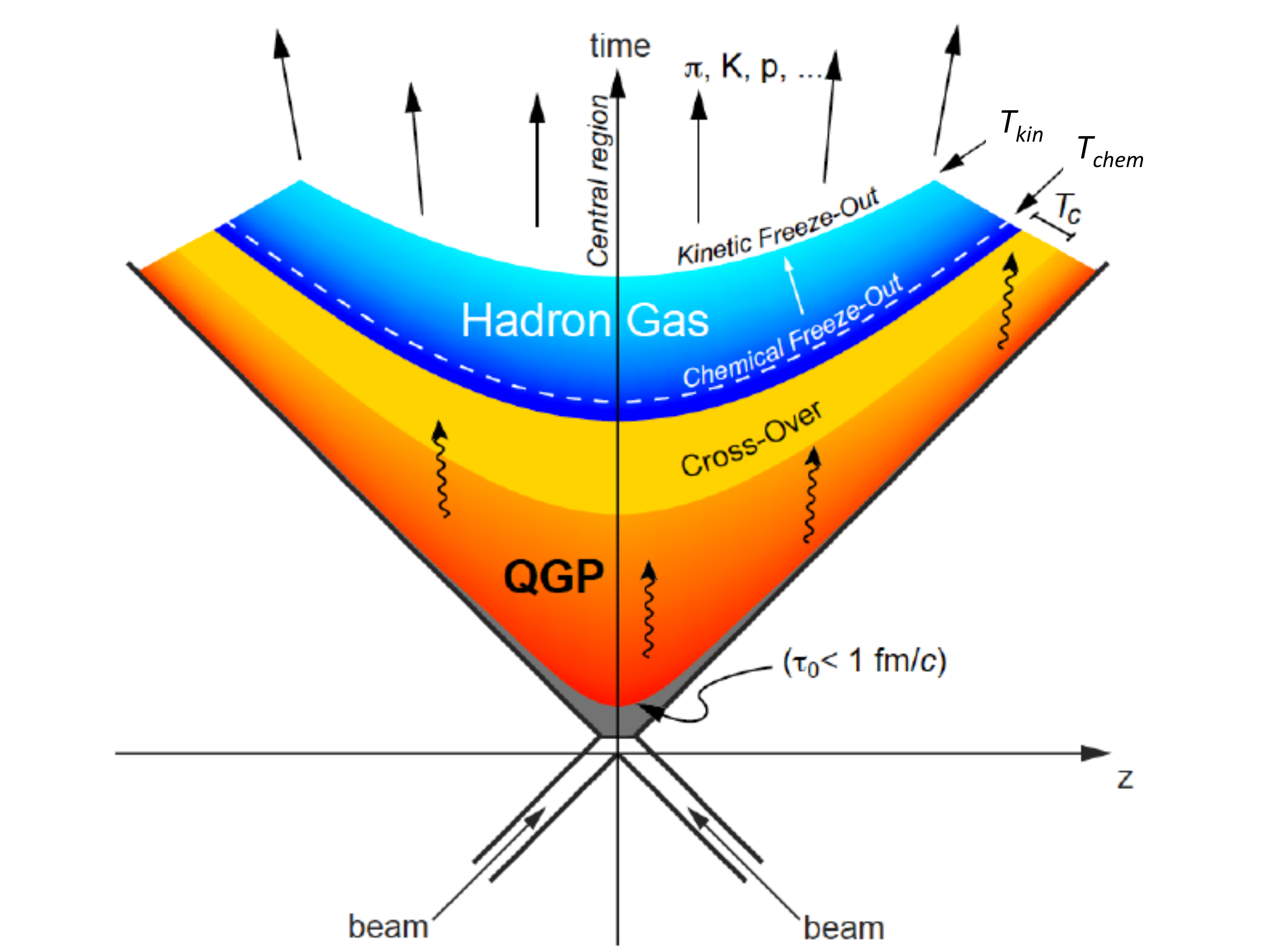}
\caption{\label{qgp} Space-time diagram of a heavy-ion collision of two nuclei colliding at time t=0 and longitudinal position z=0 (transverse direction not shown). The evolution goes from a hot-fireball in a pre-equilibrium phase through the formation of a QGP, followed by a cross-over phase transition to a hadron gas. The fireball formed in the collision emits different kinds of particles (indicated by the arrows). The temperatures crossed during the evolution are $T_{c}$, $T_{chem}$ and $T_{kin}$. For further details see text. (Figure courtesy of Boris Hippolyte).}
\end{center}
\end{figure}

A schematic picture of the evolution through the different phases of a collision of two relativistic nuclei is displayed in Fig.~\ref{qgp}. It displays the collision of two nuclei traveling with (nearly) the speed of light and exhibits the different phases of the created fireball while cooling down. In fact, the  expansion in z-direction (beam direction) is the main cooling mechanism in these collisions, and this one-dimensional space-time picture is fully analytically solvable (Bjorken model~\cite{Bjorken:1982qr}). The different phases of the fireball evolution are usually tackled by different model approaches.  The pre-equilibrium phase ($\tau_0 \leq 1$ fm/$c$) is, e.g., often modeled by quantum transport using a parton-cascade model or a Boltzmann transport approach~\cite{Heinz:2004qz,Heinz:2015arc,Liu:2015nwa,Braun-Munzinger:2015hba}. 

After a proper time $\tau = (t^2 - z^2)^{1/2}$ of about  1 fm/$c$ the fireball approaches a state of  local thermal equilibrium (the QGP) whose evolution is well described within the framework of relativistic hydrodynamics when the QGP is assumed to be a nearly ideal fluid with very low shear viscosity to entropy density ratio. The main further ingredient needed as input here is the (QCD) equation-of-state, i.e. the relation between pressure, energy density and baryon density, at a given temperature. There are several good reviews existing on the applicability of relativistic hydrodynamics for (ultra-)relativistic heavy-ion collisions~\cite{Heinz:2013wva,Heinz:2004qz,Jeon:2015dfa,Kolb:2003dz,Huovinen:2006jp,Huovinen:2003fa}.
After the temperature of the system reaches (locally) the pseudo-critical temperature $T_c$ the QGP starts to hadronise and hadrons are formed. Since  $T_c \approx T_{chem}$, see above,  where hadron yields are determined and  their yields are 'frozen', the subsequent evolution in the hadronic phase can only be described in a non-equilibrium approach. Usually it is modeled by a hadronic cascade, where further elastic and, in principle inelastic scattering can take place. The excellent agreement between thermal model predictions and data for $T = T_{chem}$ implies that inelastic rescattering must be small, see below.  

\subsubsection{Collective expansion and hydrodynamic flow}
\label{sect_flow}

In addition to the above introduced longitudinal expansion the fireball also expands in radial direction. Strong pressure gradients in the direction transverse to the beam induce  a strong flow field, which the particles experience while the fireball evolves. This radial or transverse flow field leads to a characteristic shape of the transverse momentum spectra of each particle species in heavy-ion collisions. Since the transverse momentum due to hydrodynamic flow is (essentially, up to relativistic effects) given by the product of the particle mass and common flow velocity, the heavier particles get shifted to higher mean transverse momenta, implying a characteristic mass ordering.

\begin{figure}
\centering
  \includegraphics[width=0.6\textwidth]{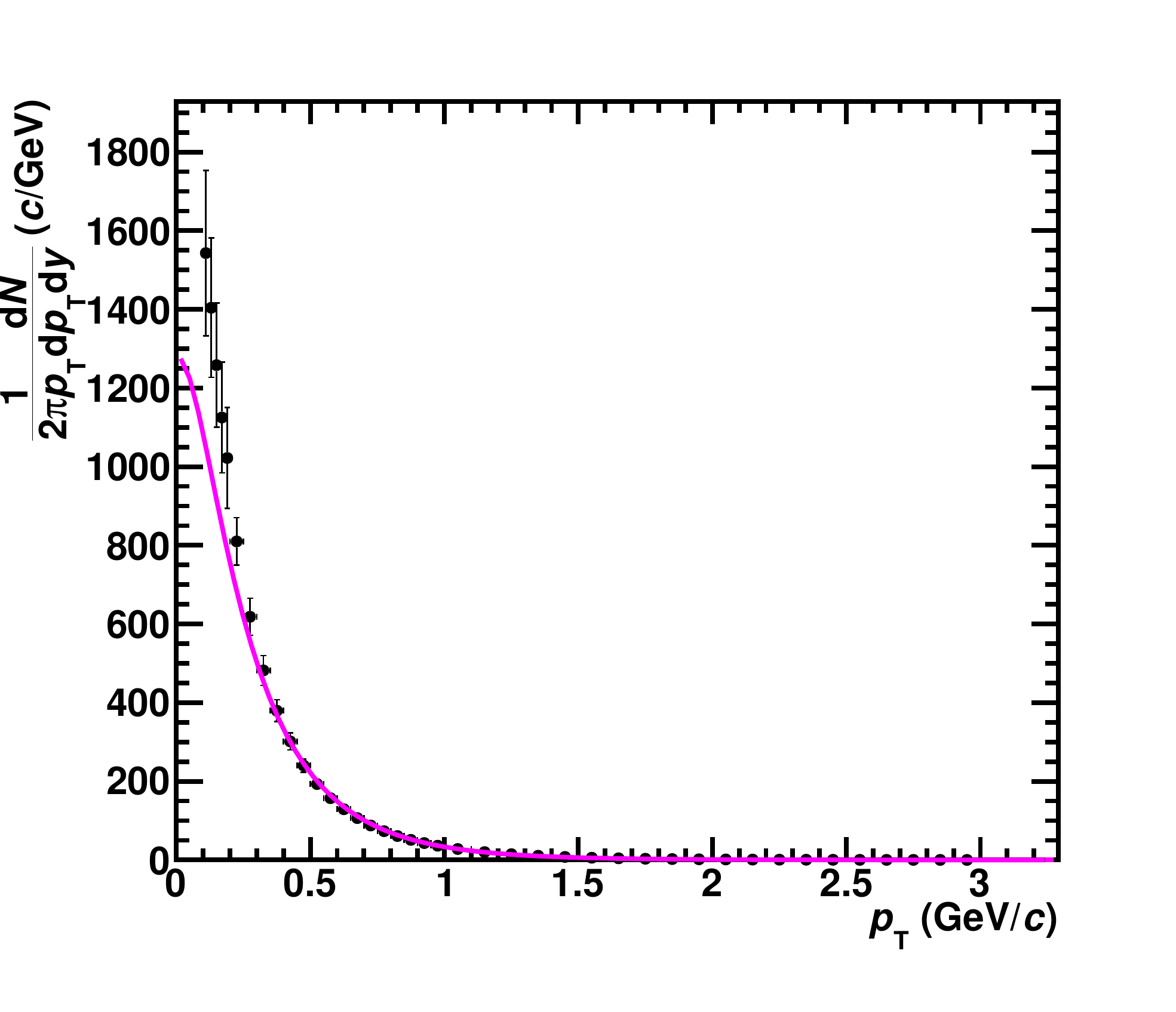}
  \includegraphics[width=0.6\textwidth]{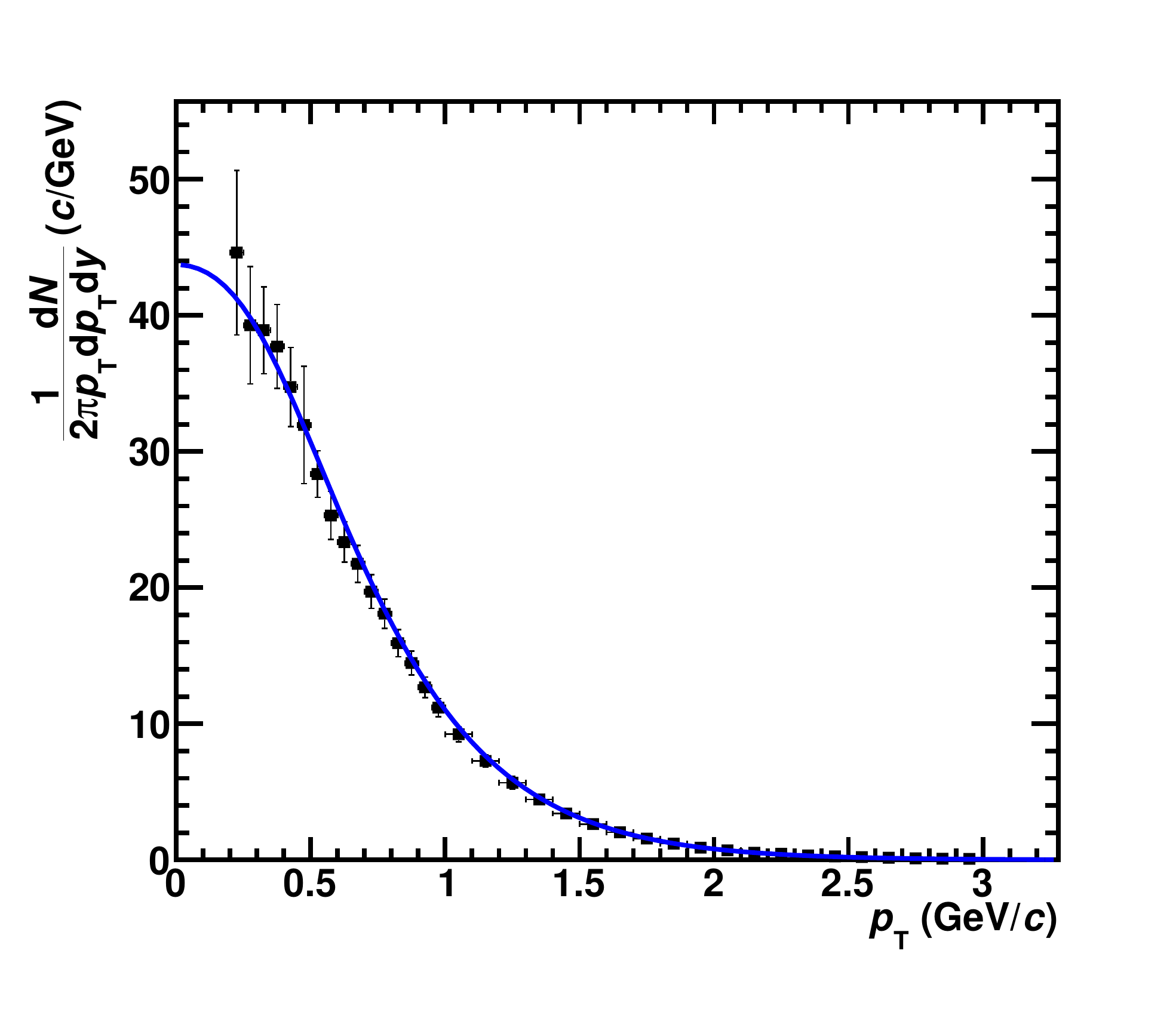}
  \includegraphics[width=0.6\textwidth]{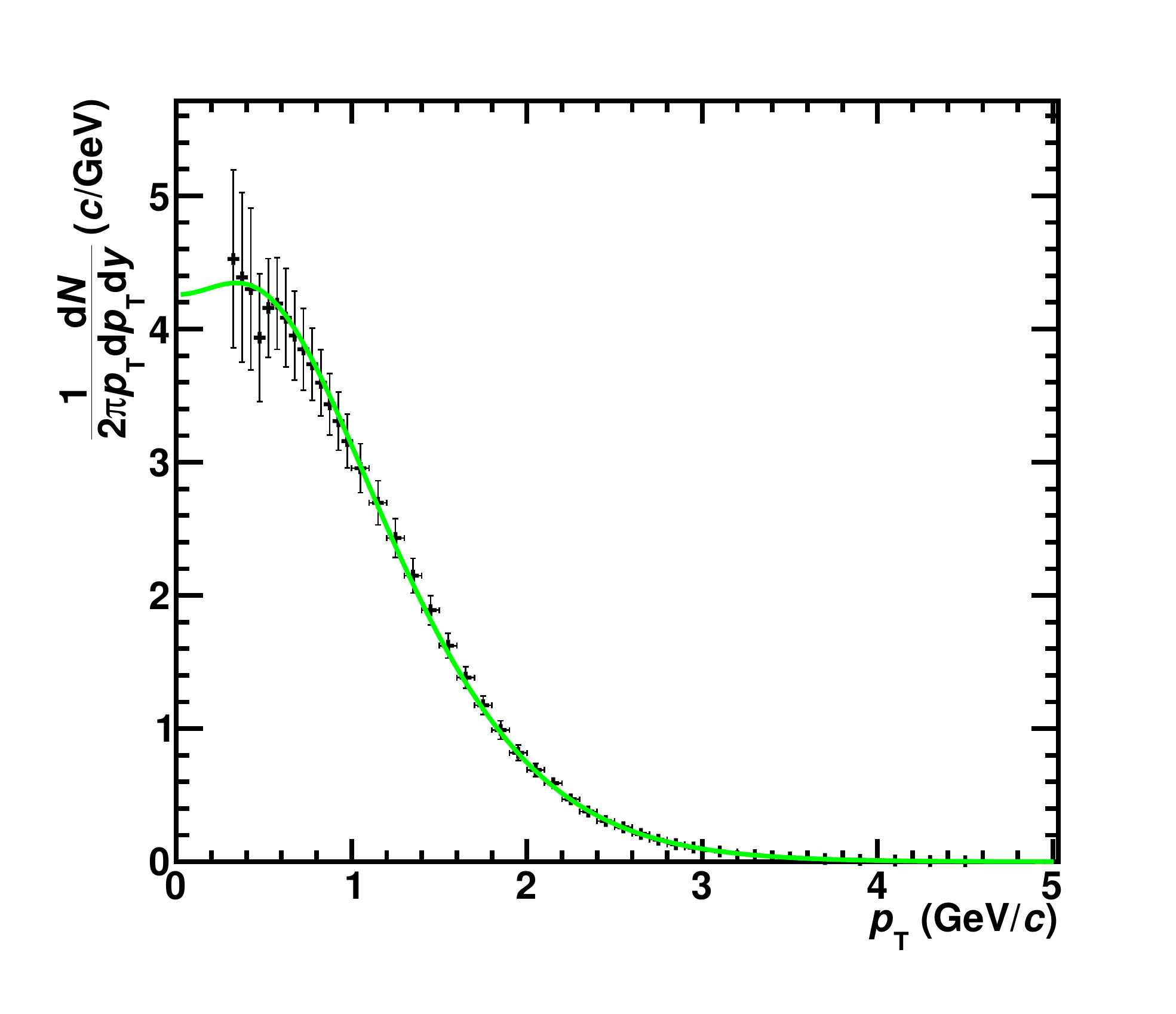}
  \caption{Blast-wave fit with a common set of parameters to pion (upper), kaon (middle), and proton (lower) ($\pi$, K, p) spectra simultaneously
in the 0-5\% centrality class. The data are from the ALICE Collaboration~\cite{pKpi_centrality} and the lines display the result of the global fit discussed there. For details see text.}
  \label{bw}
\end{figure}

\begin{figure}
\centering
\includegraphics[width=0.9\textwidth]{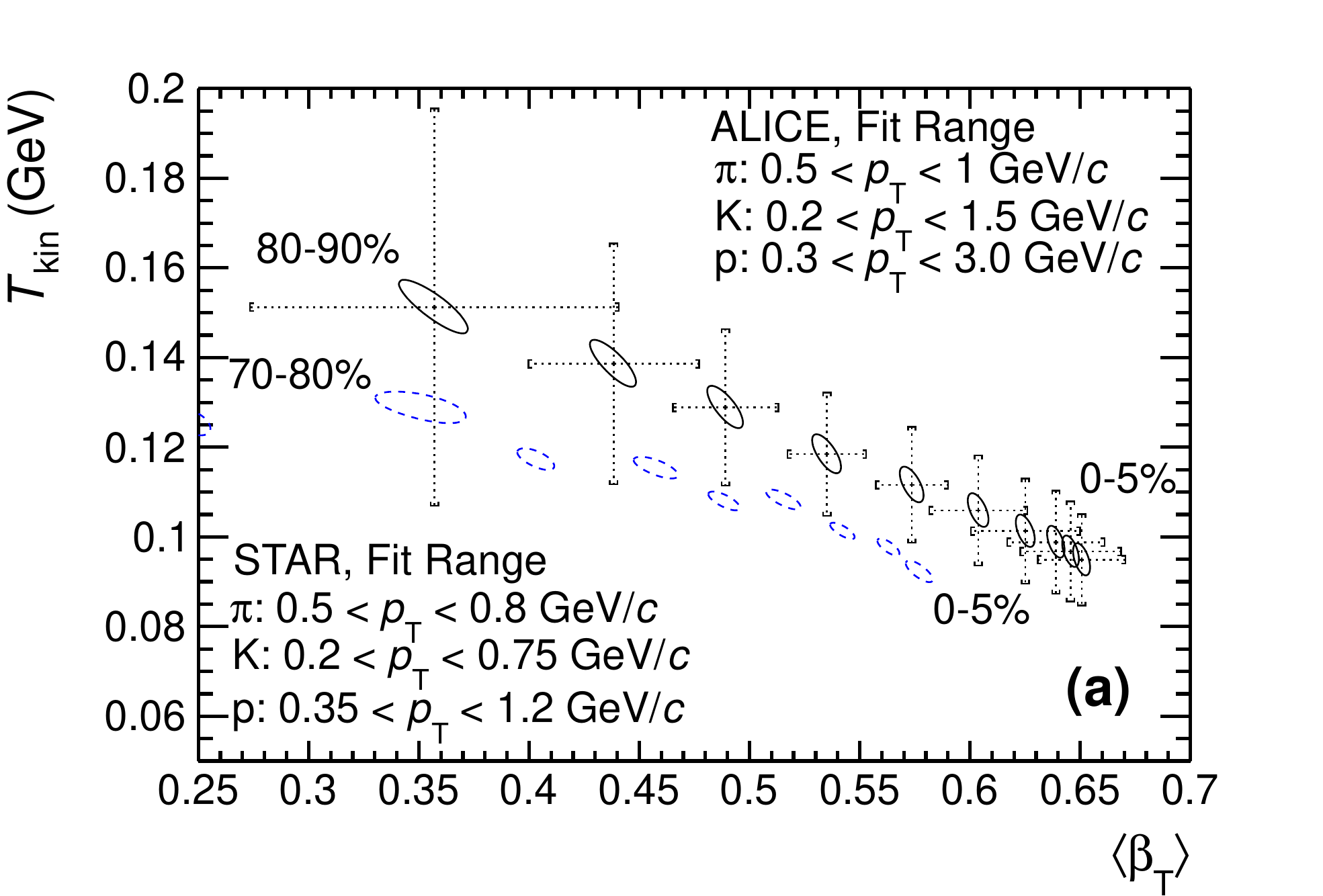}
  \caption{Resulting fit contours (1$\sigma$) for the kinetic freeze-out temperature $T_{\mathrm{kin}}$ and average transverse expansion velocity $\langle \upbeta_{\mathrm{T}} \rangle$ for different centrality bins measured in Pb--Pb collisions at LHC energy compared with the results of Au--Au collisions at RHIC. Figure from~\cite{pKpi_centrality}.}
  \label{bw_ex}
\end{figure}

A simplified version of the relativistic hydrodynamic approach is the blast-wave model within which  the collective expansion (in transverse direction) sketched above is described using a parameterized hydrodynamic flow field. 
It has three parameters: $T_{kin}$, $\beta$, $n$, i.e. the kinetic freeze-out temperature (introduced before as the temperature when the particles stop to scatter), a velocity parameter $\beta$ and a scale parameter $n$ to characterize the flow profile. A more complete description of this model can, e.g.,  be found in~\cite{Schnedermann:1993ws}.\\
The model assumes a spectrum of purely thermal sources which are boosted in transverse direction. The velocity distribution in $0 \leq r \leq R$ is assumed to be 
$$\beta_r=\left(\frac{r}{R} \right)^n \beta_s\,,$$
where $\beta_s$ is the surface velocity, a free parameter of the fit. In many applications, a linear profile is assumed and $n$ is fixed equal to unity. The quality of the fit can be improved if $n$ is considered as an additional free parameter. The resulting values for the kinetic freeze-out temperature $T_{kin}$ and $\beta_s$ are generally anti-correlated, see Fig.~\ref{bw_ex}. The so obtained spectral shape is a superposition of the contributions due to the individual thermal sources and is given by
\begin{equation}
\frac{1}{m_{\rm T}}\frac{{\rm d}N}{{\rm d}m_{\rm T}} \propto m_{\rm T}\int_0^R I_0 \left( \frac{p_{\rm T} \sinh \rho}{T_{kin}} \right) K_1 \left( \frac{m_{\rm T} \cosh \rho}{T_{kin}} \right) r\, {\rm d}r \,,
\label{equation:blastwave}
\end{equation}
where $I_0(x)$ and $K_1(x)$ are Bessel functions, $m_{\rm T}=\sqrt{p_{\rm T} ^2 + m^2}$ and $\rho = \tanh^{-1} \beta_r$. An example fit to pions, kaons and protons is shown in Fig.~\ref{bw} with a common parameter determined by the analysis of the 0-5\% centrality class of the data taken with the ALICE apparatus at $\sqrt{s_\mathrm{NN}} = 2.76$~TeV~\cite{pKpi_centrality}. The excesses at low momenta for the pions are due to feed-down from resonance decays (mainly $\rho(770) \rightarrow \pi^+\pi^-$) which are not yet included in the model. The model assumes boost-invariance which is near mid-rapidity quite well fulfilled at LHC energy. Note that Equation~\ref{equation:blastwave} is integrated over rapidity y. In principle, one should use a blast wave formula differential in $p_{\rm T}$ and y. This is briefly discussed in~\cite{Andronic:2018vqh}.\\
The  comparison of this fit with the previous results from the STAR Collaboration at RHIC is shown in Fig.~\ref{bw_ex} below. The ALICE Collaboration observes an approximately 10\% higher radial flow at LHC energies compared to that at RHIC~\cite{pKpi_centrality}.\\

Blast-wave fits allow a simple phenomenological description of spectra as the model parameters are fit to the data. The resulting distributions cannot describe the full collective properties. For this an approach using relativistic hydrodynamics including viscosity plus a detailed treatment of resonance decays is needed.
They nevertheless offer an economical way to study systematically the evolution of particle spectra with only three parameters. Additionally, they are often used to extrapolate measured particle spectra towards unmeasured $p_\mathrm{T}$-regions, namely towards low and high transverse momenta.\\

Equation~\ref{equation:blastwave} shows that the presence of transverse flow effectively leads to a characteristic modification of the spectral shape~\cite{Heinz:2004qz}. The collective flow increases the particle energies proportional to their rest mass $m_i$. Thus the spectrum at low momenta ($p_{\rm T} \ll m_i$) can be described with a correspondingly higher effective temperature $T_{\rm{eff}}$.
One directly obtains the expected scaling $T_{\rm{eff}}\approx T_{\rm{kin}} + \frac{1}{2}m_i \langle \beta_s \rangle ^2$ in the non-relativistic limit~\cite{Scheibl:1998tk}.
Another advantage of the blast-wave fits is given by the fact that the parameters resulting from a blast wave analysis determine a unique flow field which can then be used to estimate spectral shapes for other, not yet measured particles $i$ with a given mass $m_i$.
 
\begin{figure}[!htb]
\begin{center}
\includegraphics[width=0.8\textwidth]{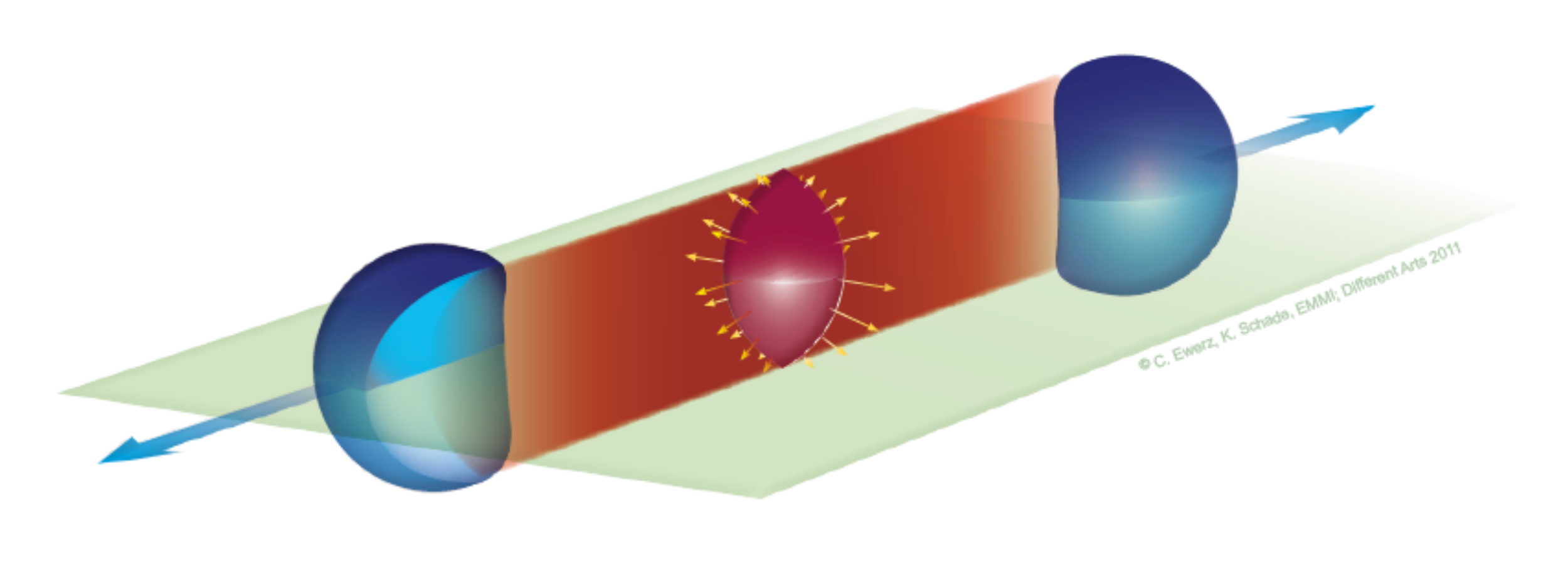}
\caption{\label{scheme_flow} Artistic view of a non-central heavy-ion collision of two nuclei, in blue. The overlap region of the collision has an almond shape as visible from the red fireball where the arrows indicate the expansion velocity. Taken from~\cite{diss_schade}.}
\end{center}
\end{figure}

Since the expansion for non-zero impact parameter collisions is generally anisotropic in azimuthal direction due to the almond shape of the overlap zone, see Fig.~\ref{scheme_flow}, one usually writes down the transverse momentum spectrum as a function of the azimuthal angle $\phi$ and expands the spectrum in form of a Fourier series as
\begin{equation} 
\label{flow_expansion}
  \frac{1}{p_{\mathrm{T}} }\frac{\dd ^{3}N}{\dd p_{\mathrm{T}}\dd y\dd \phi}=\frac{1}{2\pi p_{\mathrm{T}}} \frac{\dd ^{2}N}{\dd p_{\mathrm{T}}\dd y} \left\{ 1+2 \sum_{n=1}^{\infty} v_{n}(p_{\mathrm{T},}y) \cos[n(\phi-\Psi_R)]\right\},
\end{equation}
where the  Fourier coefficients in the sum are called flow coefficients $v_{n}$~\cite{Barrette:1994xr,Voloshin:1994mz,Voloshin:2008dg}. 

The second flow coefficient $v_2$ is usually referred to as the elliptic flow parameter. In a non-central heavy-ion collision this coefficient has a large contribution in the decomposition.
This can be understood from the Fig.~\ref{scheme_flow} showing a non-central collision which leaves an almond shape fireball in the overlap region. The spatial anisotropy visible as almond shape leads to a momentum anisotropy in the expansion of this fireball: the part of the almond with large curvature, lying in the so called reaction plane (indicated in green), will be pushed away stronger than the part being out of the reaction plane (in equation~\ref{flow_expansion} described by $\Psi_R$). This anisotropy is also indicated by the momentum arrows displayed in yellow.    
The $v_2$ coefficient exhibits a similar mass ordering as the transverse flow, namely an increase with mass of the measured $v_2$ as a function of $p_{\mathrm{T}}$. 

\subsection{Statistical hadronisation model} \label{sect:stat_hadron}
\subsubsection{Concepts}
Strong interactions are quantitatively described in the framework of
quantum chromodynamics (QCD) which describes the interactions among
the basic constituents of QCD, the coloured quarks and gluons. 
The corresponding QCD Lagrange density is hence formulated
entirely in terms of these fundamental particles. On the other hand,
all hadrons are colour-less. This makes a direct connection between the
QCD Lagrangian and hadron observables difficult. Here, the lattice QCD
framework, in which the Lagrangian is expressed in a thermodynamical
partition function $Z(V,T,\mu)$ comes to the rescue. Indeed, it
was realized recently \cite{PhysRevD.90.094503,Borsanyi2010},
that $Z$ can be well approximated with the partition function for
the hadron resonance gas, provided that the temperature $T$
stays below $T_c$, the transition temperature to the QGP. 

The partition function of the hadron resonance gas is, in this low
temperature, low density regime, usually evaluated in the
non-interacting limit~\cite{Andronic:2017pug}. Sometimes, repulsive interactions are modeled
with an 'excluded volume' prescription, see, e.g.,
\cite{Andronic:2012ut,Vovchenko:2016rkn,Vovchenko:2017drx} and references therein. As long as all hadrons have
the same excluded volume this correction leads to a reduction of the
total particle density but does not change the relative densities of
individual hadrons. In the references below the excluded volume
correction has been used exclusively in this spirit.  In the absence
of any reliable knowledge of such interactions we consider it
inappropriate to go further and use different excluded volumes for
specific hadrons, especially also since there are no experimental data
which would warrant such a step.

From this partition function of the hadron resonance gas all
thermodynamical quantities for hadrons can be computed. Specifically,
one can compute, for each hadron, its density $n(T,\mu,V)$. If all
hadrons are produced from a state of thermodynamical equilibrium then,
at a given beam or center-of-mass energy, the measured hadron yield
for hadron $j$, ${\rm d}N_j/dy$
at a given rapidity $y$ but integrated over transverse momentum, should be
reproduced as ${\rm d}N_j/dy = V \cdot n(T,\mu,V)$. In practice, a fit is
performed at each energy to the measured yield data to determine the 3 parameters $T,
\mub, V$. Note that $\mu_Q$ and $\mu_S$ are fixed by strangeness and
charge conservation.

Since 1994 a very large body of data on hadron yields produced in
ultra-relativistic nuclear collisions has been collected. From an
analysis of these data in the spirit of the above approach convincing
evidence has been obtained
\cite{BraunMunzinger:2001ip,BraunMunzinger:2003zd,Becattini:2005xt,Andronic:2005yp,Stachel:2013zma,Becattini:2016xct,Andronic:2017pug} 
that the yields of all hadrons produced in central (nearly head-on)
collisions can indeed be very well described, yielding the complete
energy dependence of the parameters $T, \mub, V$
\cite{Becattini:2005xt,Andronic:2005yp}, see in particular also the
recent fit to the precision LHC data \cite{Andronic:2017pug}. For 
recent reviews see \cite{Braun-Munzinger:2015hba,Andronic:2017pug}. Since the yields
of particles are frozen at these parameters the corresponding
temperature is also called chemical freeze-out temperature $T_{chem}$, as already indicated above.

Of particular interest for the present review is that the fit includes
also loosely-bound states such as the deuteron (and anti-deuteron),
and even the very weakly bound hyper-triton (and anti-particle). This
implies that particle production takes place at rather low
temperatures and densities (for LHC energy the temperature is
$T_{chem} = 156.5 \pm 1.5$ MeV, implying a total particle density $n_{tot} \leq
0.45~\mathrm{fm}^{-3}$). After chemical freeze-out the density must be even much lower for the loosely bound hyper-triton to survive, see the detailed discussion below.

\subsubsection{Application}

In the following we will present and discuss some examples of the analysis of particle production data using this model. The specific physics connected to the production of nuclei and hyper-nuclei will be discussed in sections~\ref{sect:nuclei} and~\ref{sect:hypernuclei} below.

\begin{figure}[!htb]
\begin{center}
\includegraphics[width=0.8\textwidth]{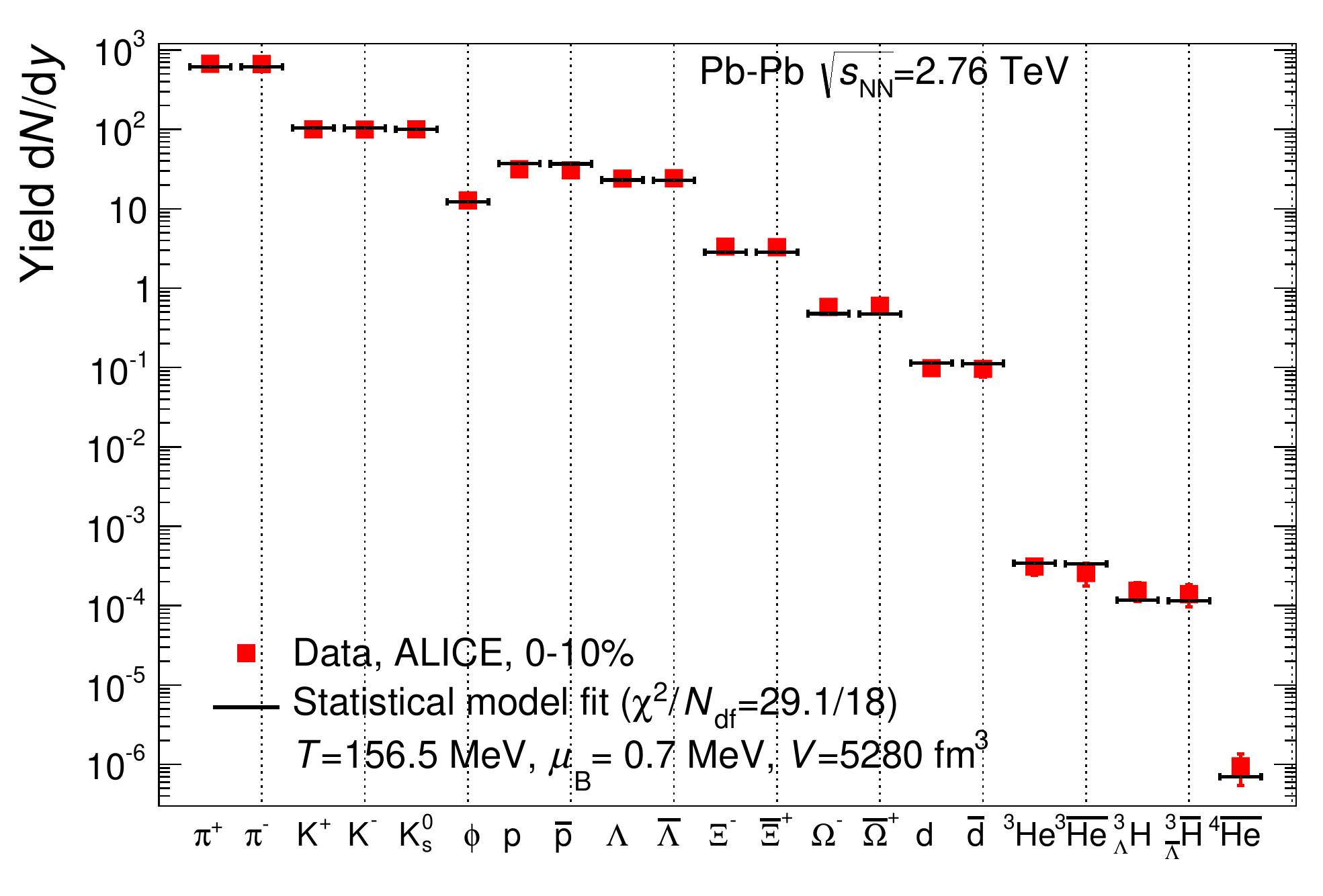}
\caption{\label{thermal_fit_lhc} Thermal model description of the production yields (rapidity density) of different particle species in heavy-ion collisions at the LHC for a chemical freeze-out temperature of 156.5 MeV (from~\cite{anton_sqm2016}).}
\end{center}
\end{figure}

We begin with the comparison of hadron production     
at the LHC with the statistical hadronisation model.  In Figure~\ref{thermal_fit_lhc} the result is shown of  a thermal model analysis of the data collected by the ALICE Collaboration using the GSI-Heidelberg model~\cite{thermalModel,pbm,pbm1,anton_thermal,Stachel:2013zma,anton_sqm2016,Andronic:2017pug}.  Very good agreement is obtained for $T_{chem} = 156.5 \pm 1.5$ MeV over the 9 orders of magnitude in particle  production yields.

At LHC energy, the baryo-chemical potential $\mu_B$ which is a measure of the difference of production probabilities for baryons and anti-baryons is expected to be close to zero, since the LHC c.m. energy exceeds twice the baryon mass by more than a factor of $10^3$. The value presented in the  figure from the fit  is $0.7 \pm 3.8$ MeV, in excellent agreement with this expectation. The nearly vanishing baryo-chemical potential leads to equal yields of baryons and anti-baryons and in consequence also to equal yields of nuclei and anti-nuclei for the different species.  This also implies that measurements of particle production at LHC energies are relevant for the understanding of the evolution of the early universe. In fact, different from the situation for nuclear collisions at LHC energy, the production of nuclei in the early universe can not happen when the baryons are produced because the photons, still in equilibrium with the baryons, would destroy all formed nuclei immediately. Thus, the formation of nuclei happens in the early universe at a much later time after the temperature has dropped sufficiently, such that no thermal photons are left to destroy the formed deuterons. From this point on, the process n + p $\rightarrow$ d + $\gamma$ is dominating the detailed balance, deuterons are produced and the backward reaction is energetically suppressed.  

\begin{figure}[!htb]
\begin{center}
\includegraphics[width=0.7\textwidth]{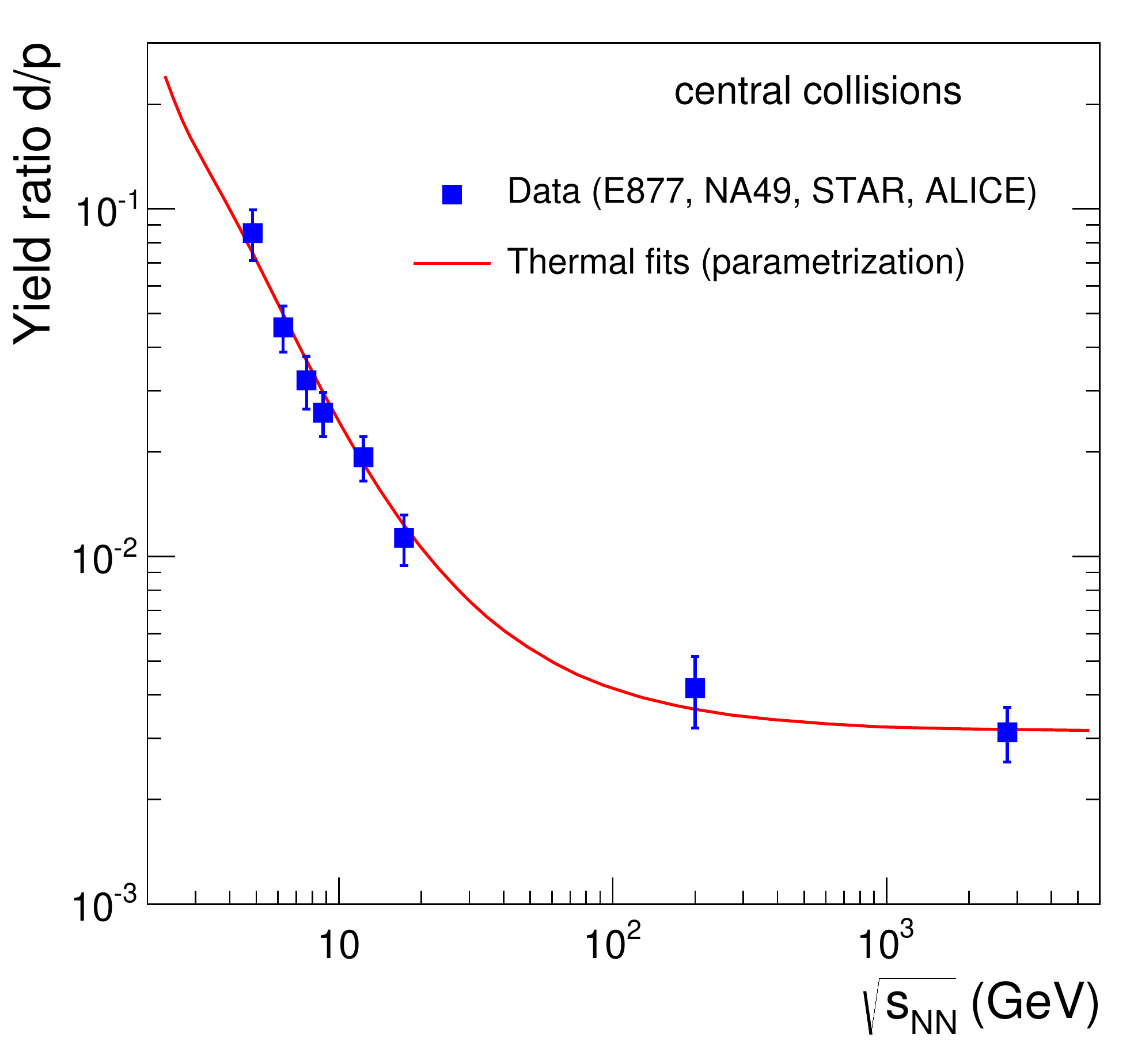}
\caption{\label{d_to_p_sqrts} Deuteron-to-proton ratio as measured in central nuclear collisions at different centre-of-mass energies $\sqrt{s_{\mathrm{NN}}}$. The data points are compared with predictions based on the thermal model (parameterised in the red line).}
\end{center}
\end{figure}

Since, in this review, we are in particular interested in loosely-bound states we show in Figure~\ref{d_to_p_sqrts} the deuteron-to-proton ratio in relativistic nuclear collisions as a function of centre-of-mass energy, bridging data from the SPS to RHIC to the LHC. Assuming thermal production of deuterons according to the particles mass and spin reproduces the data very well, implying that the statistical hadronisation model is a useful tool to estimate production yields also for loosely-bound states as developed in \cite{pbm,pbm1,thermalModel}.
The application of the parameterization of the energy-dependence of $T_{chem}$ and $\mub$~\cite{Becattini:2005xt,Andronic:2005yp} within the framework of the statistical hadronisation model leads to an impressive description of all hadron production data. In fact, yields for the production of loosely-bound states at LHC energy were successfully predicted in  \cite{thermalModel} before data taking. This shows that the production of nuclei is quantitatively well reproduced within the framework of the statistical hadronisation model, implying that the same parameters ($T_{chem}, \mub, V$) governing light hadron production yields also determine the production of  light composite objects, with only the particle mass and quantum numbers and not structural parameters such as binding energy or radius as input.
 
\begin{figure}[!htb]
\begin{center}
\includegraphics[width=0.7\textwidth]{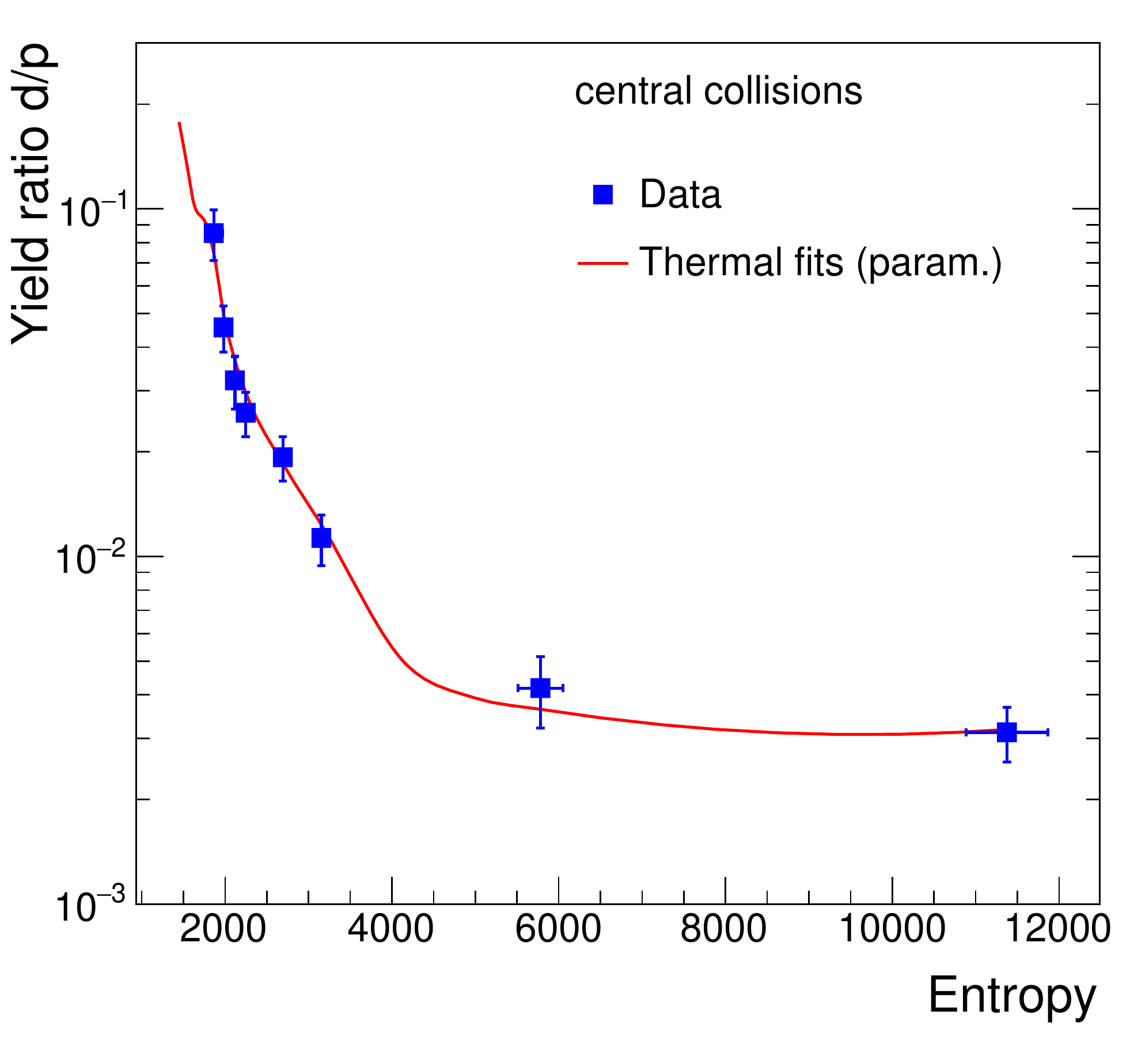}
\caption{\label{d_to_p_entropy} Deuteron-to-proton ratio as measured in central nuclear collisions versus produced entropy. The same data points as in Figure~\ref{d_to_p_sqrts} are compared to calculations of the entropy obtained using the thermal model (red line). }
\end{center}
\end{figure}

Another way to look at the deuteron-proton ratio is displayed in Figure~\ref{d_to_p_entropy} extracted from the thermal model~\cite{Andronic:2005yp}. In this Figure, the d/p ratio is shown as function of the entropy per unit of rapidity in the collision. As naively expected, increasing the entropy leads first to a precipitous drop of the ratio, as the entropy/baryon scales $\propto -\ln{(\mathrm{d/p})}$, \cite{Siemens:1979dz,Nagamiya:1984vk}. Above $\sqrt{s_{\mathrm{NN}}} \approx 20$ GeV the chemical freeze-out temperature saturates at around 160 MeV, implying that the entropy density stays constant. The main entropy increase is then due to the volume expansion of the fireball at freeze-out, implying that the d/p ratio approaches a constant value of $\approx  3 \cdot 10^{-3}$. 

No detailed microscopic description for the production of loosely-bound objects exists to-date. The results for deuteron production and, in fact, for the production of other loosely-bound states, see below, could be connected with the assumption that the total entropy is conserved after chemical freeze-out at each collision energy ($\sqrt{s_{\mathrm{NN}}}$). This would imply a very dilute phase directly after chemical freeze-out. We will discuss another possibility, especially keeping in mind that also very extended objects such as the hyper-triton need to be considered. 

\subsection{Coalescence model}

A different approach for the production of composite objects such as deuterons and light nuclei in nuclear and hadronic collisions is the coalescence model. It was first established for the description of data collected at the proton synchrotron at CERN, when for the first time a 25 GeV proton beam was used to study particle production in collisions with  a variety of different targets~\cite{Cocconi:1960zz}. In view of the surprisingly large cross sections observed for deuteron production in p-nucleus collisions
 a mechanism was proposed~\cite{butler_pearson61,butler_pearson63}, in which deuterons are formed by protons and neutrons which are close in phase-space. This picture was further developed to describe the yields of clusters in heavy-ion collisions at different energies. The first time it was used in heavy-ion collisions was at the Bevalac at Lawrence Berkeley Laboratory starting in the 70s~\cite{PhysRevLett.37.667, Gosset:1976cy,PhysRevC.24.971,Nagamiya:1984vk,PhysRevC.28.1552,PhysRevLett.74.2646}. It was further used as the model applied to data obtained at the Alternate Gradient Synchrotron (AGS) at Brookhaven National Laboratory (BNL) where several different experiments (E802/E866, E814, E864, E877, E878) have results on the production of light nuclei~\cite{PhysRevC.49.3211,PhysRevC.50.1024,PhysRevC.50.1077}. Furthermore, at the CERN SPS it was used for the interpretation of heavy-ion data at three different experiments  (NA44, NA49, NA52)~\cite{SIMONGILLO1995483,Bearden:1999hv,Bearden:1999iq,PhysRevLett.69.2345,Afanasiev200022,
PhysRevC.85.044913,
PhysRevC.69.024902,
PhysRevC.94.044906,Appelquist:1996qy,0954-3899-28-7-347,Ambrosini:2000qk,Arsenescu:2003mf,1367-2630-5-1-150}. 
The model was also successfully applied to describe the yields of nuclei at RHIC~\cite{star,Agakishiev:2011ib,Adler:2001uy,Abelev:2009ae,Afanasiev:2007tv,Nygaard:2007zz,Arsene:2010px}.   

In the following, we briefly summarise important aspects of this approach. 
An empirical coalescence model based on the above pioneering publications was developed for the analysis of light nucleus production data from relativistic nuclear collisions at the Berkeley Bevalac, see, e.g., the review in~\cite{Csernai:1986qf} and references given there. Such collisions typically lead to the complete disintegration of the overlap zone of the colliding nuclei into their constituent nucleons. In such a situation, the production cross-section of a light nucleus with mass number $A$ is given by the probability that $A$ of the 'produced' nucleons have relative 
momenta less than an empirical parameter $p_0$, to be determined by comparison with measured yields.  
This model relates the production cross-section of the (light) nucleus, having a momentum $p_A$, to a scaled power  of the production cross-section for nucleons (in practice protons since neutrons are typically not measured)  which have a momentum $p_{\mathrm{p}}$:
\begin{equation}
E_A \frac{\mathrm{d}^3N_A}{\mathrm{d}^3p_A} = B_A \left(E_{\mathrm{p}} \frac{\mathrm{d}^3N_{\mathrm{p}}}{\mathrm{d}^3p_{\mathrm{p}}}\right)^A,
\label{eq:coal}
\end{equation}
whereby $p_A = Ap_{\mathrm{p}}$. This leads to the interesting fact that, for a given nucleus, the coalescence parameter $B_A$ should not depend on momentum or centrality of the collision but only on the cluster parameters:
\begin{equation}
\label{eq:coal_mass}
B_A =  \left( \frac{4 \pi}{3} p_0^3 \right) ^{(A-1)} \frac{M}{m^A}
\end{equation}
where $M$ and $m$ are the nucleus and the proton mass, respectively, and $\frac{4 \pi}{3} p_0^3$ is the coalescence volume in momentum space. With this approach a reasonable description of the Bevalac data was obtained, see for instance~\cite{Csernai:1986qf}.
In fact, already at the Bevalac measurements it was observed that using this formalism one gets different coalescence radii $p_0$ for different nuclei (d,t,$^3$He). They differ by about 20-30\% for the different species.
Equation \ref{eq:coal_mass} actually demonstrates an independence on the momentum of the particles and in fact at the Bevalac and in elementary collisions such a behaviour is observed, namely $B_A$ is little dependent on the collision energy and  the multiplicity in the events.

The coalescence model approach can also be connected with a thermodynamic treatment, as for instance discussed in~\cite{PhysRevLett.38.640,PhysRevC.17.1051,MEKJIAN1978491,DasGupta:1981xx,Kapusta:1980zz}. From this approach one can get the proportionality
\begin{equation}
\label{eq:coal_volume}
B_A \propto  \left( \frac{1}{V}  \right)^{(A-1)}
\end{equation} 
where $V$ is now the volume in coordinate space. Thus often in the first approaches one either used a coordinate space or a momentum space approach.

In the 1990ties, data obtained at the Brookhaven AGS and CERN SPS accelerators at much higher energies 
provided, however, clear evidence for a momentum and centrality/multiplicity dependence of $B_A$. Furthermore, it was realised that the production of bound objects from their free constituents violates energy and momentum conservation. To address these issues and to provide a more systematic theoretical description of the coalescence process, new approaches were developed which also took into account the temporal evolution of the fireball formed in the collision, see., e.g. ~\cite{Roepke-1990,Danielewicz:1992pei,Scheibl:1998tk,Beyer:1999xv}. 

Nevertheless, for a full model description one has to calculate the coalescence process itself which has several drawbacks.
The transverse kinetic energies of particles produced in ultra-relativistic heavy-ion collisions lie with some hundreds of MeV to several GeV significantly above the relevant binding energies of the multi-baryon objects (2.2 MeV for deuterons, 8.48 MeV for tritons and 7.72 MeV for $^3$He nuclei). This fact is used in transport models (e.g. UrQMD) to argue that one can neglect the structure and intrinsic dynamics of such nuclei altogether. To describe the production of nuclei one usually uses different classes of coalescence models:

\begin{itemize}
\item Momentum-space coalescence: pure momentum-space analysis is done and particles below a lower limit, smaller as a given cut-off momentum are treated as part of a (formed) nucleus~\cite{Schwarzschild:1963zz,Sato:1981ez,PhysRevC.44.1636,PhysRevLett.37.667,Botvina:2014lga,Botvina:2016wko,Botvina:2017yqz},
\item Phase-space coalescence: analysis in momentum- and coordinate-space \cite{PhysRevC.31.1770,BALTZ19947,PhysRevLett.73.1219},
\item Phase-space coalescence with treatment of the potential forces: coordinate-space and momentum-space parameters are related to known or predicted potentials of the bound states~\cite{GARCIA1994597}, 
\item Generalised phase-space coalescence: projection of the n-particle phase-space-distribution in the final state onto a corresponding multi-particle wave function of the bound state~\cite{REMLER1975295,REMLER1975455,REMLER1981293,GYULASSY1983596,PhysRevLett.58.1926,PhysRevC.35.1291,PhysRevLett.74.2180},
\item Models using a statistical fragmentation: assuming a chemical and thermal equilibrium~\cite{PhysRevLett.38.640,PhysRevC.17.1051,MEKJIAN1978491,0305-4616-7-10-006,HAHN1988718},
\item Special case (very often understood in the heavy-ion community as coalescence, since it is also applicable for hadron formation): Coalescence from quarks instead of nucleons: usually connected to the description of flow in heavy-ion collisions~\cite{PhysRevLett.90.202302,Greco2004202,PhysRevC.80.064902,PhysRevC.92.064911,exhic,exhic1,Cho:2017dcy,Sun:2016rev,Sun:2017ooe}.
\end{itemize}

All these model implementations have still a profound problem, they violate momentum and energy conservation since they assume a formation of a nucleus from its nucleons without any possible recoil partner. 
If one assumes a dense medium after the chemical freeze-out one could easily find a partner to conserve energy and momentum in the coalescence process. This is not the case as can be seen by the particle yields which should have different spectra and thus much lower yields if a dense and interacting medium would exist throughout the expansion and further cooling  period following the chemical freeze-out. In fact, the notion of chemical freeze-out implies a phase of non-equilibrium (see for instance~\cite{Beyer:1999xv}) for temperatures below $T_{chem}$, which fits nicely to the the observations coming from the statistical thermal model that the entropy per baryon is fixed at chemical freeze-out and  thus the yields of nuclei are not modified while the system is expanding isentropically.

In addition, a full treatment would need a detailed knowledge of the wave function of the nuclei under consideration. A recent discussion is for instance given in~\cite{Sun:2017ooe}. In this approach, the coalescence yield is proportional of the square of the n-body-wave function of the state formed by coalescence, which is usually approximated by a Gaussian function, which is far away from the true distribution (although a recent study showed that the usage of a more realistic function, i.e. the Hulth\'{e}n wave function leads to similar results~\cite{Nagle:1996vp}). In practice this n-body-wave function is adjusted such that the corresponding rms radius ($\sqrt{\langle r^2\rangle}$) agrees with the size of the nucleus of consideration. This is still a crude approximation but at least takes account of the global parameters such as reduced mass and binding energy. 

The different binding energies are also reflected in the rms radii which are summarised in Tab.~\ref{tab:radii}, where the the measured radii of the light nuclei are taken from~\cite{nuclei:datatable} and the hypertriton rms radius comes from a theoretical calculation of the wave function discussed in~\cite{Nemura:1999qp}. One sees that the radii drop by 28\% going from deuteron to $^4$He, whereas the hypertriton rms radius is at least a factor 2.2 larger than the deuteron.

\begin{table}
\begin{center}
 \begin{tabular}{l | c }
    \hline
    {\bf nucleus} & {\bf rms radius (fm)}  \\
    \hline
    deuteron & 2.1421$ \pm $0.0088 \\
    triton & 1.7591$ \pm $0.0363 \\
    $^3$He & 1.9661$ \pm $0.0030 \\  
    $^4$He &  1.6755$ \pm $0.0028\\
    $^3_\Lambda$H   &  4.9\\
    \hline 
  \end{tabular}
 \caption{\label{tab:radii}Radii of light (hyper-)nuclei. All from~\cite{nuclei:datatable}, except the value of the hypertriton which is not measured but obtained from a calculation~\cite{Nemura:1999qp}.}
\end{center}
\end{table} 

These considerations notwithstanding, most actual data analyses are based on the simple momentum space coalescence picture. In~\cite{Shah:2015oha}, e.g.,  the authors use this approach to describe the apparent thermal ordering observed for the production of light nuclei at different RHIC energies. In this approach, the exponential mass dependence with a parameter $p$ (usually named penalty factor) is introduced indirectly through the coalescence parameter $B_A$. To reach such an exponential behaviour, which one could call thermal-like, one has to put in by hand a thermal distribution of the nuclei (which is already the case if you assume a blast-wave distribution for the baryons you use in the coalescence calculation). If this is not done one cannot easily reproduce the described observations, while the exponential behaviour comes out naturally in a thermal model as discussed in section~\ref{sect:stat_hadron}.  

A slightly different and rather new approach~\cite{Mrowczynski:1987oid,Mrowczynski:1989jd,Scheibl:1998tk,Mrowczynski:2016xqm,Bazak:2018hgl} uses the size of the fireball to cope with the above mentioned centrality dependence of the $B_A$ which was observed first at the AGS and the SPS. The size of the fireball is typically measured in high-energy collisions by the technique proposed by Hanbury Brown and Twiss in 1956~\cite{hbt54,hbt56a,hbt56b} to estimate the size of stars. This technique was then applied to elementary collisions by Goldhaber et al. in 1960~\cite{GoldhaberHBT} and is nowadays one of the first physics measurements in heavy-ion collisions, since the measurement can be done using small statistics and using only charged pions which are abundantly produced. The measurement uses the fact that one can construct a correlation function from the two bosons, in the latter case pions, from calculating only their relative momenta. Quantum mechanics encodes the spatial information in this quantity by interference effects. The measurement is widely called intensity interferometry and by some model assumptions one can extract the spatial extension and the time evolution of the fireball. The spatial extension is usually identified by the volume of homogeneity. For an informative review see~\cite{Baym:1997ce}.

This is used as an input for the model by Scheibl and Heinz~\cite{Scheibl:1998tk}, where they develop a coalescence approach from phase-space and quantum mechanical aspects of nuclei formation based on~\cite{Danielewicz:1992pei}. They end up at a formalism which takes into account the probability of formation depending on the size of the fireball, whose expansion is modeled in a semi-realistic way. This allows to predict the $B_A$ as a function of transverse momentum, or as it is done in the paper as function of transverse mass $m_\mathrm{T} = \sqrt{p_\mathrm{T}^2 + m^2}$.

The approach by Scheibl and Heinz~\cite{Scheibl:1998tk} was used by Blum et al.~\cite{Blum:2017qnn,Blum:2017iwq} to estimate the production probability of anti-nuclei (anti-deuterons and anti-$^3$He) by cosmic-ray interactions. They calculate the $B_A$ from all existing data and from that the production probability of anti-matter in the universe by standard processes. This production mechanism leads to background for anti-matter production from exotic processes such as decay of heavy dark matter particles constructed to explain the anti-matter candidates observed in the AMS02 experiment~\cite{BATTISTON2008227,PhysRevLett.110.141102}, for more details see~\cite{PhysRevD.62.043003,Donato:2008yx,Brauninger:2009pe,Kadastik:2009ts,1475-7516-2005-12-008,Fornengo:2013osa,Ibarra:2012cc,Aramaki:2015pii,vonDoetinchem:2015yva}.  

\section{Results of (multi-)strange baryon production measurements}
\label{strange_baryons}

\begin{figure}[!htb]
\begin{center}
\includegraphics[width=0.7\textwidth]{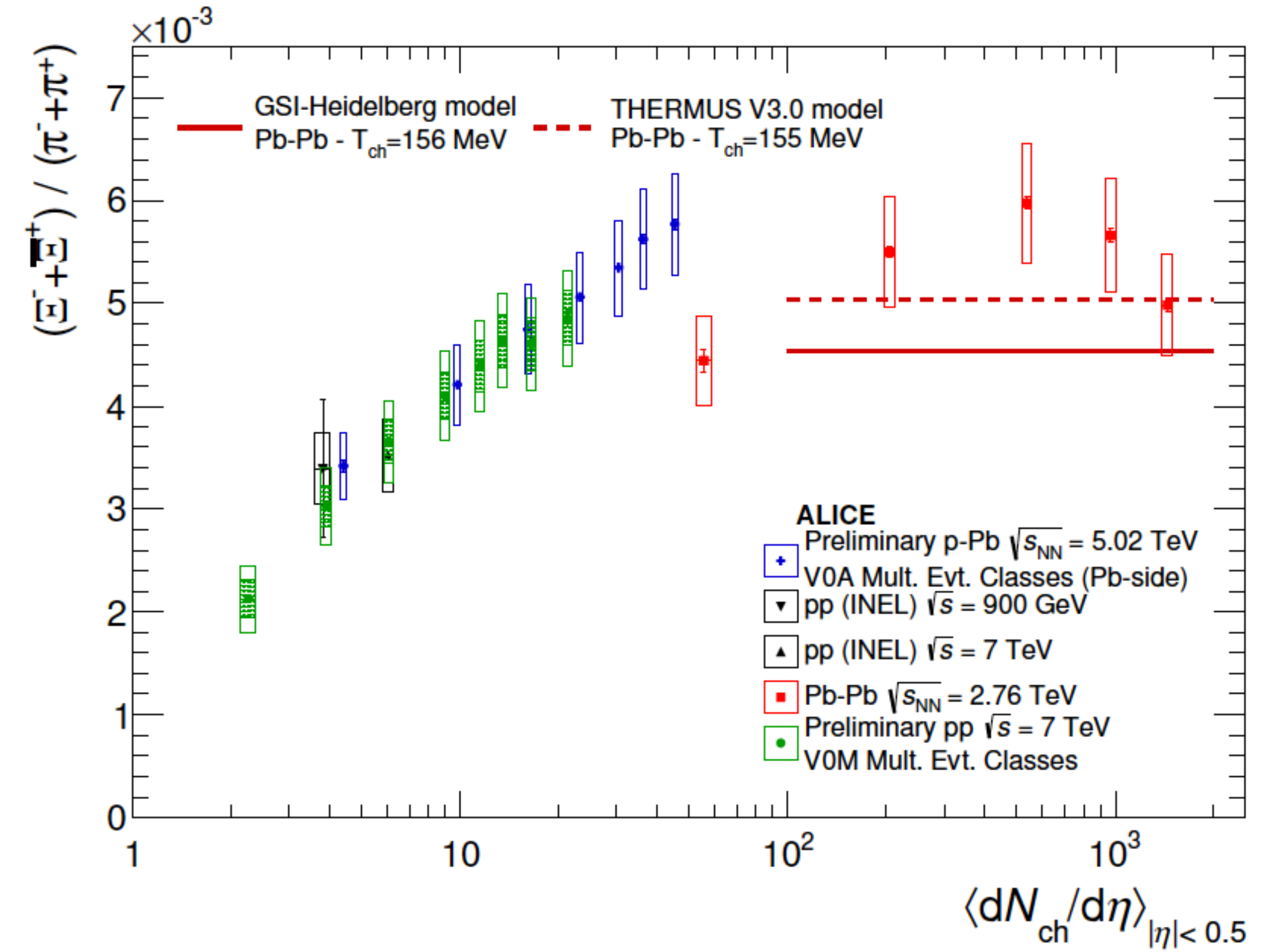}
\caption{\label{xi_over_pi} Ratio of ($\Xi^{-} + \bar{\Xi}^{-}$) over ($\pi^{-} + \pi^{+}$) vs. multiplicity for different collision systems. The red horizontal lines are from slightly different thermal model implementations. From~\cite{Bianchi:2016szl}.}
\end{center}
\end{figure}

An interesting observation is described in this section, obtained by comparing the results for strangeness production for different collision systems, namely pp, p--Pb and Pb--Pb. Since the system size dependence of particle production is relevant for the understanding of the production mechanism of loosely-bound nuclei and hyper-nuclei,  this is connected with the main topic of the review. The charged particle multiplicities reached in minimum bias pp collisions are about d$N_\mathrm{ch}$/d$\eta \approx 6$, whereas high multiplicity events in p--Pb collisions lead to d$N_\mathrm{ch}$/d$\eta \approx 45$ and attain d$N_\mathrm{ch}$/d$\eta \approx 1500$ in central (0-5\%) Pb--Pb collisions. This means one can span three orders of magnitude in multiplicity at the LHC going from one system to the other. When one uses the data in the different systems and does a careful study for instance of the production yield ratio of $\Xi^-$ to $\pi^-$ versus the multiplicity, one can see~\cite{multistrangePPB} a rather smooth increase in the ratio going from pp to p--Pb until an approximately constant plateau is reached in Pb--Pb collisions. This is displayed in Figure~\ref{xi_over_pi} as the ratio of the sum of $\Xi^{-}+\bar{\Xi}^{+}$ and $\pi^{-}+\pi^{+}$. This can be qualitatively understood as a lifting of the canonical strangeness suppression, valid for pp and p--Pb collisions, until the suppression becomes completely lifted in peripheral Pb--Pb collisions and the grand-canonical limit is reached~\cite{multistrangePPB,hamieh2000canonicalSuppresion,redlich2002canonicalSuppresion}. In detail, the canonical suppression effect is not fully understood, as the $\phi$ meson which carries not net strangeness but is dominantly an $s \bar s$ state, is also following the suppression trend. This could imply that the $\phi$ meson is dominantly made at the phase boundary from uncorrelated $s$ and $\bar s$ quarks similar to the observations in the charm sector~\cite{Andronic:2017pug}.

The ALICE Collaboration did a detailed study with fine multiplicity bins for all accessible strange hadrons here~\cite{Donigus:2017gom,ALICE:2017jyt}. The main focus of this work is to establish the trends with associated multiplicity  of the production of strangeness in elementary collisions.

\section{Recent results of (anti-)nucleus production measurements}
\label{sect:nuclei}

\begin{figure}[!htb]
\begin{center}
\includegraphics[width=0.7\textwidth]{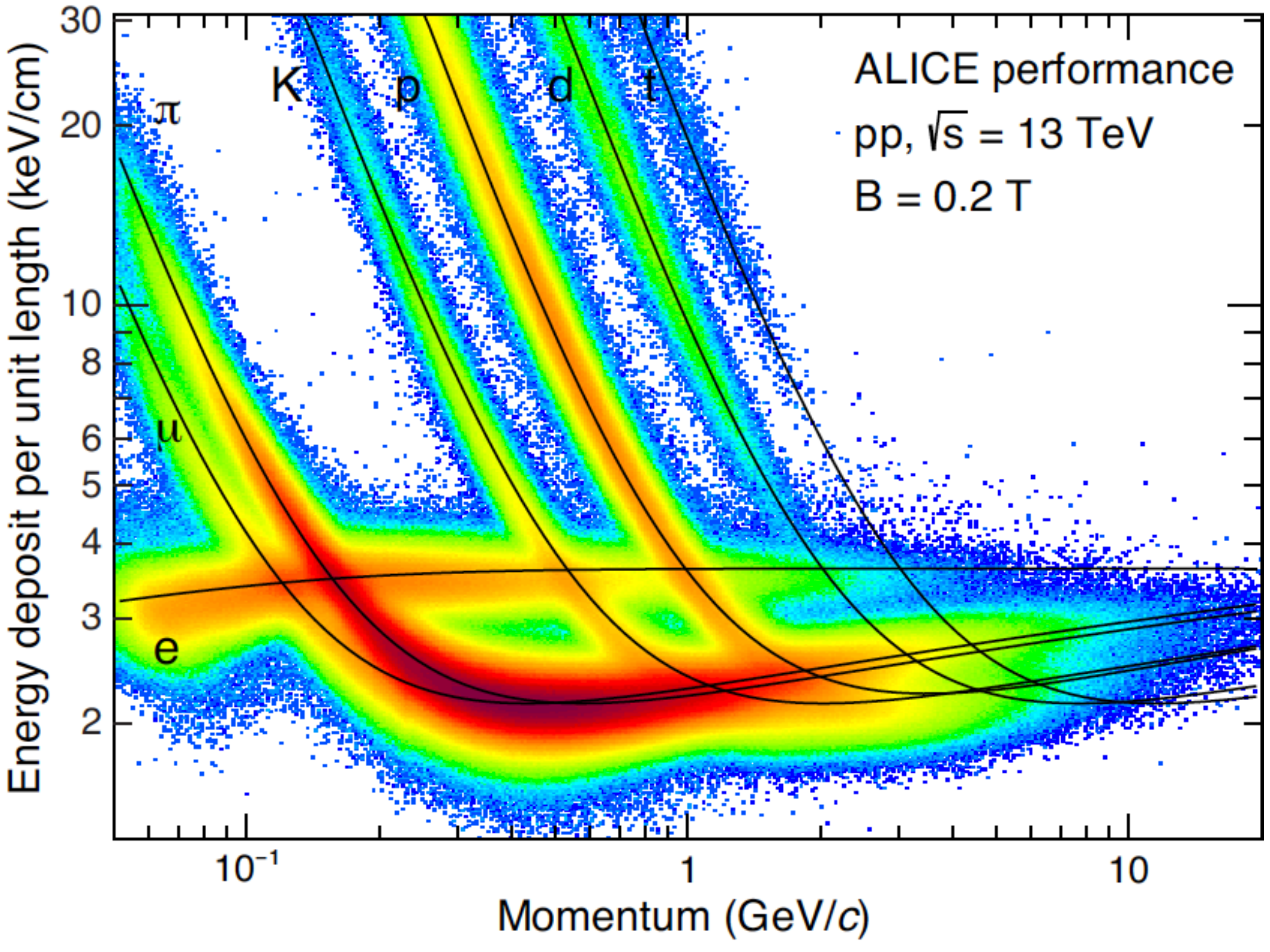}
\caption{\label{bethe-bloch-13tev} Energy deposit per unit length of each track versus rigidity $p/z$ in the TPC shown for a dedicated run of the 13 TeV data taking with a magnetic field of 0.2 T applied. Plot prepared for~\cite{Olive:2016xmw}.}
\end{center}
\end{figure}

\begin{figure}[!htb]
\begin{center}
\includegraphics[width=0.49\textwidth]{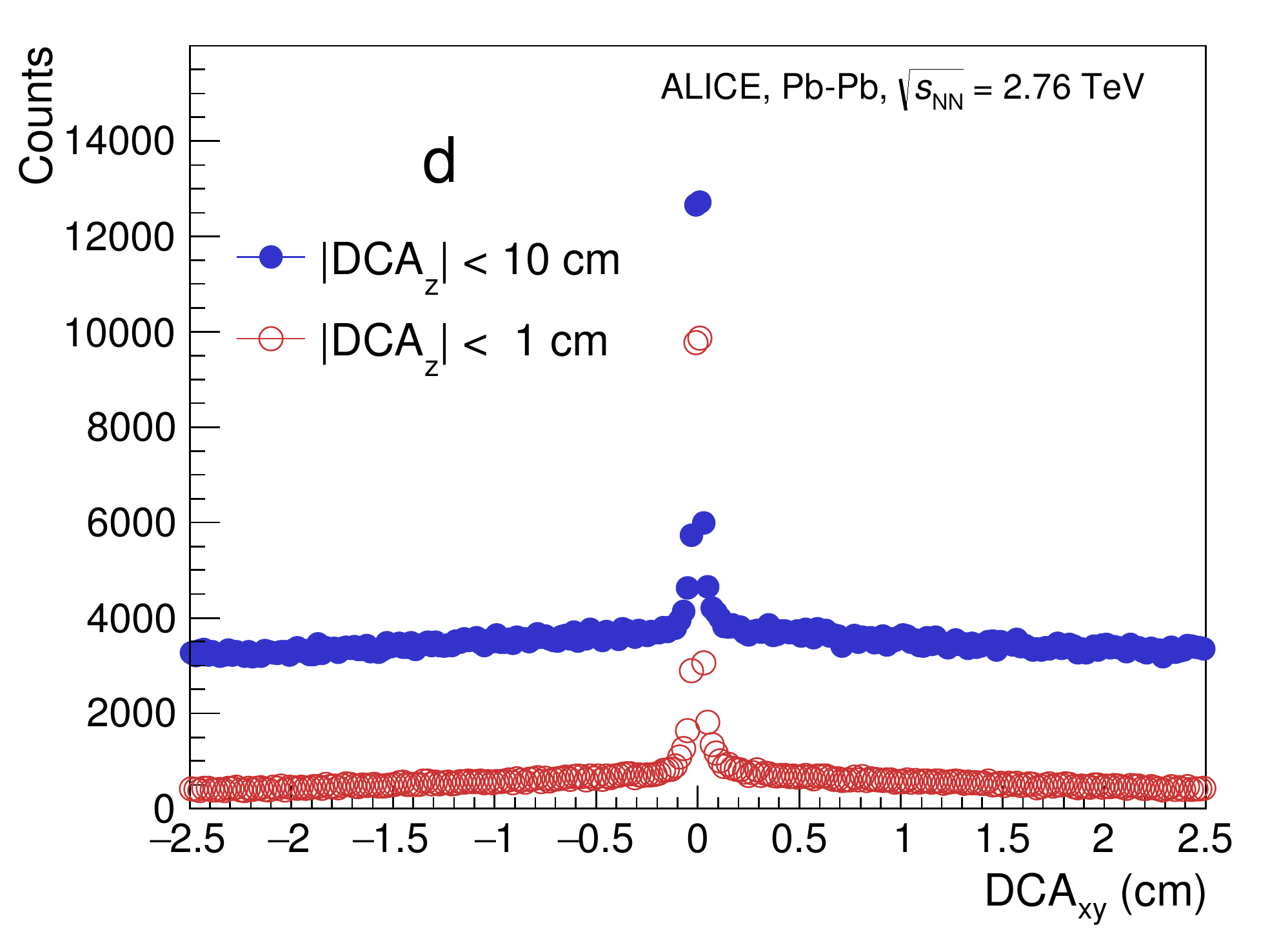}
\includegraphics[width=0.49\textwidth]{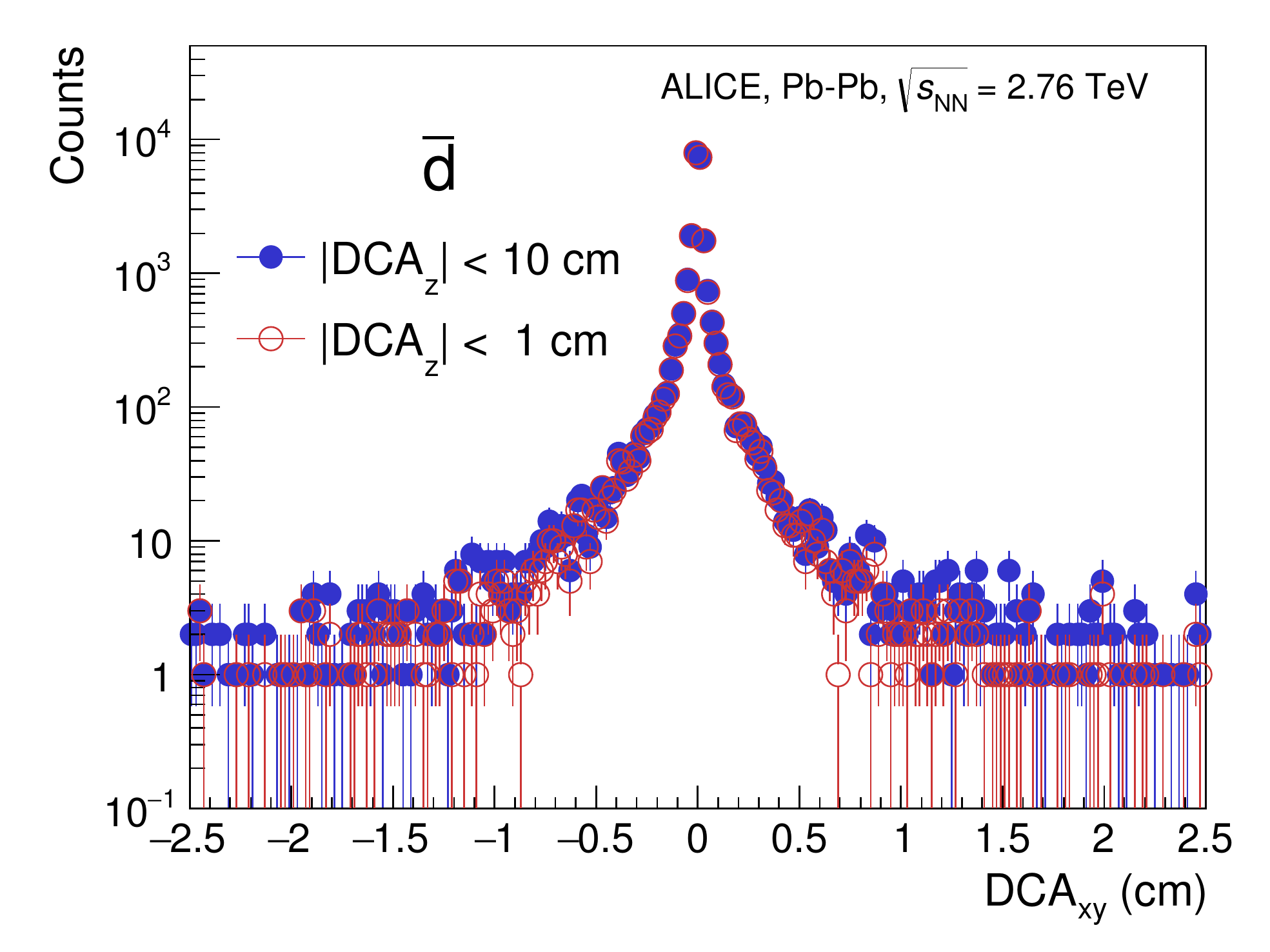}
\caption{\label{dca_xy} Distribution of the Distance-to-Closest Approach (DCA$_{xy}$) for deuterons (left) and anti-deuterons (right) in the transverse plane (${xy}$), with two different cut values on the DCA$_z$ applied. The y-axis for anti-deuterons is shown in a logarithmic scale to allow for the visibility of the difference between the two DCA$_z$ cuts. Figure taken from~\cite{nuclei}. For details see text.}
\end{center}
\end{figure}

Light nuclei and anti-nuclei such as the deuteron, $^3$He and the triton are loosely-bound objects, with binding energies and, in particular, nucleon separation energies much smaller than $T_c$. For mass number $A \leq 3$ they are rather copiously produced at LHC energies and can be directly measured and identified with the ALICE detector. Figure~\ref{bethe-bloch-13tev} shows the signal in the TPC versus rigidity for particles with charge number 1. For momenta less than 2 GeV/c the lines for deuterons and tritons are well separated from those of the lighter hadrons. Experimentally it turns out to be much easier to measure the production of anti-nuclei as nuclei are produced also by knock-out processes from secondary particles traversing the beam-pipe and the detector material of the ITS.  In fact, since the beam pipe is made from Beryllium, the main process releasing the light nuclei observed are from spallation processes due to pions interacting with detector material. Such processes lead to the production of nuclei such as  $^{7}$Li, $^{4}$He, $^{3}$He and d;  these particles are mainly observed at low momentum values. Such  "knock-out" nuclei have to be separated from the nuclei produced in the fireball of the initial collision. The selection criterion to suppress these nuclei is the so-called distance-of-closest-approach (DCA) in the beam direction (DCA$_z$) and in transverse direction (DCA$_{xy}$). An example of such DCA$_{xy}$ distributions for nuclei and anti-nuclei is shown in Figure~\ref{dca_xy} for two different DCA$_{z}$ cut settings ($|$DCA$_z| < $10 cm and $|$DCA$_z| < $1 cm). For the deuterons (left panel) one clearly sees a strong reduction of counts when changing the cut. For anti-deuterons, which are shown on the right panel, the reduction is much smaller and only becomes visible if a logarithmic scale is used, since there is much reduced background on the anti-matter side compared to the strong effect on the matter side.

\begin{figure}[!htb]
\begin{center}
\includegraphics[width=0.7\textwidth]{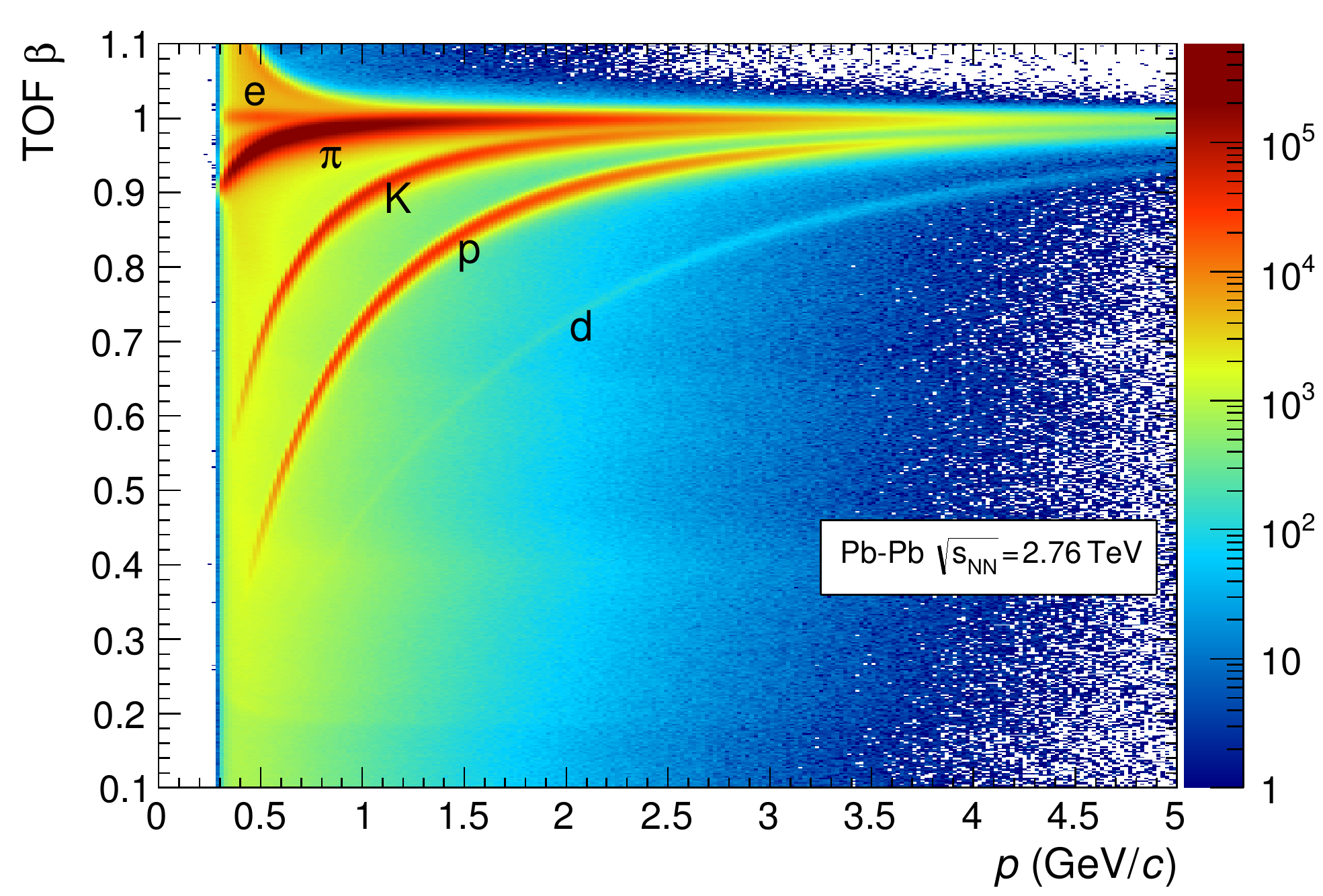}
\caption{\label{tof_perf} Measured velocity $\beta$ as function of rigidity $p/z$. The clear separation of light flavoured hadrons over a wide momentum range is visible. Taken from~\cite{alice_performance}.}
\end{center}
\end{figure}

At higher momenta, when the Bethe-Bloch curves of the nuclei start to merge with those of protons and light hadrons (around 1.5 GeV/$c$ for deuterons and 2 GeV/$c$ for tritons), one can additionally use the TOF detector to remove the contamination from lighter particles. The separation in the TOF detector is shown as a velocity $\beta$ over $p/z$ in Figure~\ref{tof_perf}. 

\begin{figure}[!htb]
\begin{center}
\includegraphics[width=0.49\textwidth]{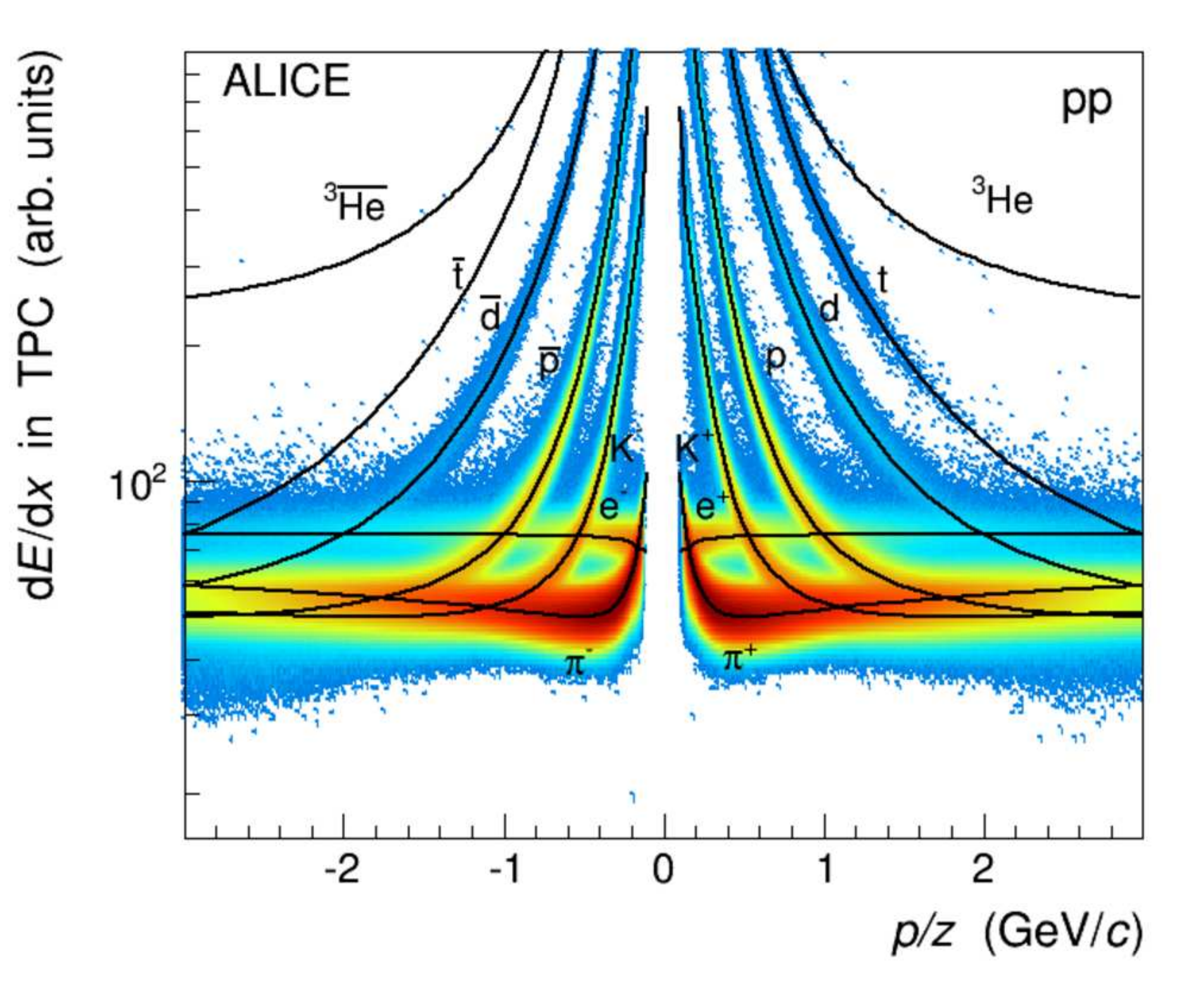}
\includegraphics[width=0.49\textwidth]{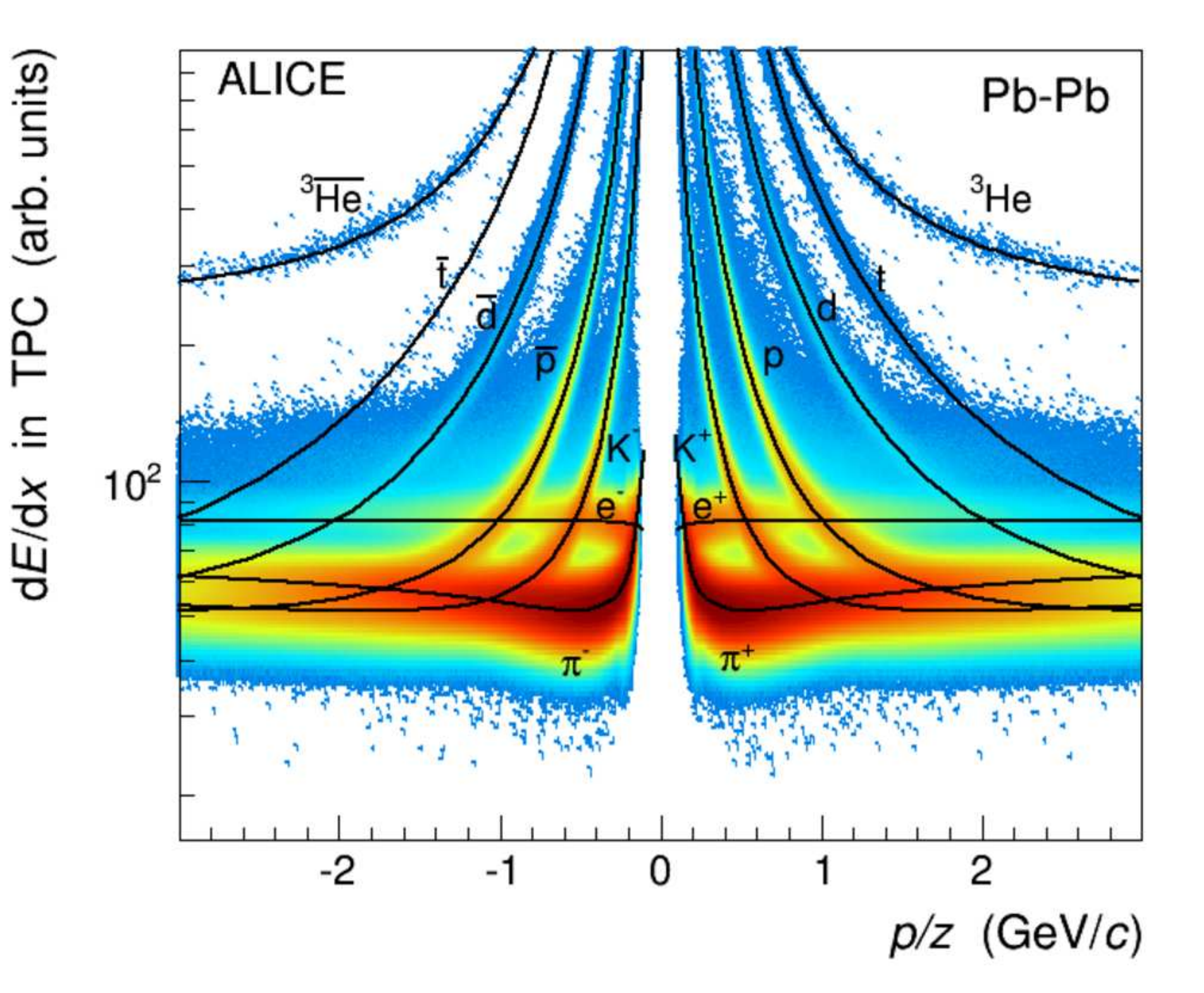}
\caption{\label{tpc_comparison} TPC d$E$/d$x$ signal as function of rigidity for pp (left) and Pb--Pb (right) collisions. Figures clearly show the asymmetry between particles and anti-particles (positive and negative rigidities), caused by enhancement of particles at low rigidity, whereas anti-nuclei at similar rigidities have a higher probability to be absorbed. From~\cite{nuclei}. For details see text.}
\end{center}
\end{figure}

\begin{figure}[!htb]
\begin{center}
\includegraphics[width=0.7\textwidth]{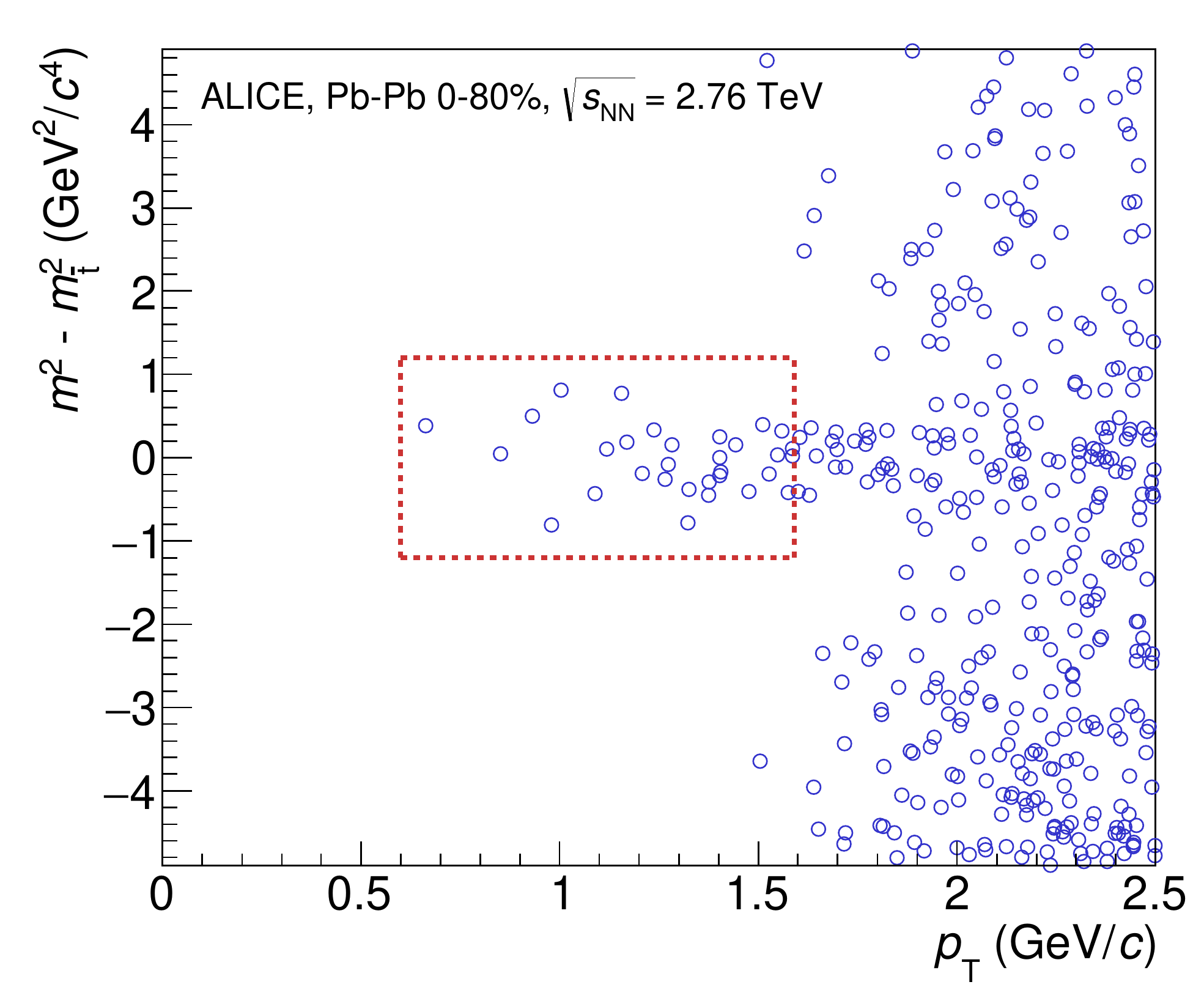}
\caption{\label{triton_candidates} Anti-triton candidates measured by TPC and TOF shown in a plot of expected triton mass versus transverse momentum, taken from~\cite{nuclei}. For details see text.}
\end{center}
\end{figure}

The above mentioned knock-out effect for nuclei is also strongly visible when the energy loss in the TPC for particles and anti-particles is compared directly as in Figure~\ref{tpc_comparison}, where the effect is easily spotted by eye, thus a difference between the rigidity distribution for all nuclei species (deuterons, tritons, helium-3) is seen. Another interesting observation that usually leads to confusion is only visible in the right panel of Figure~\ref{tpc_comparison}. We first note that primary production of the iso-dublets $^3$He and triton should be very similar at LHC energies. This is also shown explicitly in Fig.~\ref{spectra_a3_pp} below. However, because of the very different rigidities ($p/z$) of these two particles due to their different charge, the produced particles appear at different positions in the figure and the produced bulk of particles appears to be different.
At equal rigidity ($p/z$) $^3$He and tritons have similar tracking efficiencies.  Nuclei with mass number 3 produced in the fireball have typical momenta of 2-3 GeV/c, see, e.g.,Fig.~\ref{spectra_hypertriton} below.  The associated mean rapidities are 1.25 and 2.5 for production of $^3$He and t, respectively.  Tritons with charge $z=1$ start to become indistinguishable from light hadrons, protons, and deuterons at such high rapidity. For  $^3$He with $z$ = 2 the most likely momenta lead to a reduced rigidity. In addition, their specific energy loss is larger by a factor of 4. Both factors lead to clean separation of $^3$He by energy loss and momentum measurements alone. To clearly identify (anti-)tritons one needs additionally the time-of-flight measurement as displayed in Fig.~\ref{triton_candidates} as expected mass (measured mass minus world average) versus $p_\mathrm{T}$.
 
In this Fig., the mass, measured combining the TPC and TOF information, is shown vs. $p_\mathrm{T}$ relative to the expected mass for anti-tritons. Clearly 31 anti-tritons can be identified, reaching up to 1.6 GeV/$c$ in transverse momentum. For more likely higher momenta the identification becomes difficult. This yield coincides with the expectation for $^3\overline{\mathrm{He}}$ in the corresponding $p_\mathrm{T}$ window.

\begin{figure}[!htb]
\begin{center}
\includegraphics[width=0.53\textwidth]{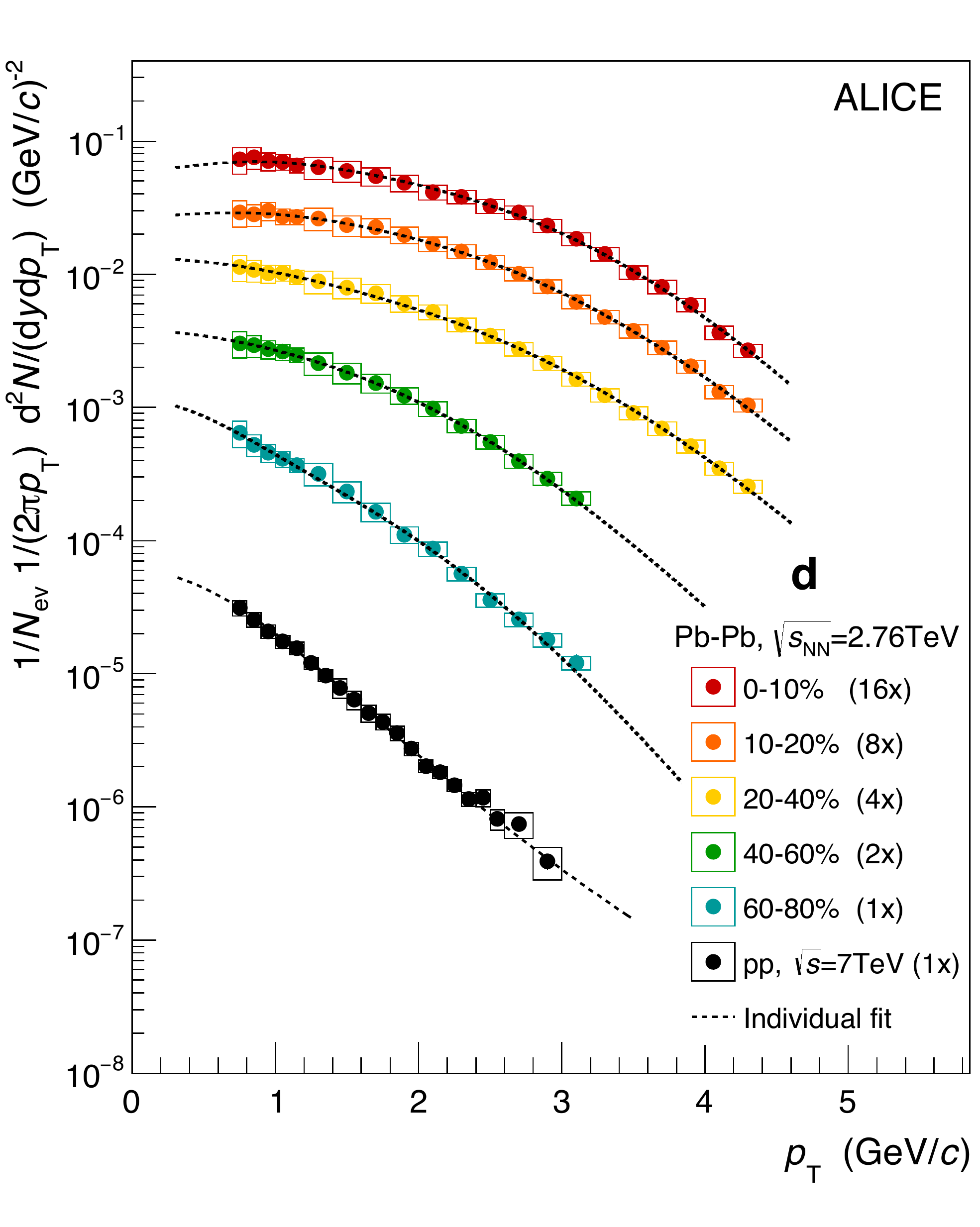}
\includegraphics[width=0.45\textwidth]{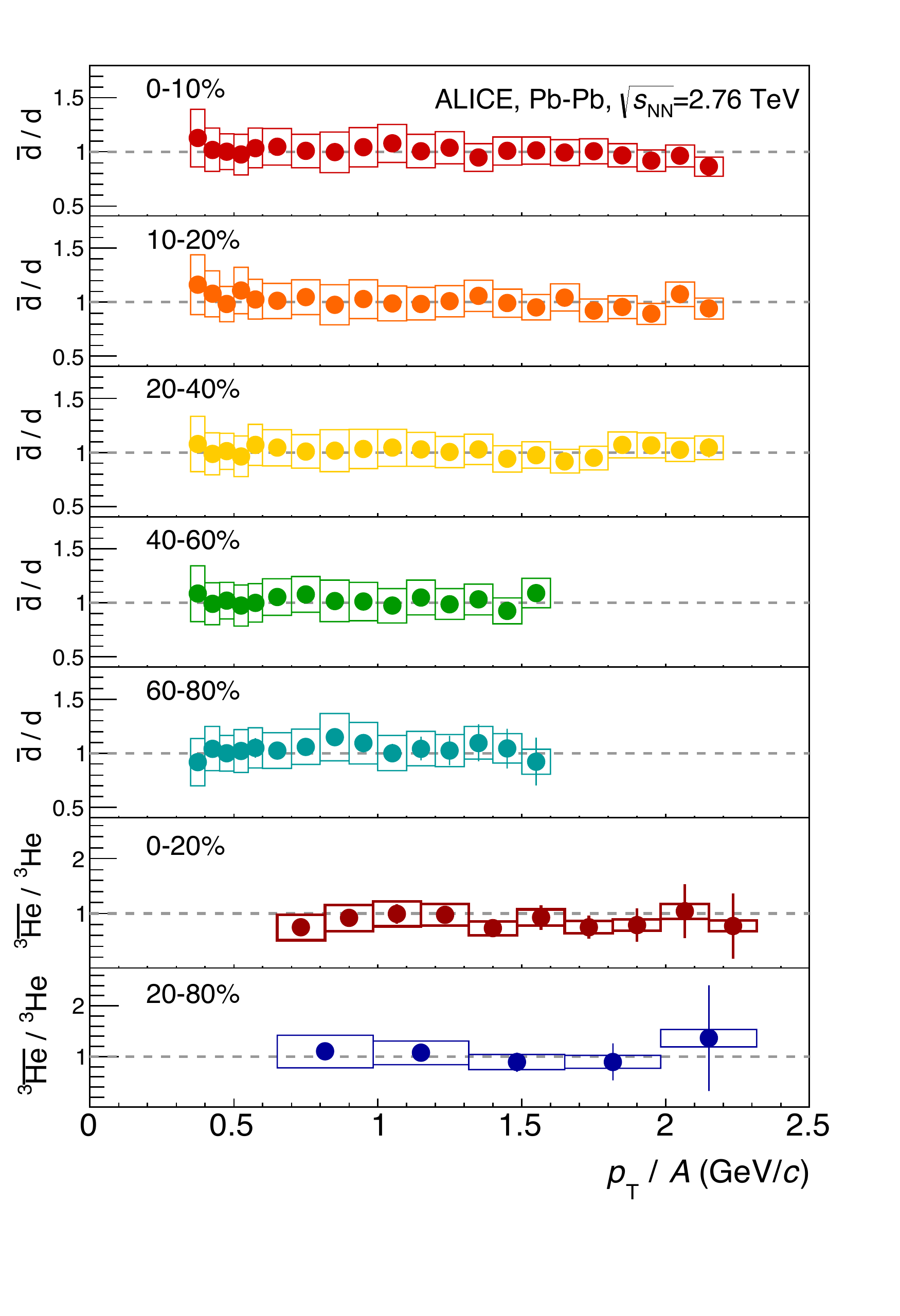}
\caption{\label{nuclei_spectra} Invariant transverse momentum spectra for deuterons in pp and five centrality classes in Pb--Pb (left panel) and ratio between anti-nuclei and nuclei ($\bar{\mathrm{d}}$/d and $^3\overline{\mathrm{He}}/^{3}$He) for the different centrality classes in the right panel. Figures taken from~\cite{nuclei}.}
\end{center}
\end{figure}

\begin{figure}[!htb]
\begin{center}
\includegraphics[width=0.7\textwidth]{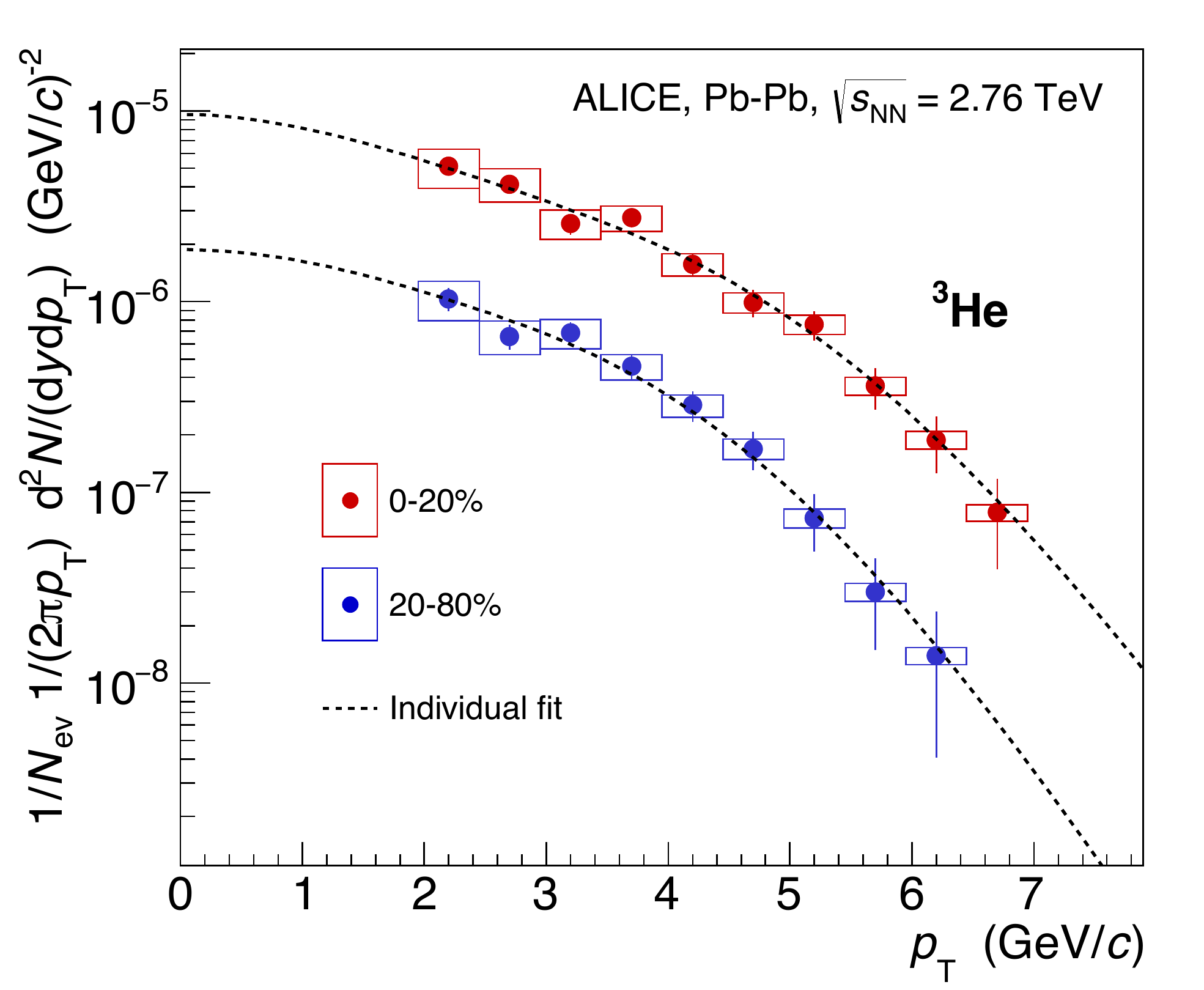}
\caption{\label{3he_spectra} Invariant transverse momentum spectra for $^3$He nuclei measured for two centrality classes in Pb--Pb collisions, figure from~\cite{nuclei}.}
\end{center}
\end{figure}

\begin{figure}[!htb]
\begin{center}
\includegraphics[width=0.8\textwidth]{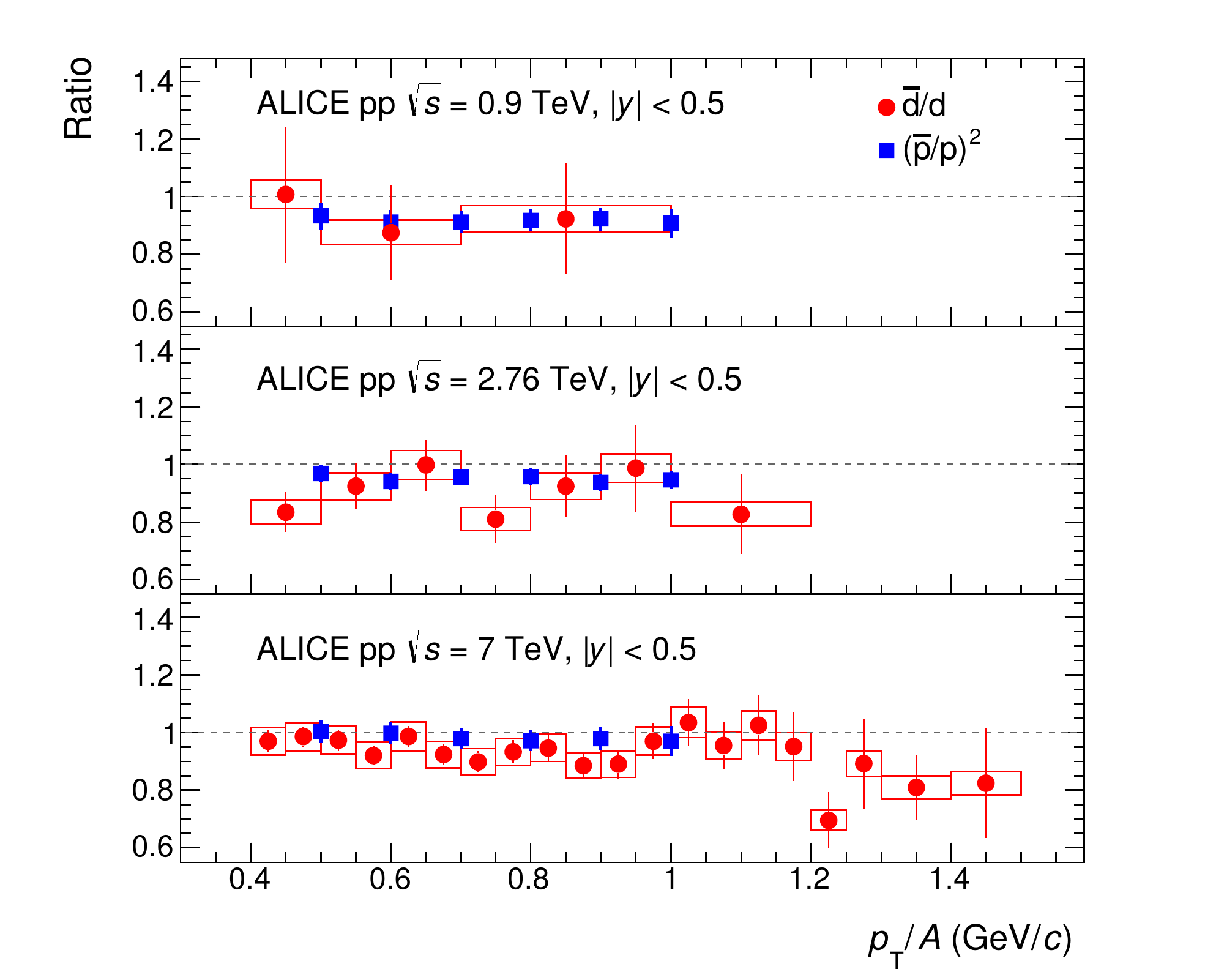}
\caption{\label{ratio_pp} $\bar{\mathrm{d}}$/d ratio as function of transverse momentum scaled by baryon number $A$ for the different pp collision energies available at the LHC ($\sqrt{s} = 900$ GeV, 2.76 TeV and 7 TeV), compared with the squared ratio of $\bar{\mathrm{p}}$/p which is expected in models. Figure from~\cite{Acharya:2017fvb}. For more details see text.}
\end{center}
\end{figure}

After a careful analysis of the data for deuterons in transverse momentum slices and correcting the data for acceptance and efficiency one obtains the spectra shown in Figure~\ref{nuclei_spectra} for pp minimum bias data and five centrality intervals and for two centrality intervals for $^3$He in Figure~\ref{3he_spectra}. Instead of giving the spectra for particles and anti-particles, the ALICE Collaboration decided to show the spectra for particles and the ratio between anti-particles and particles as depicted in the right panel.
These ratios are also of interest because, in the framework of the thermal model and the coalescence model, one expects them to be close to unity at LHC energies. In addition, Figure~\ref{ratio_pp} shows the same ratio for pp collisions at the three different energies available. 

The spectra of nuclei show a hardening going from more peripheral to more    
central events. In addition, a clear difference is visible when the shape of pp and Pb--Pb spectra are compared. The Pb--Pb spectra show a particular shape which is caused by the radial flow, originating from the radial expansion of the fireball. This particular shape of the spectra is usually modeled by the blast-wave approach~\cite{Schnedermann:1993ws} discussed in section~\ref{sect_flow} above.
To estimate the total production yield for specific particles this function is also used to extrapolate the spectra towards low $p_\mathrm{T}$, where the measurement is not possible because of the low acceptance and efficiency. 

\begin{figure}[!htb]
\begin{center}
\includegraphics[width=0.7\textwidth]{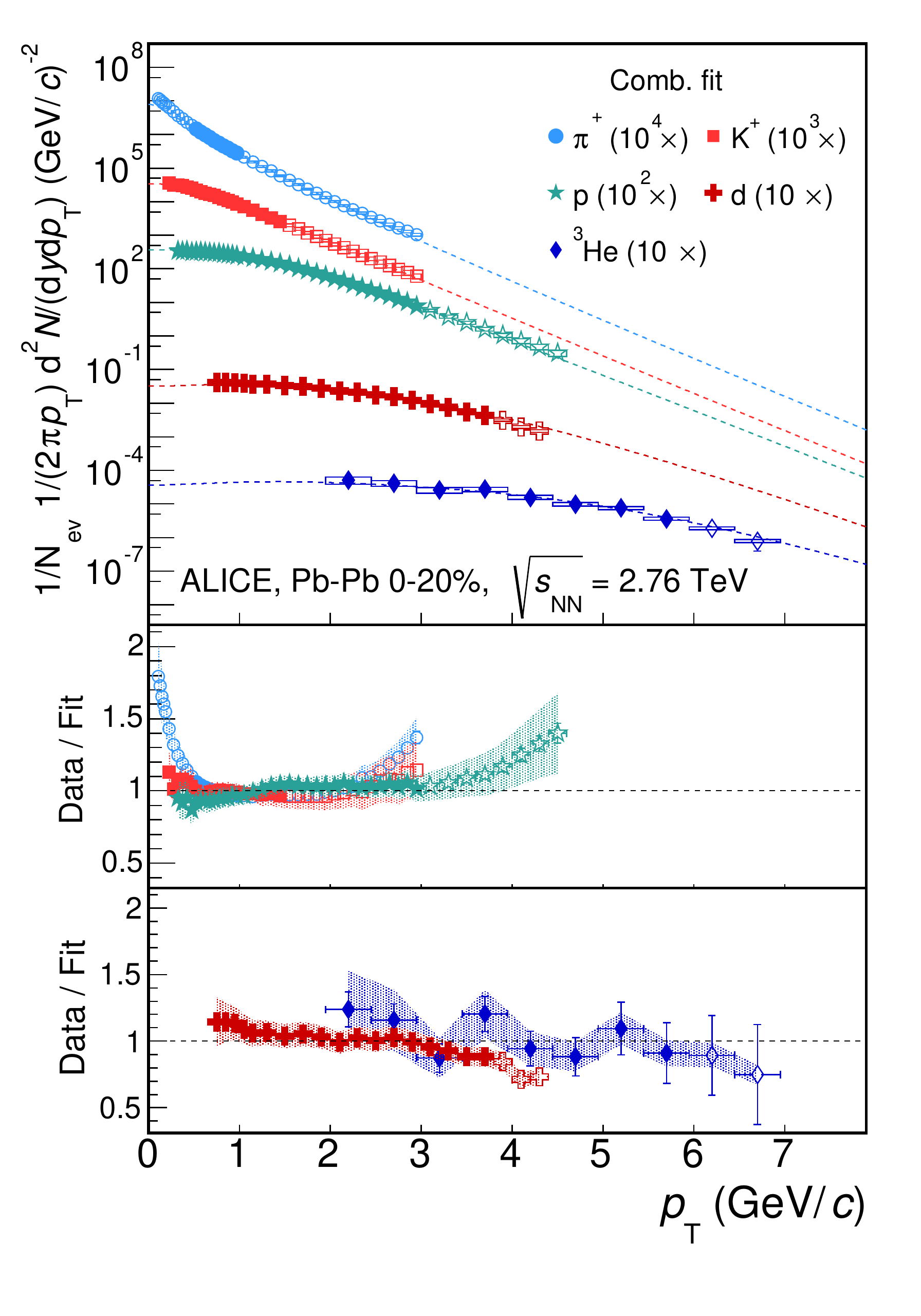}
\caption{\label{combined_bw} Invariant transverse momentum spectra for $\pi^{+}$, K$^+$, p, d, $^3$He in Pb--Pb collisions at 0-20\% centrality fitted with one global set of blast-wave parameters. The lower panels show the comparison of model with data. From~\cite{nuclei}. For more details see text.}
\end{center}
\end{figure}

It is impressive to see how well the blast-wave approach reproduces the shape of transverse momentum spectra for different particle species at a given centrality with one common set of parameters. This is shown in Figure~\ref{combined_bw} for $\pi$, K, p, d and $^3$He in Pb--Pb collisions at $\sqrt{s_{\mathrm{NN}}} = 2.76$ TeV and 0-20\% central events. Only the particle mass was changed in the blast-wave formula. This strongly supports the hydrodynamic flow picture with one  common kinetic freeze-out temperature $T_{kin}$ and radial expansion velocity $\langle\beta\rangle$. In the lower panels of Figure~\ref{combined_bw} the common fit is compared with the measured spectra. The deviation at low $p_\mathrm{T}$ visible for $\pi^-$ is mainly caused by resonance decays which feed  into the spectra in this region. At higher $p_\mathrm{T}$ values the spectra deviate from the fit because the blast-wave model is no longer a good approximation and the expected power-law behaviour from hard-scattering processes becomes visible. 

\begin{figure}[!htb]
\begin{center}
\includegraphics[width=0.7\textwidth]{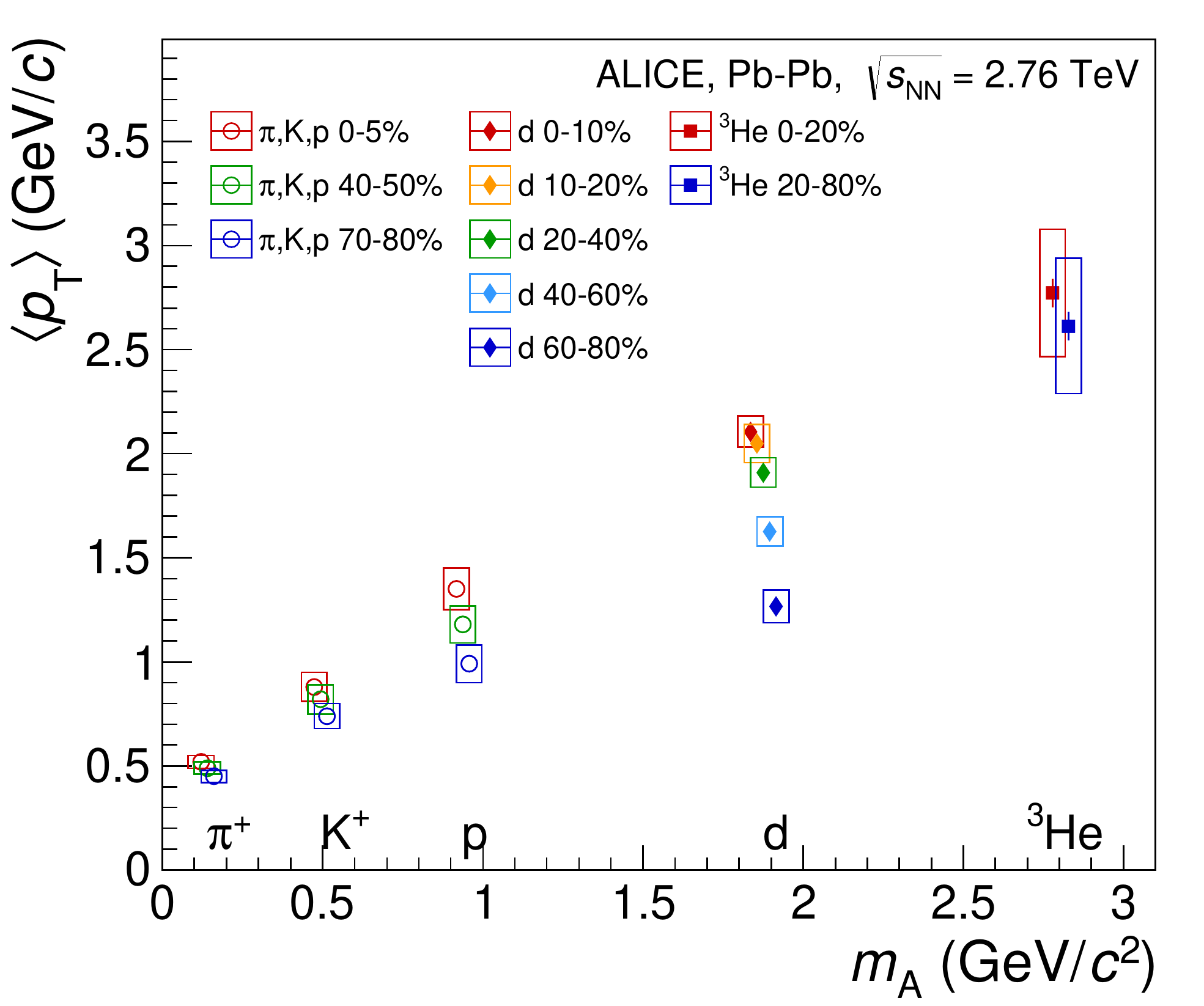}
\caption{\label{mean_pt} Measured mean transverse momentum versus particle mass for different centrality intervals. Taken from~\cite{nuclei}.}
\end{center}
\end{figure}

Clearly also the loosely-bound deuteron and $^3$He particles participate in the flow.  Their mean transverse momentum can also be extracted with the help of the blast-wave model. The mean $p_{\mathrm{T}}$ (Figure~\ref{mean_pt}) shows a clear increase with the particle mass corresponding to the previously discussed radial flow which at common expansion velocity implies a mass ordering. At the same time this also gives additional confirmation about the briefly discussed behaviour of anti-tritons and anti-$^3$He yields which have nearly equal masses and thus have the same $\langle p_{\mathrm{T}}\rangle$.   

\begin{figure}[!htb]
\begin{center}
\includegraphics[width=0.9\textwidth]{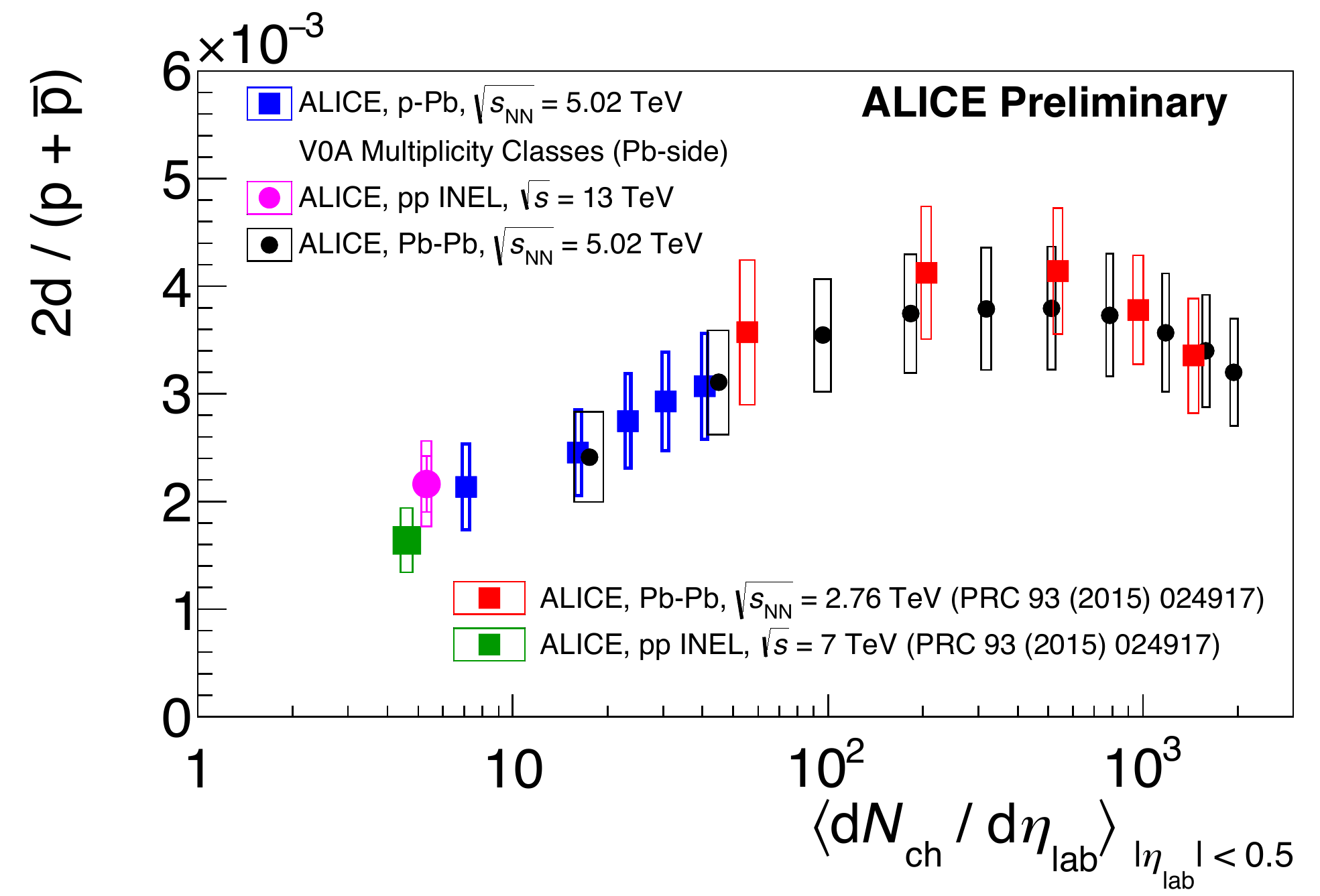}
\caption{\label{d_over_p} Ratio of the production yields of deuterons to protons (2d/(p+$\bar{\mathrm{p}}$) as a function of the mean multiplicity $\langle$d$N_\mathrm{ch}$/d$\eta\rangle$ in pp, p--Pb and Pb--Pb collisions. The green and magenta points are extracted from the minimum bias pp measurement at different energies, the p--Pb results are preliminary results from the ALICE Collaboration and the Pb--Pb corresponds to the five centrality classes discussed before. From~\cite{ALICE-PUBLIC-2017-006}.}
\end{center}
\end{figure}

When the spectra are extrapolated to low $p_\mathrm{T}$, as discussed before, the production yield per rapidity unit, namely the rapidity density d$N$/d$y$, can be extracted. The ratio $\frac{2\mathrm{d}}{\mathrm{p} + \overline{\mathrm{p}}}$  of these rapidity densities is presented in Figure~\ref{d_over_p} for different multiplicities going from pp, over p--Pb to Pb--Pb collisions. A linear increase can be seen going from pp to p--Pb, until a maximum is reached in Pb--Pb collisions, similar to the previously discussed case of the lifting of strangeness suppression. The linear increase of the d/p ratio is expected for na\"{i}ve coalescence models. In such models one would, however,  expect the ratio to  increase further for more central Pb--Pb collisions. Instead, the values reach a plateau at the level predicted using the thermal model. In fact, the behaviour can be described completely by a canonical statistical-thermal model approach as shown in~\cite{Vovchenko:2018fiy}.

\begin{figure}[!htb]
\begin{center}
\includegraphics[width=0.7\textwidth]{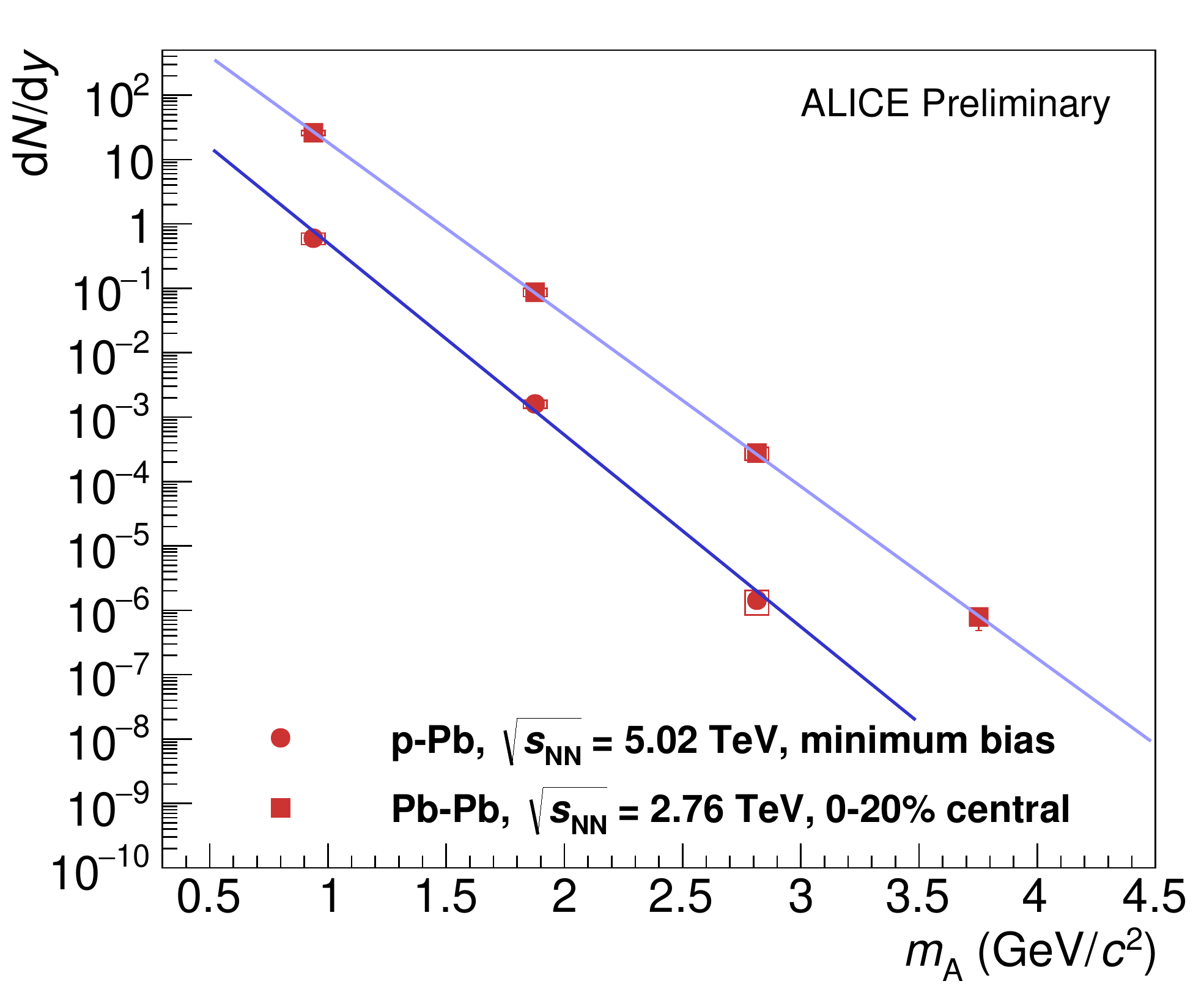}
\caption{\label{mass_ordering} Production yields of different nuclei as a function of particle mass in p--Pb and Pb--Pb collisions, as shown in~\cite{SHARMA2016461}. The lines are exponential fits. }
\end{center}
\end{figure}

\begin{figure}[!htb]
\begin{center}
\includegraphics[width=0.45\textwidth]{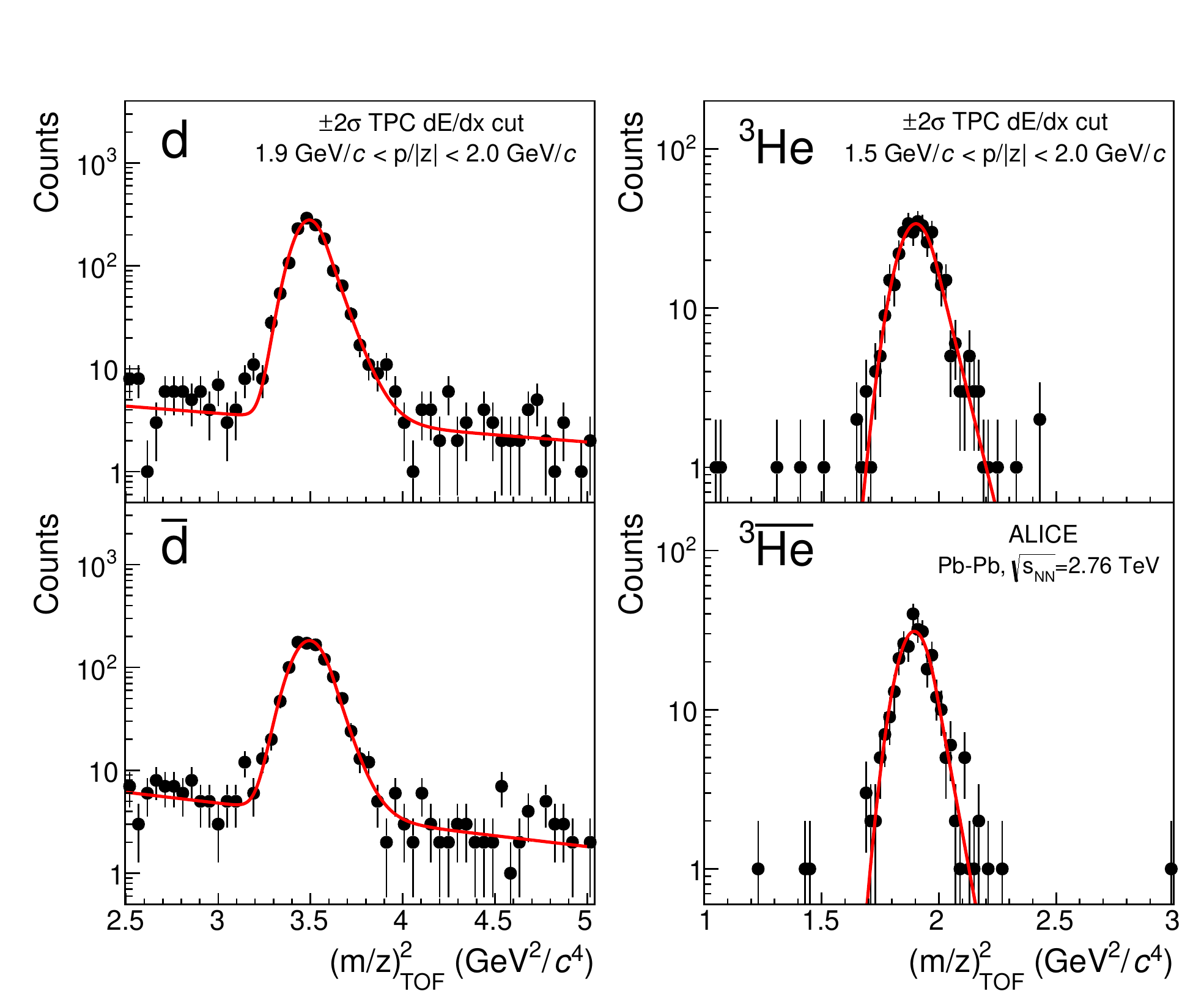}
\includegraphics[width=0.53\textwidth]{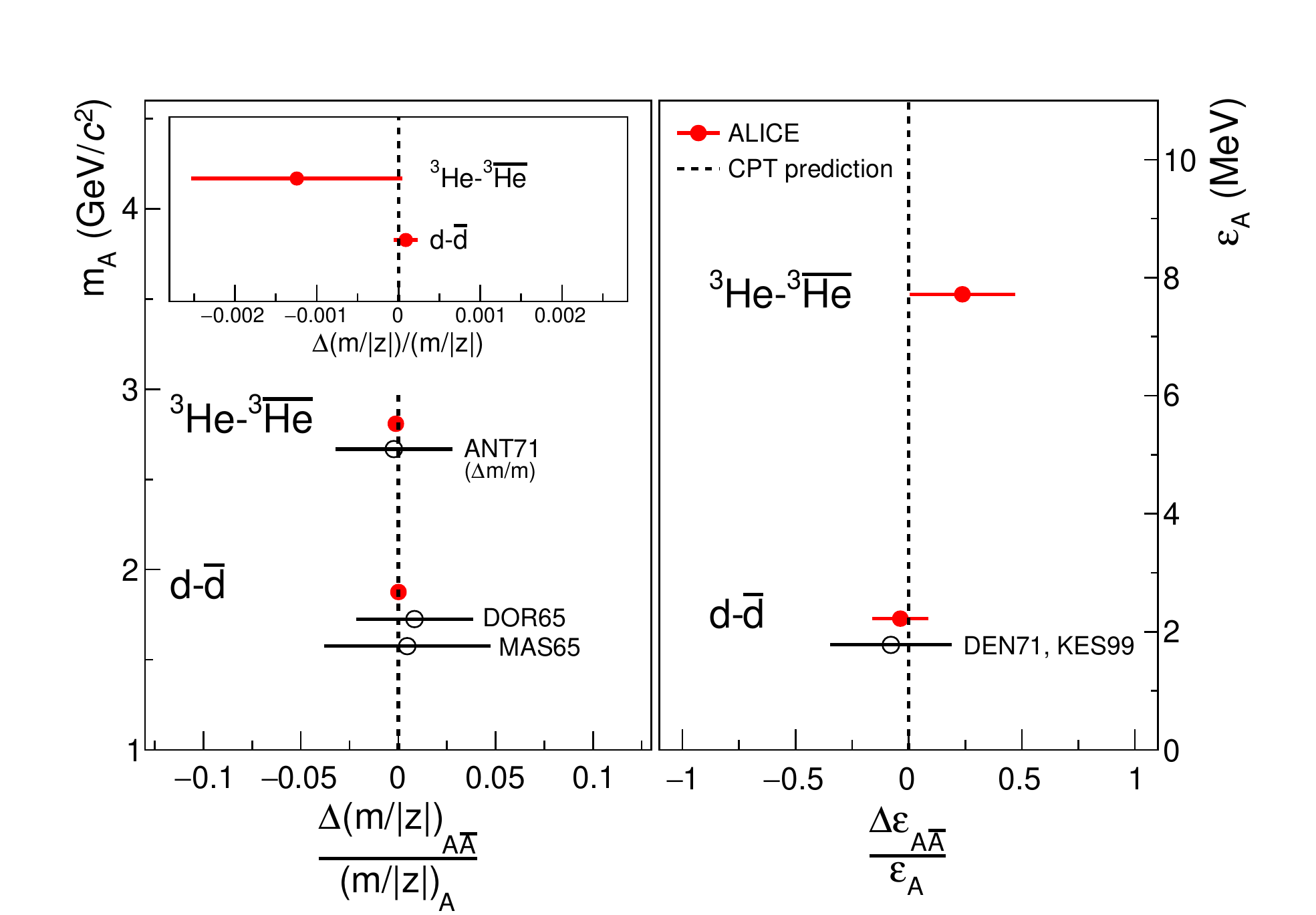}
\caption{\label{mass_difference} The left panel shows the  squared ratio of mass and charge number $(m/z)^2$ distributions for d, $\bar{\mathrm{d}}$, $^3$He and $^3\overline{\mathrm{He}}$. The right panel depicts the normalised mass difference $\frac{\Delta(m/|z|)_{A\bar{A}}}{(m/|z|)_A}$ and the normalised binding energy difference $\frac{\Delta\epsilon_{A\bar{A}}}{\epsilon_A}$, from~\cite{mass_difference}. For details see text.}
\end{center}
\end{figure}

The measured rapidity densities exhibit an exponential mass ordering, as shown for p, d and $^3$He in p--Pb and Pb--Pb collisions (in Pb--Pb also anti-$^4$He is drawn) in Figure~\ref{mass_ordering}. The exponential decrease in the rapidity density d$N$/d$y$ when another baryon is added to the system is called penalty factor~\cite{BraunMunzinger:2001mh}. The value of the penalty factor is about 300 for Pb--Pb collisions and 600 for p--Pb collisions. The lines shown in Figure~\ref{mass_ordering} are exponential fits for the two different systems, separately. The exponential falling of the rapidity density as a function of mass  is a natural prediction of the thermal model.    

\begin{figure}[!htb]
\begin{center}
\includegraphics[width=0.7\textwidth]{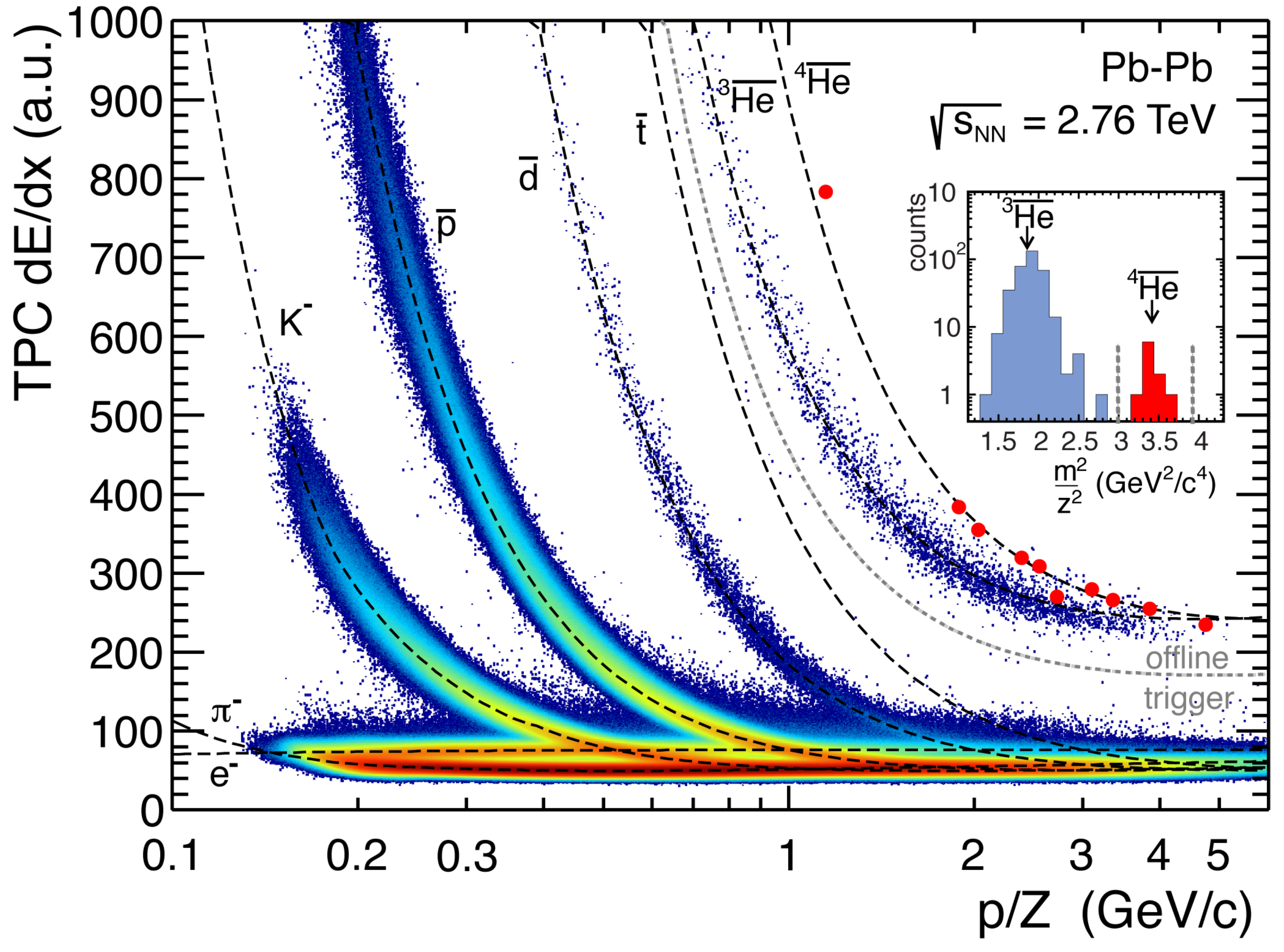}
\caption{\label{anti_alpha} Measured d$E$/d$x$ signal in the ALICE TPC versus magnetic rigidity, together with the expected curves for negatively-charged particles in the data set of 2011, taken at $\sqrt{s_\mathrm{NN}}=2.76$~TeV. The inset panel shows the TOF mass measurement which provides additional separation between anti-$^3$He and anti-$^4$He for tracks with $p/z > 2.3$ GeV/$c$ (from \cite{DONIGUS2013547c}). See text for more details.}
\end{center}
\end{figure}

The anti-alpha yield shown in Figure~\ref{mass_ordering} is extracted by a combined measurement using TPC and TOF to clearly identify the anti-alphas with the two detectors as shown in Figure~\ref{anti_alpha}, where the 10 identified $^4\overline{\mathrm{He}}$ in the analyzed sample. The figure only shows data from events which contain a signal above the line indicated by the the gray line named offline trigger in the figure, thus every every event used in this figure contains at least a $|z|=2$ particle.    

The high abundance of deuterons and anti-deuterons as well as $^3$He and anti-$^3$He, allows for a precise comparison of the particle to anti-particle masses. This comparison is at the same time a CPT test in the nuclei sector. The result of this mass measurement~\cite{mass_difference}, expressed as squared ratio of mass over charge number, of the different (anti-)particle species, is depicted in Figure~\ref{mass_difference} together with the result for the mass difference and the difference in the binding energies. 
The results represent the highest precision direct measurements of particle-antiparticle  mass differences for nuclei. They improve by one to two orders of magnitude results originally obtained more than 4 decades ago for the anti-deuteron. 
See~\cite{mass_difference} for more details.

\begin{figure}[!htb]
\begin{center}
\includegraphics[width=0.7\textwidth]{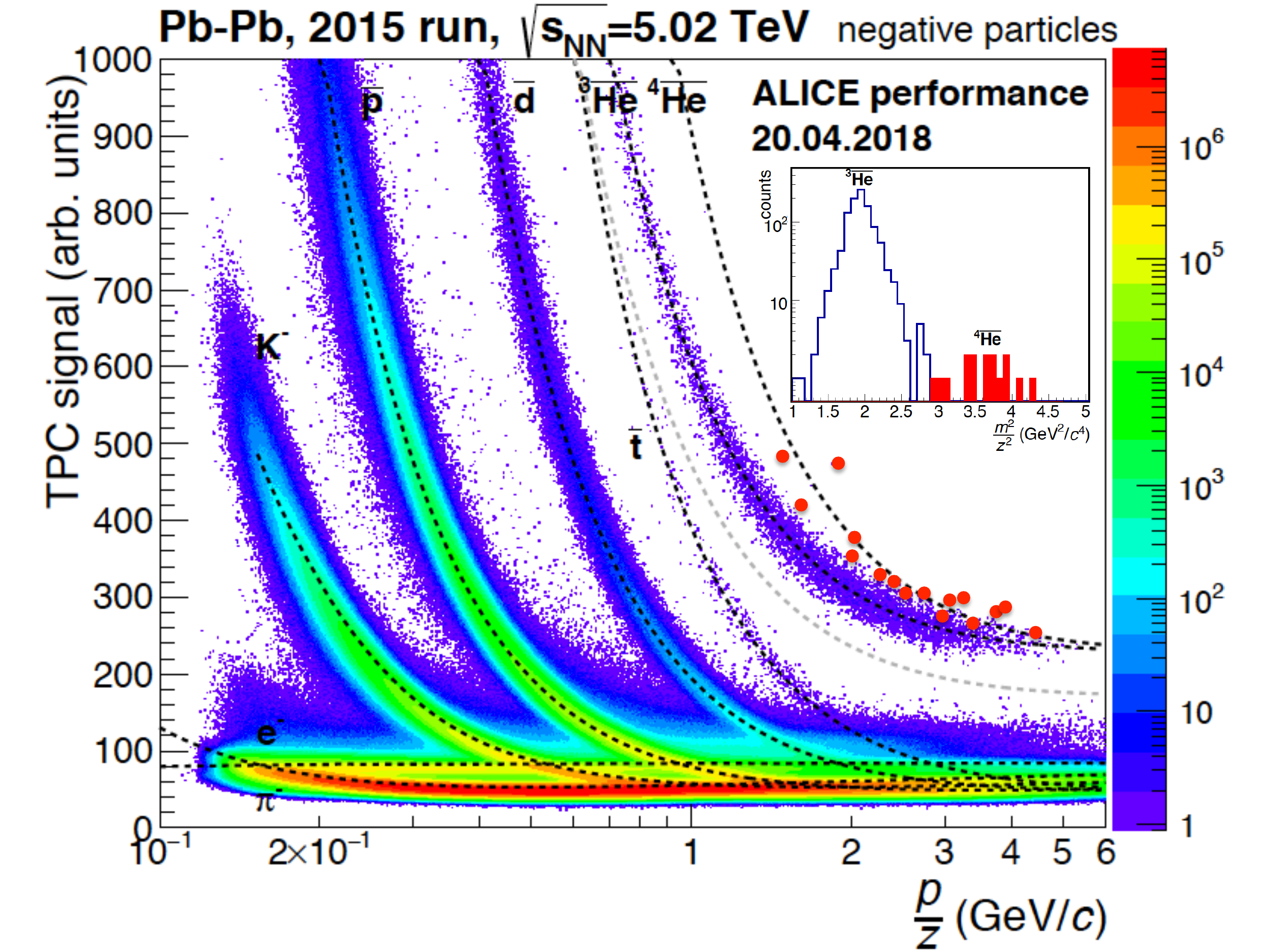}
\caption{\label{anti_alpha_5TeV} Measured d$E$/d$x$ signal in the ALICE TPC versus magnetic rigidity, together with the expected curves for negatively-charged particles in the data set of 2015, taken at $\sqrt{s_\mathrm{NN}}=5.02$~TeV. The inset panel shows the TOF mass measurement which provides additional separation between anti-$^3$He and anti-$^4$He for tracks with $p/z > 2.3$ GeV/$c$. As shown at QM2018 conference by the ALICE Collaboration.}
\end{center}
\end{figure}

\begin{figure}[!htb]
\begin{center}
\includegraphics[width=0.9\textwidth]{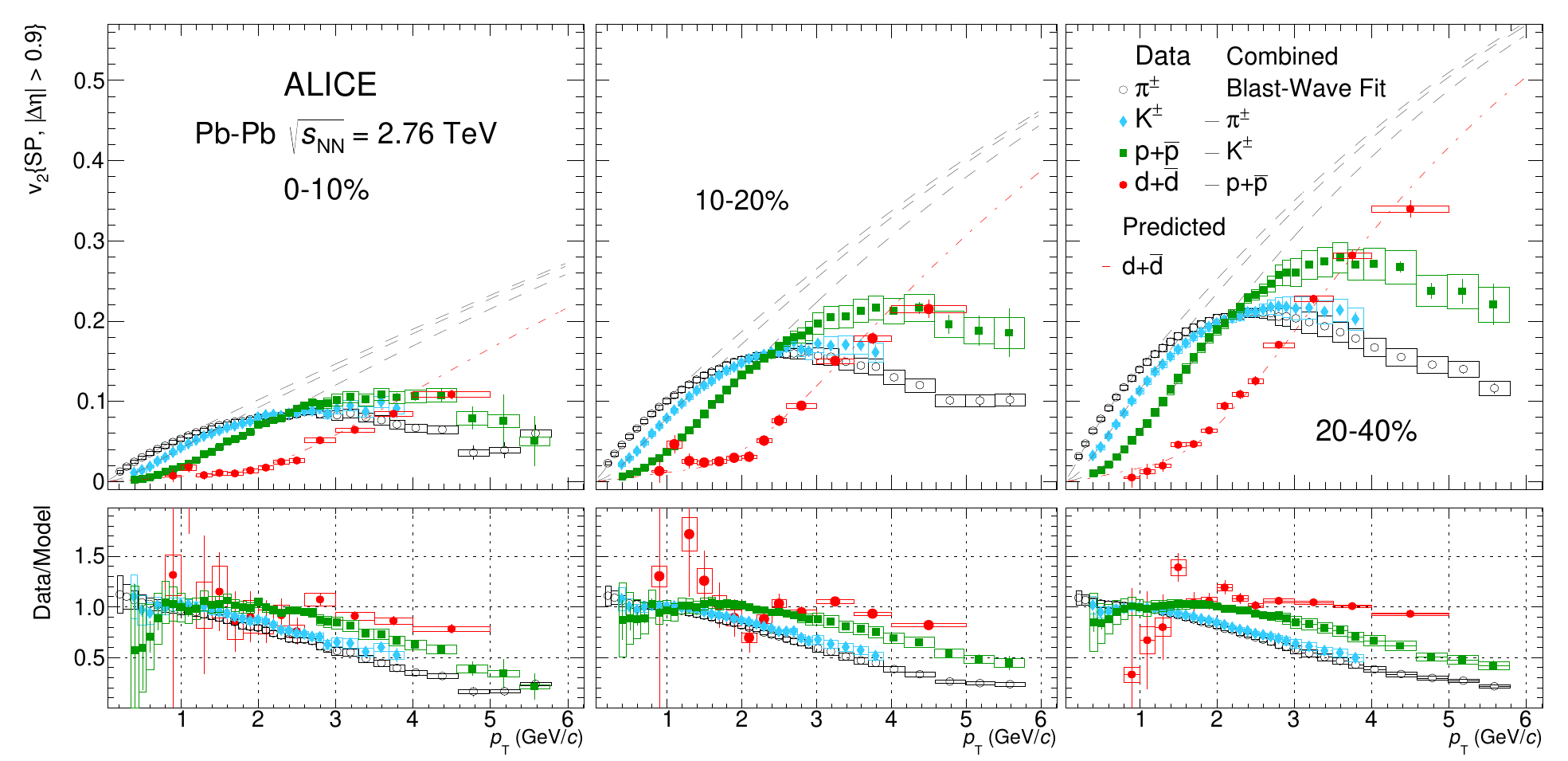}
\caption{\label{v2_bw} Elliptic flow coefficient $v_2$ for pions, kaons, protons and deuterons for three different centrality classes (left: 0-10\% centrality; middle: 10-20\%; right: 20-40\%). The measured data points are compared with a blast-wave fit of pions, kaons and protons. The parameters of this fit is used for the prediction for deuterons. As shown in~\cite{Acharya:2017dmc}.}
\end{center}
\end{figure}

\begin{figure}[!htb]
\begin{center}
\includegraphics[width=0.7\textwidth]{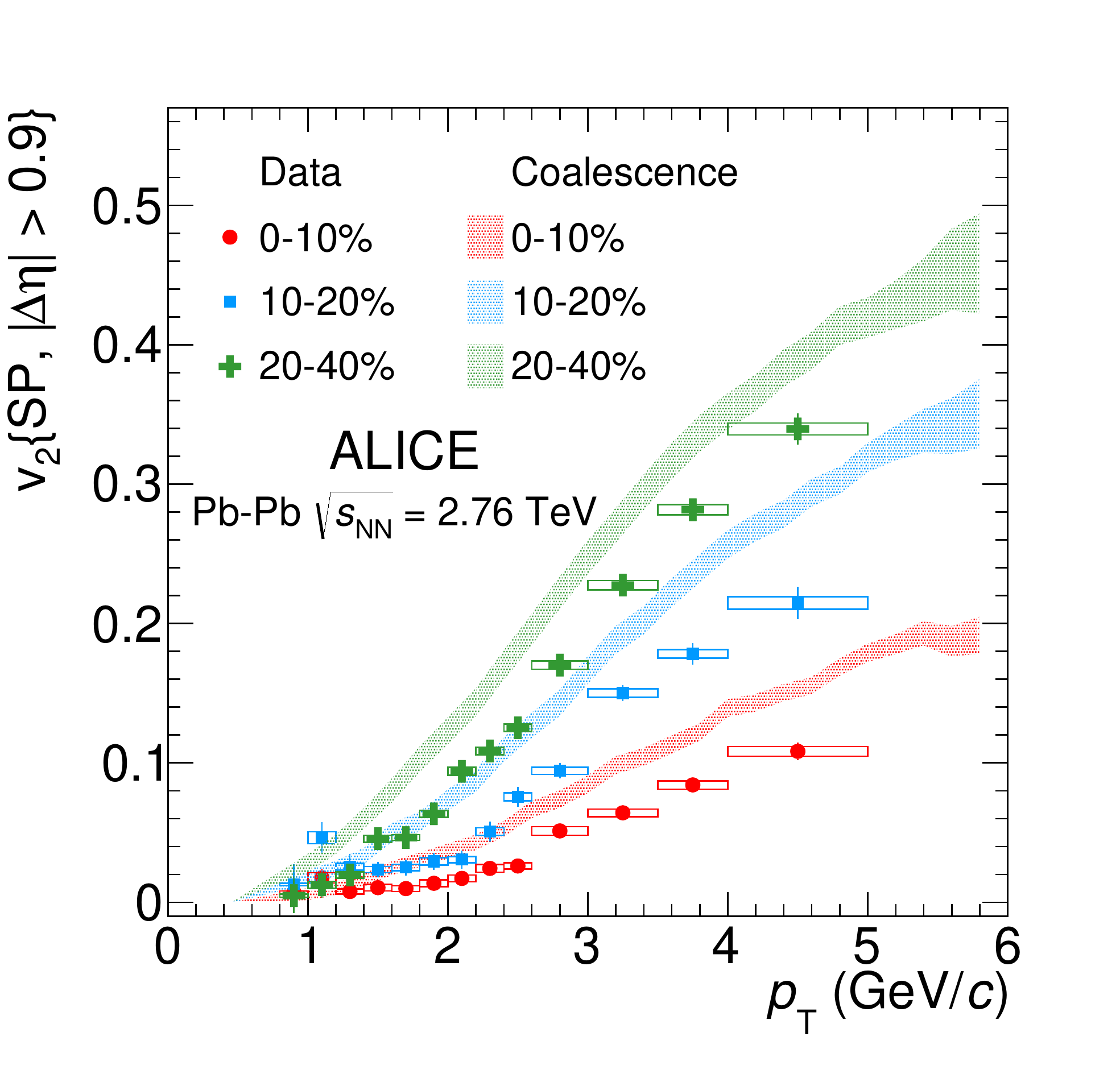}
\caption{\label{v2_coal} Elliptic flow coefficient $v_2$ of deuterons for three different centrality classes. The measured data points are compared with predictions of a na\"{i}ve coalescence model assuming a simple scaling of the $v_2$ of protons by a factor of 2 to get the $v_2$ of deuterons. As shown in~\cite{Acharya:2017dmc}.}
\end{center}
\end{figure}

\begin{figure}[!htb]
\begin{center}
\includegraphics[width=0.9\textwidth]{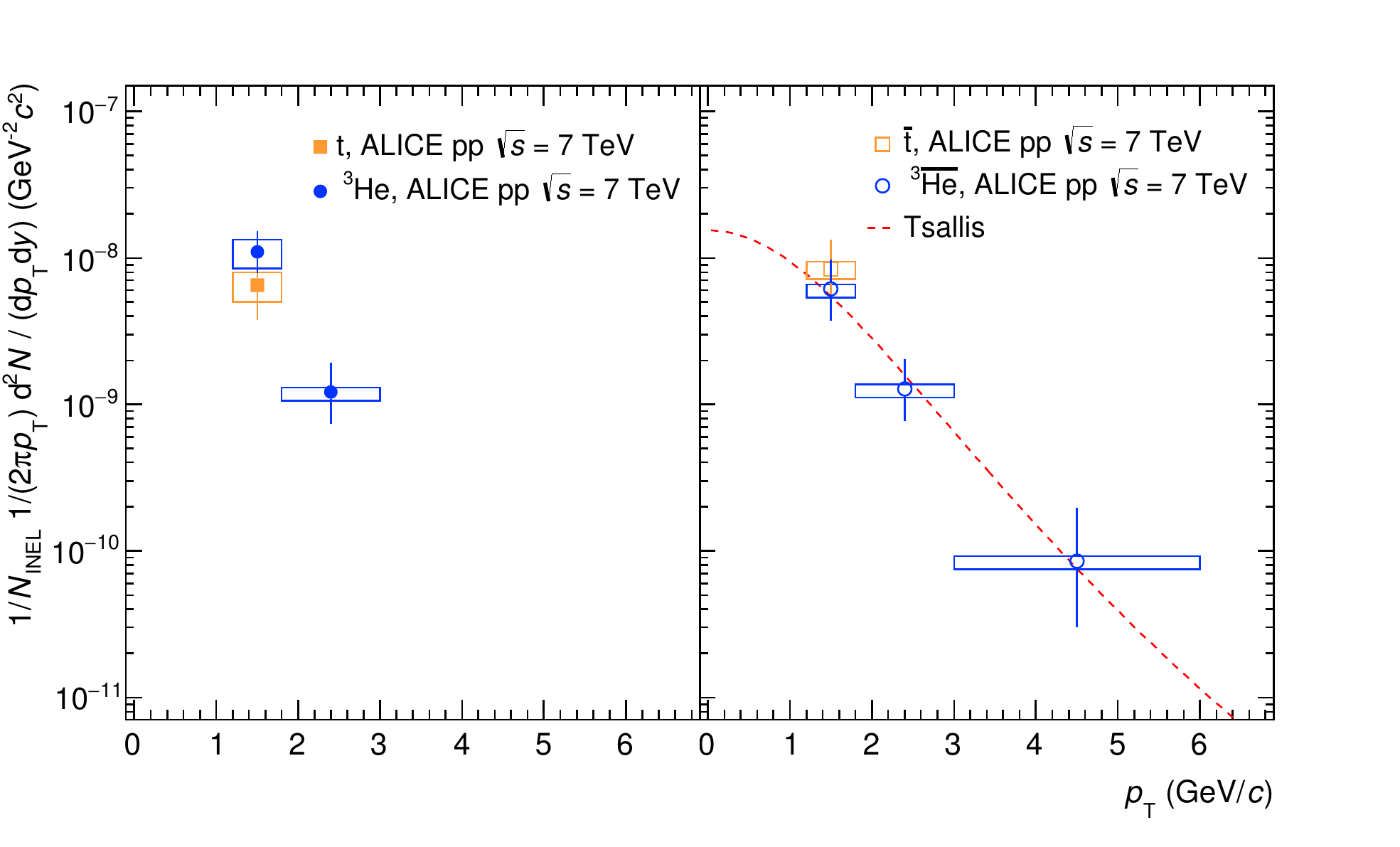}
\caption{\label{spectra_a3_pp} Invariant differential yields of tritons and $^3$He nuclei (left panel) and their anti-nuclei (right panel) in inelastic pp collisions at $\sqrt{s} = 7$~TeV, as shown in~\cite{Acharya:2017fvb}. The dashed line is a Tsallis fit to the data, see text for details. }
\end{center}
\end{figure}

\begin{figure}[!htb]
\begin{center}
\includegraphics[width=0.8\textwidth]{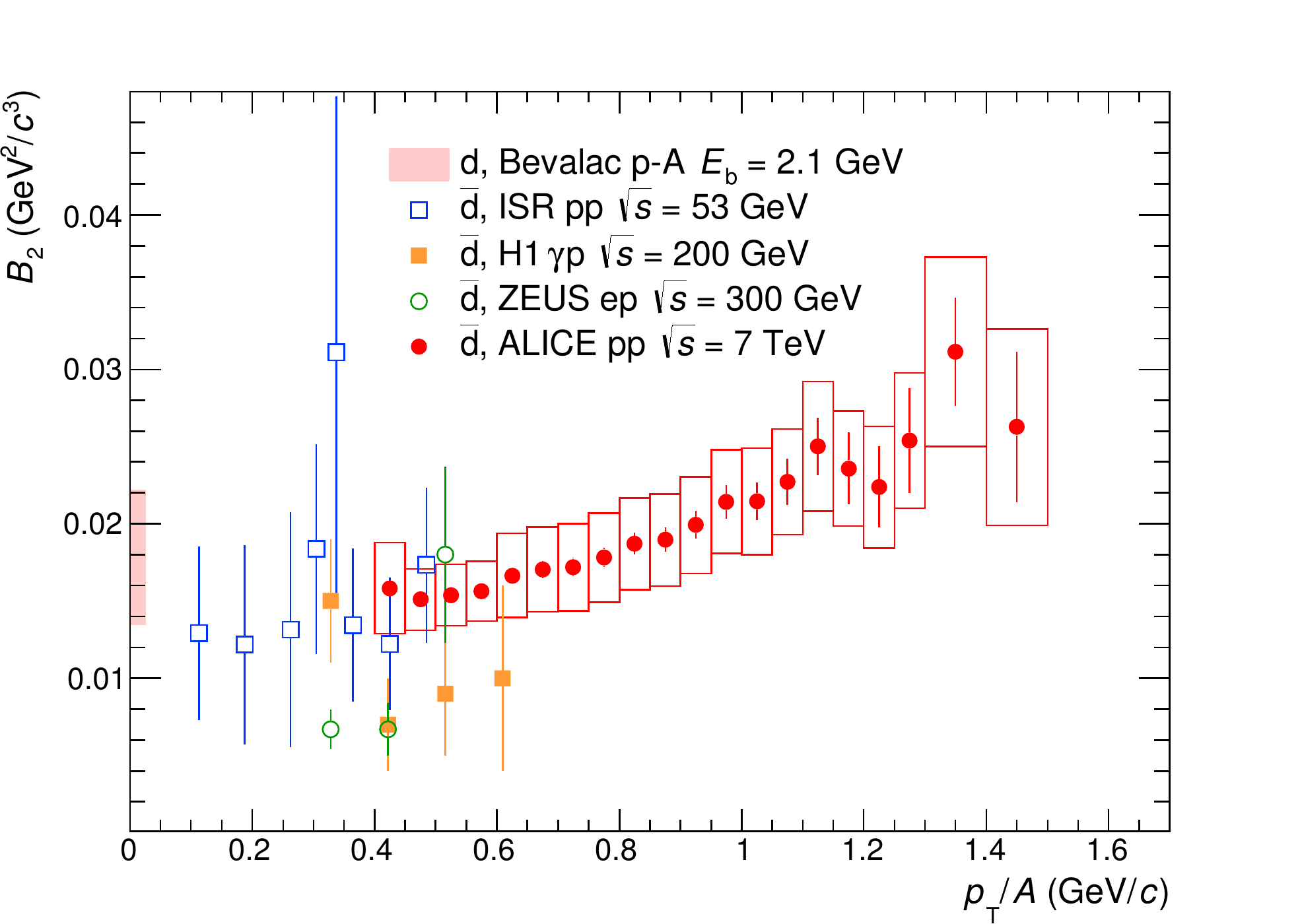}
\caption{\label{coal_b2_pp} Coalescence parameter $B_2$ of anti-deuterons in inelastic pp collisions at $\sqrt{s} = 7$~TeV, compared to the values measured at lower energies in pp, $\gamma$p, ep, in p--Cu and p--Pb collisions. As shown in~\cite{Acharya:2017fvb}.}
\end{center}
\end{figure}

\begin{figure}[!htb]
\begin{center}
\includegraphics[width=0.9\textwidth]{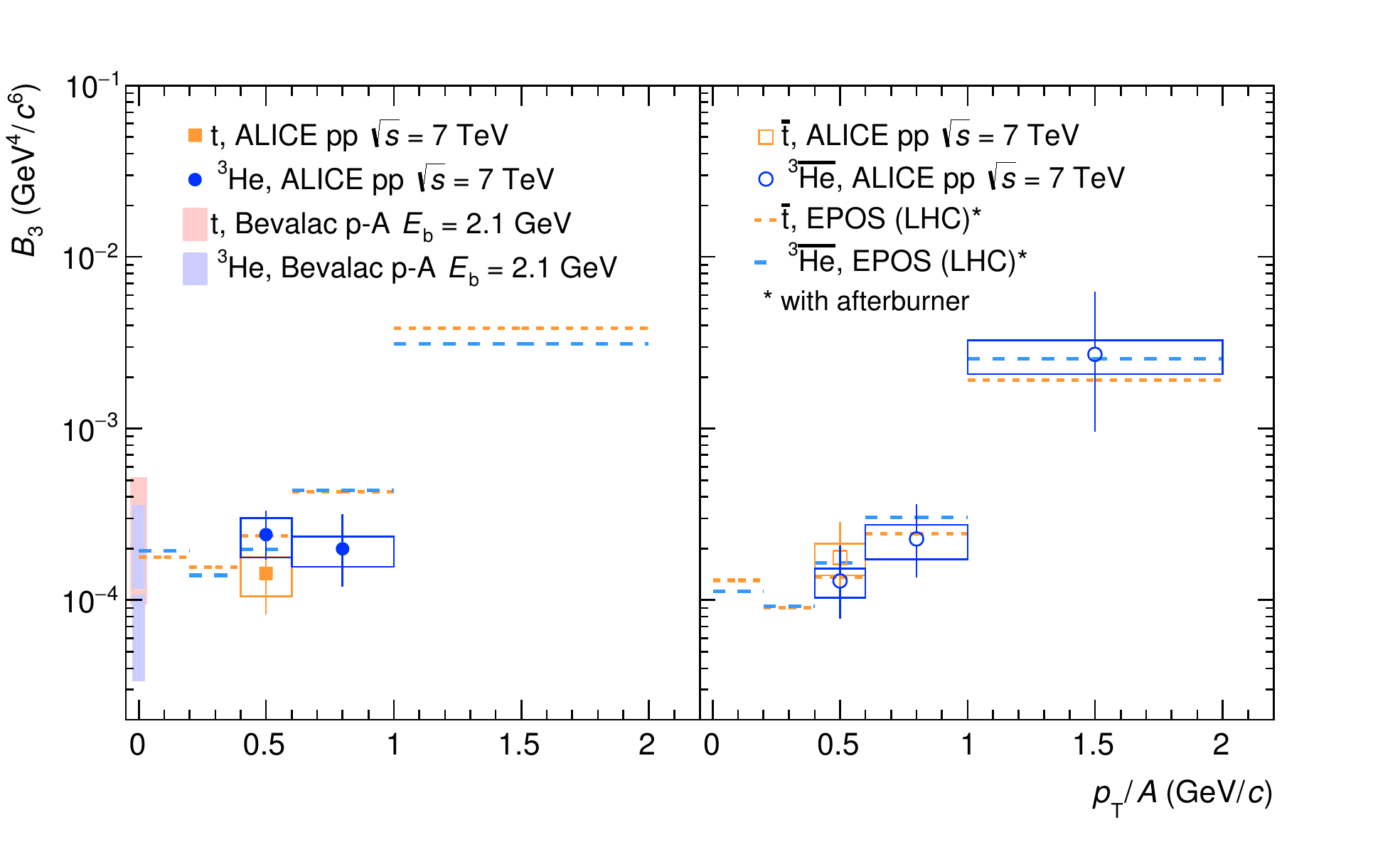}
\caption{\label{coal_b3_pp} Coalescence parameter $B_3$ of tritons and $^3$He nuclei (left panel) and their anti-nuclei (right panel)  in elastic pp collisions at $\sqrt{s} = 7$~TeV. The Bevalac measurements in p--C, p--Cu and p--Pb collisions are shown as bands at low momentum. Dashed lines indicate the values obtained with EPOS-LHC using a simple afterburner for the coalescence. From~\cite{Acharya:2017fvb}.}
\end{center}
\end{figure}

Another observable which can help to understand the production mechanisms of loosely-bound systems is the azimuthal or elliptic flow characterized by the flow coefficient $v_2$, introduced above. 
The dependence of this second Fourier coefficient $v_2$ on $p_\mathrm{T}$ is shown in Figure~\ref{v2_bw} for pions, kaons, protons and deuterons. A na\"{i}ve simple coalescence model would assume a scaling of the $v_2$ of protons by a factor 2 to estimate the $v_2$ of the deuterons, as shown in Fig.~\ref{v2_coal}. This is not fulfilled as seen from the data compared with the dashed regions being the scaled proton $v_2$. Instead the data are reasonably well described by a blast-wave model calculation, when the parameters of the blast-wave function are determined by a fit to the $\pi$, K, p and then only inserting the mass of the deuteron, for details see~\cite{Acharya:2017dmc}. This implies that deuterons 'feel' the anisotropic hydrodynamic expansion velocity in very similar ways as the nearly point-like pions, kaons, and protons. The result is in agreement with the recent findings by~\cite{Yang:2018ghi}.

In passing we note that the current data also allow the investigation of the validity of isospin symmetry in the production process. This  in principle could be done by comparing production yields of protons and neutrons. Since neutrons are not detectable with the ALICE apparatus this test is performed indirectly through the comparison of the production yields for triton and $^3$He nuclei. As discussed above it is rather difficult to extract a full transverse momentum spectrum for tritons. Therefore the test is currently only done in a limited  $p_{\rm{T}}$ range. Figure~\ref{spectra_a3_pp} shows how well this works for the transverse momentum spectra of triton and $^3$He nuclei, together with their anti-particles, for pp collisions at $\sqrt{s} = 7$~TeV. Within the experimental uncertainties isospin symmetry is observed. From these spectra and the corresponding proton and deuteron $p_\mathrm{T}$ spectra one can also extract the $B_2$ and $B_3$ coefficients as depicted in Fig.~\ref{coal_b2_pp}  and Fig.~\ref{coal_b3_pp}. 

\begin{figure}[!htb]
\begin{center}
\includegraphics[width=0.8\textwidth]{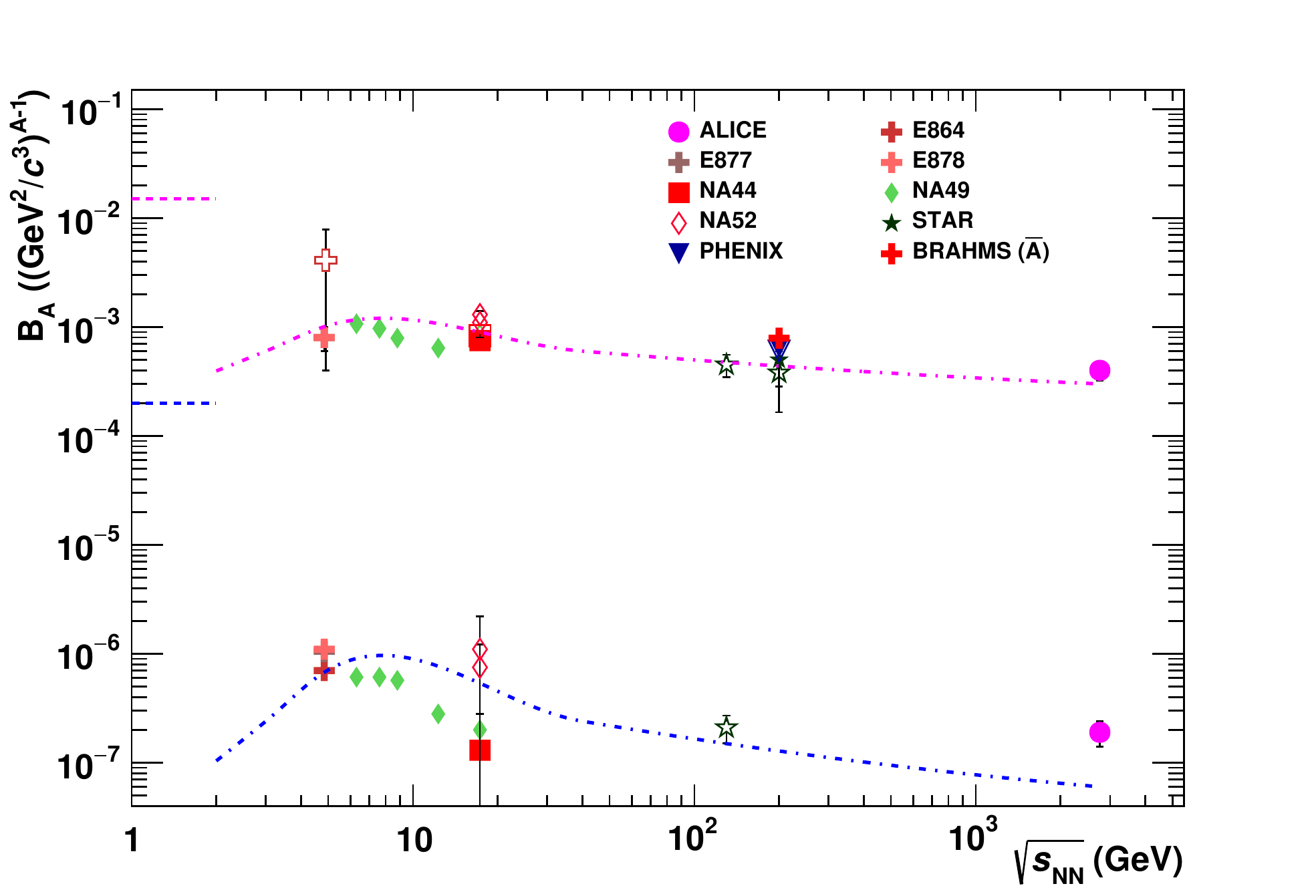}
\caption{\label{b2_vs_sqrts} Coalescence parameters $B_2$ and $B_3$ from different heavy-ion collision experiments as a function of $\sqrt{s_{\mathrm{NN}}}$. Data from heavy-ion collisions, where open symbols represent the anti-nucleus measurement. The horizontal dashed lines at low energies indicate the $B_2$ and $B_3$ values in elementary collisions as pp, p$\bar{\mathrm{p}}$, p--A and $\gamma$A but also the Bevalac heavy-ion data is close to it. The dashed-dotted lines show a simple model assuming $B_A\propto 1/V^{A-1}$, where the volume $V$ is taken from HBT radius measurements by STAR at their beam energy scan~\cite{Adamczyk:2014mxp}. Please note that the ALICE $B_3$ measurement from $^3$He nuclei is in a broader centrality interval (0--20\%) as the corresponding $B_2$ (0--10\%).}
\end{center}
\end{figure}

Figure~\ref{b2_vs_sqrts} shows the experimentally extracted $B_2$ and $B_3$ (mainly from central collisions of these publications~\cite{PhysRevC.60.064903,PhysRevC.60.064901,PhysRevC.61.064908,PhysRevC.61.044906,PhysRevC.65.034907,Ambrosini1998202,Bearden2002,Afanasiev200022,PhysRevC.85.044913,PhysRevC.69.024902,PhysRevC.94.044906}) as a function of the center-of-mass energy in the heavy-ion collisions. One finds that the coalescence parameters are rather insensitive on the collision energy in heavy-ion collisions since all data points of $B_2$ lie around $10^{-3}$ and $B_3$ between $10^{-7}$ and $10^{-6}$. Nevertheless, the coalescence parameters in elementary collisions are are found to be significantly higher as indicated by the constant dashed lines at low energies. The latter is also visible in figures~\ref{coal_b2_pp} and~\ref{coal_b3_pp}. The dashed-dotted line shows a simple model description of these data assuming $B_A$ to be proportional to $(1/V)^{A-1}$ (see Eq.~\ref{eq:coal_volume}), where the volume $V$ is obtained from a parameterisation of STAR HBT radii measured at different energies~\cite{Adamczyk:2014mxp}. The volume from the STAR measurement agrees well with the volume from the thermal model~\cite{Andronic:2014zha}. 

One important observation in Fig.~\ref{b2_vs_sqrts} is that there is a significant jump between elementary collisions and central heavy-ion collisions. This is mainly connected to the different multiplicities in these collision systems, but rather insensitive to the collision energy since the data span an energy range from some GeV up to 13 TeV, see Fig.~\ref{coal_b2_pp}. If one plots for instance $B_2$ as a function of the mean number of particles (e.g. $\langle \mathrm{d}N_\mathrm{ch}/\mathrm{d}\eta \rangle$) as in Fig.~\ref{d_over_p} for the d/p ratio there is a smooth transition visible from pp towards Pb--Pb as shown recently by the ALICE Collaboration at QM2018.  

\section{Results of (anti-)hypernucleus production measurements}
\label{sect:hypernuclei}

Hypernuclei, as bound states of nucleons and hyperons, are of particular interest. Their study provides an interesting testing ground of the baryon-hyperon interaction. In their ground states they generally decay weakly, i.e. have lifetimes of the order of 1-10$\times10^{-10}$ s. Generally they are produced and identified by $(K^-,\pi^-)$ or $(\pi^+,K^+)$ reactions on stable nuclear targets~\cite{Hashimoto:2006aw}. In relativistic nuclear collisions 
their signal can be reconstructed by an invariant mass analysis, using the decay products.  For instance the hypertriton $^{3}_\Lambda$H decays into $^3$He + $\pi^-$ and $^3$He and $\pi$ with displaced vertex (see below) can be well identified with detectors of STAR at RHIC and ALICE at the LHC. 

The hypertriton  is a bound state of a proton, a neutron and a $\Lambda$ baryon. The separation energy of the $\Lambda$ from the p and n inside the hyper-nucleus is only about a few hundred keV, which leads to a rough estimate for its rms radius (distance of $\Lambda$ to d) of between 5 and 10 fm. In addition, the very low binding energy implies only a small modification of the wave function of the $\Lambda$ inside the hyper-nucleus As a consequence, we expect the lifetime of the hypertriton to be very close to that of the free $\Lambda$ as discussed later in more detail. 

\subsection{Transverse momentum spectra}
\begin{figure}[!htb]
\begin{center}
\includegraphics[width=0.49\textwidth]{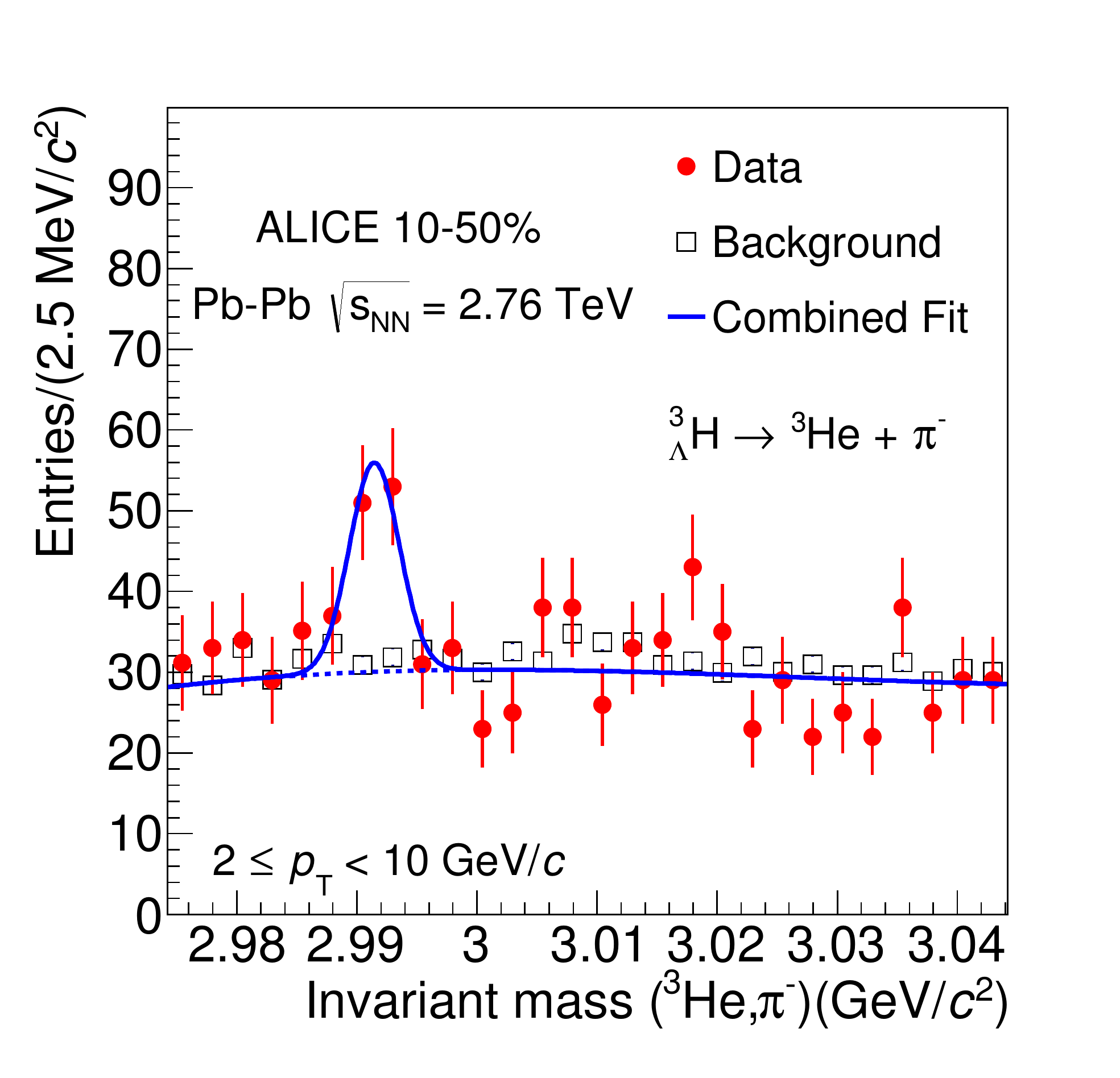}
\includegraphics[width=0.49\textwidth]{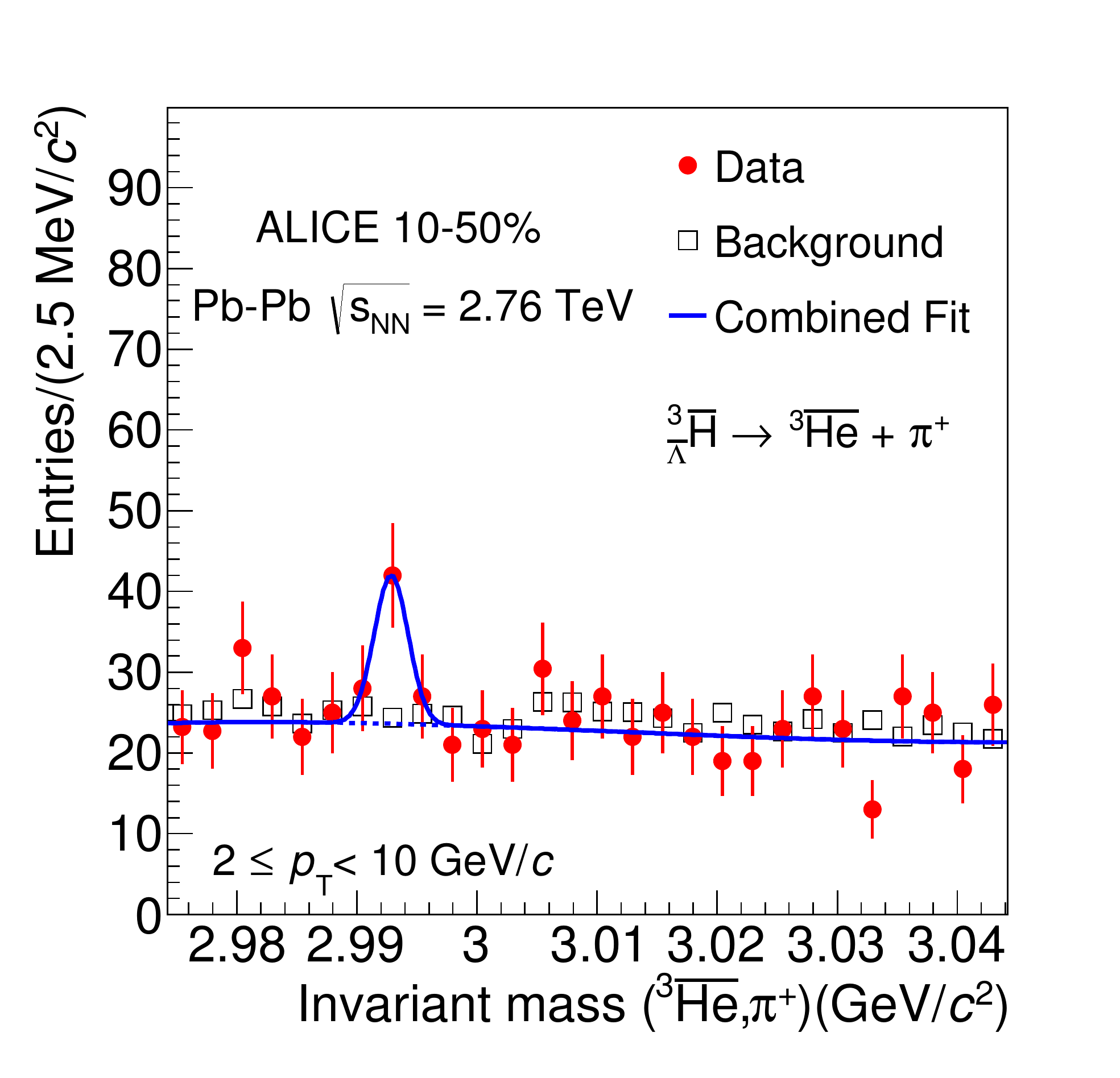}
\caption{\label{inv_mass_hypertriton} Invariant mass of hypertriton (left) and anti-hypertriton (right) for events with 10-50\% centrality in the hypertriton $2 \leq p_\mathrm{T} < 10$ GeV/$c$ interval. The data points are shown as filled circles, while the squares represent the background distribution. The line indicates the function used to perform the fit and used to evaluate the background and the raw signal. Figure originates from~\cite{hypertriton}.}
\end{center}
\end{figure}

\begin{figure}[!htb]
\begin{center}
\includegraphics[width=0.7\textwidth]{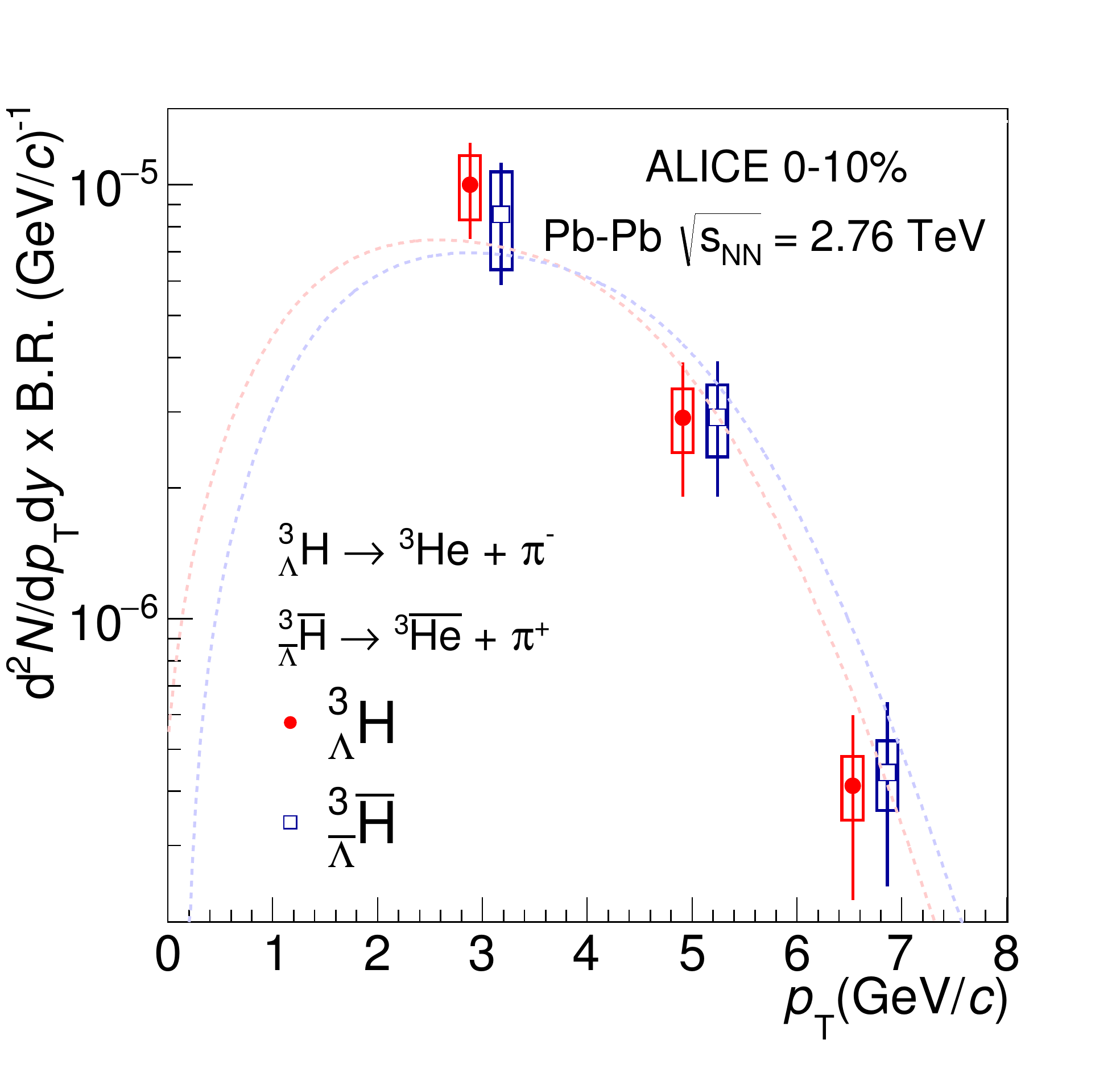}
\caption{\label{spectra_hypertriton} Transverse momentum spectra multiplied by the branching ratio B.R. of the charged two-body decay for hypertriton (filled circles) and anti-hypertriton (squares) for the most central Pb--Pb collisions for $|y| < 0.5$. The indicated dashed lines are the blast-wave curves used to extract the particle yields integrated over the full $p_\mathrm{T}$-range. Figure taken from~\cite{hypertriton}.}
\end{center}
\end{figure}

The signal extracted in a centrality interval of 10-50\% in a transverse momentum bin of $2 \leq p_\mathrm{T} < 10$ GeV/$c$ for hypertriton and anti-hypertriton using the ALICE setup is displayed in Figure~\ref{inv_mass_hypertriton}. The statistics gathered in the 2011 Pb--Pb collision data taking period by the ALICE Collaboration allowed for a split in two centrality classes and three transverse momentum bins for hypertriton, respectively anti-hypertriton. The transverse momentum spectrum x branching ratio is depicted in Figure~\ref{spectra_hypertriton} compared with a scaled blast-wave function. 

The measurement of transverse momentum distributions for the hypertriton is of key importance for the understanding of its production mechanism. Already the measurement of deuteron transverse momentum spectra clearly showed that the shape of these distributions agrees well the notion that the deuteron participates in the hydrodynamic expansion of the fireball along with all other hadrons. This is recognised by a hydrodynamic analysis based on the 'blast wave' approach.  In this approach the overall expansion parameters are collected as discussed above in a formula describing the transverse expansion. 

It already came as a surprise that this  blast-wave function can be used to describe the measured deuteron distributions  under the assumption that all blast wave parameters except the mass remain constant, implying full participation in the expansion. It would be even more surprising if the very loosely bound hypertriton 'flows' along with the deuteron, as well as with the tightly bound other hadrons. First results are visible in Figure~\ref{spectra_hypertriton} and indeed lend support to this picture. Further results will come from the LHC Run2 data taking during 2015 - 2018. A precision test of this assumption by a high statistics measurement of hypertriton p$_T$ spectra is planned for Run3 of ALICE data taking. If the flow hypothesis is confirmed this would be a challenge for coalescence models but be naturally explained in the multi-quark production hypothesis discussed below.

\subsection{Rapidity distributions} 

\begin{figure}[!htb]
\begin{center}
\includegraphics[width=0.7\textwidth]{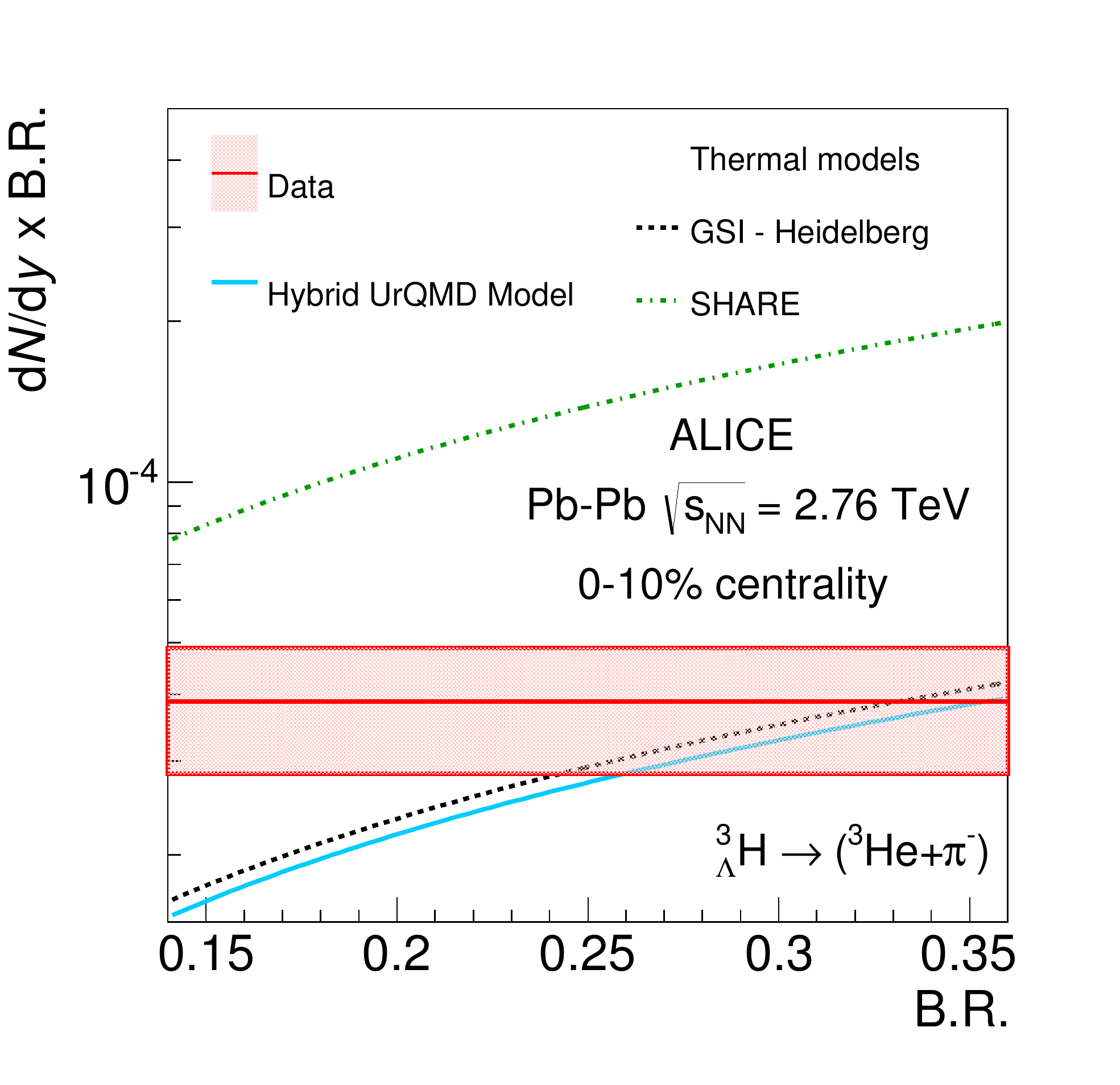}
\caption{\label{dndy_br} $p_\mathrm{T}$-integrated rapidity density times branching ratio as a function of branching ratio (d$N$/d$y~\times~$B.R. 
vs B.R.). The horizontal line is the measured d$N$/d$y~\times~$ B.R. and the band around it represents the quadratic sum of statistical and systematic uncertainties. The dashed lines indicate different theoretical expectations. Taken from~\cite{hypertriton}.}
\end{center}
\end{figure}

From this also an integrated rapidity density x branching ratio can be extracted as shown in Figure~\ref{dndy_br}. It is compared to model predictions as a function of the branching ratio. An exact measurement of all decay modes and their branching ratios is still needed. The values 0.15 and 0.35 are limits set by experimental knowledge of ratios of branching ratios. The upper limit for instance is determined by the ratio of the $^3$He + $\pi^-$ decay channel to all decay channels containing a $\pi^-$. The most referenced theoretical calculation expects a branching ratio of about 25\%, which is also used to correct the experimental data~\cite{star}. 
The same thermal model which is used to predict the light nuclei yields describes also the (anti-)hypertriton yield rather well around the expected branching ratio.

\begin{figure}[!htb]
\begin{center}
\includegraphics[width=0.7\textwidth]{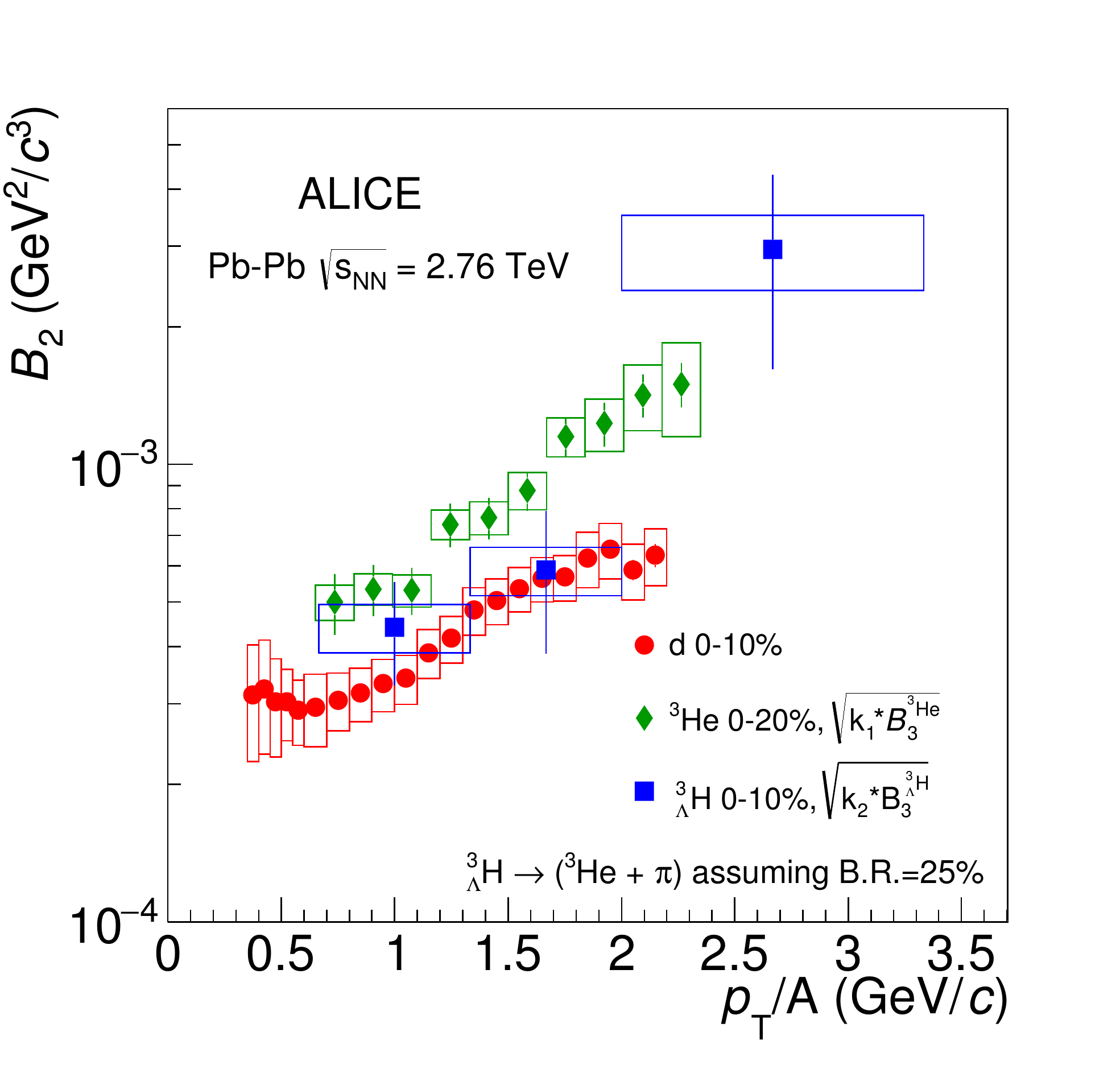}
\caption{\label{b2_with_b3} Comparison of the coalescence parameter $B_2$ for d, $^3$He and $^3_\Lambda$H. The $B_2$ values of $^3$He and $^3_\Lambda$H were calculated by scaling the $B_3$ parameter. Taken from~\cite{hypertriton} and thus more details can be found there.}
\end{center}
\end{figure}

One can further compare the coalescence parameters of different light nuclei with those of the hypertriton, as shown in Figure~\ref{b2_with_b3}. This is done by scaling the $B_3$ value determined for $^3$He and $^3_\Lambda$H to the $B_2$, to allow for a comparison (using the mass scaling given by equation~\ref{eq:coal_mass}). 

%Currently, no prediction of the $^3_\Lambda$H and anti-$^3_\Lambda$H yields from non-trivial dynamical coalescence models is available for LHC energies.
%Nevertheless, a simple coalescence model can be used to estimate parameters which can be compared to data to test if hypernuclei are formed through a coalescence mechanism.

\subsection{Impact on thermal analysis}
\label{impact}

In the thermal approach the production yield of loosely-bound states is entirely determined by mass, quantum
numbers and fireball temperature while the yield in the framework of coalescence should significantly depend on the relevant wave functions.
The hypertriton and $^3$He have very different wave functions but have essentially equal production yields, as we will see later on.

In contrast to what was discussed before, the energy conservation needs to be taken into account when forming objects with
baryon number $A$ from $A$ baryons, since the coalescence of off-shell nucleons does not help as the density must be much lower than nuclear
matter density.
To quantify the delicate balance between formation and destruction one can calculate the maximum momentum
transfer onto the hypertriton before it breaks up, which is of the order $Q_\mathrm{max} < 20$ MeV/$c$, whereas typical pion
momenta  are $p_\pi > 250$ MeV/$c$, and the typical hadronic momentum transfer in the fireball is  $\langle Q \rangle > 100$ MeV/$c$.
This means the hypertriton interaction cross-section with pions or nucleons at thermal freeze-out is of order $\sigma \approx 70$ fm$^2$. For the majority of hypertritons to survive, the mean-free path $\lambda$ has to exceed the system size at thermal freeze-out which is estimated~\cite{Andronic:2017pug} to be about 10 fm. Taking $\lambda > 15 $fm for a rough estimate this would lead to a density of the fireball at formation of hypertriton of $n < 1/(\lambda\sigma) = 0.001$ fm$^{-3}$. This is completely inconsistent with a formation at kinetic freeze-out,
where typically $n = 0.05$ fm$^{-3}$.
In addition to that, the description of the centrality dependence of spectra and d/p ratio as a function of multiplicity is not consistent with current coalescence predictions.

Interestingly, the above mentioned facts are not raising any troubles for the thermal model. The only scale there (at LHC energy and below) is the temperature $T < 160$ MeV.
At such a scale, the momentum transfer $q = T$, and the form factors of hadrons are sampled at $q^2 = T^2$. This implies that sizes of hadrons  $d < 2$ fm cannot be resolved
since $G(q) \propto 1- q^2R^2/6$, and since all (rms) radii for nuclei with $A$ = 2, 3, and 4 are smaller than 2 fm,
the correction due to the finite size of nuclei will not exceed 35\%.
As such the actual change from this on thermal model results should be much less as
only the relative change between normal hadrons and light nuclei matters, the
overall change only leads to a volume correction, so the correction for nuclei is
estimated to be less than 25\%.
On the other hand, the hypertriton has a radius exceeding 5 fm, while the measured yield of hypertriton and $^3$He is well compatible with the thermal model
prediction, even though their wave functions are very different. Because of its large size, however,   hypertriton production  should not be described thermal model calculations.
In fact, the agreement of the yields of baryon number 3 states with the predictions using the thermal model supports the notion that, at the low thermal scale of about 155 MeV, their wave function matters little for the production process, in contrast to expectations  within the framework of coalescence models. We will provide, in section~\ref{discussion} an interesting but speculative way out of this dilemma based on the assumption that loosely-bound states are formed at chemical freeze-out as compact multi-quark states.

\begin{figure}[!htb]
\begin{center}
\includegraphics[width=0.7\textwidth]{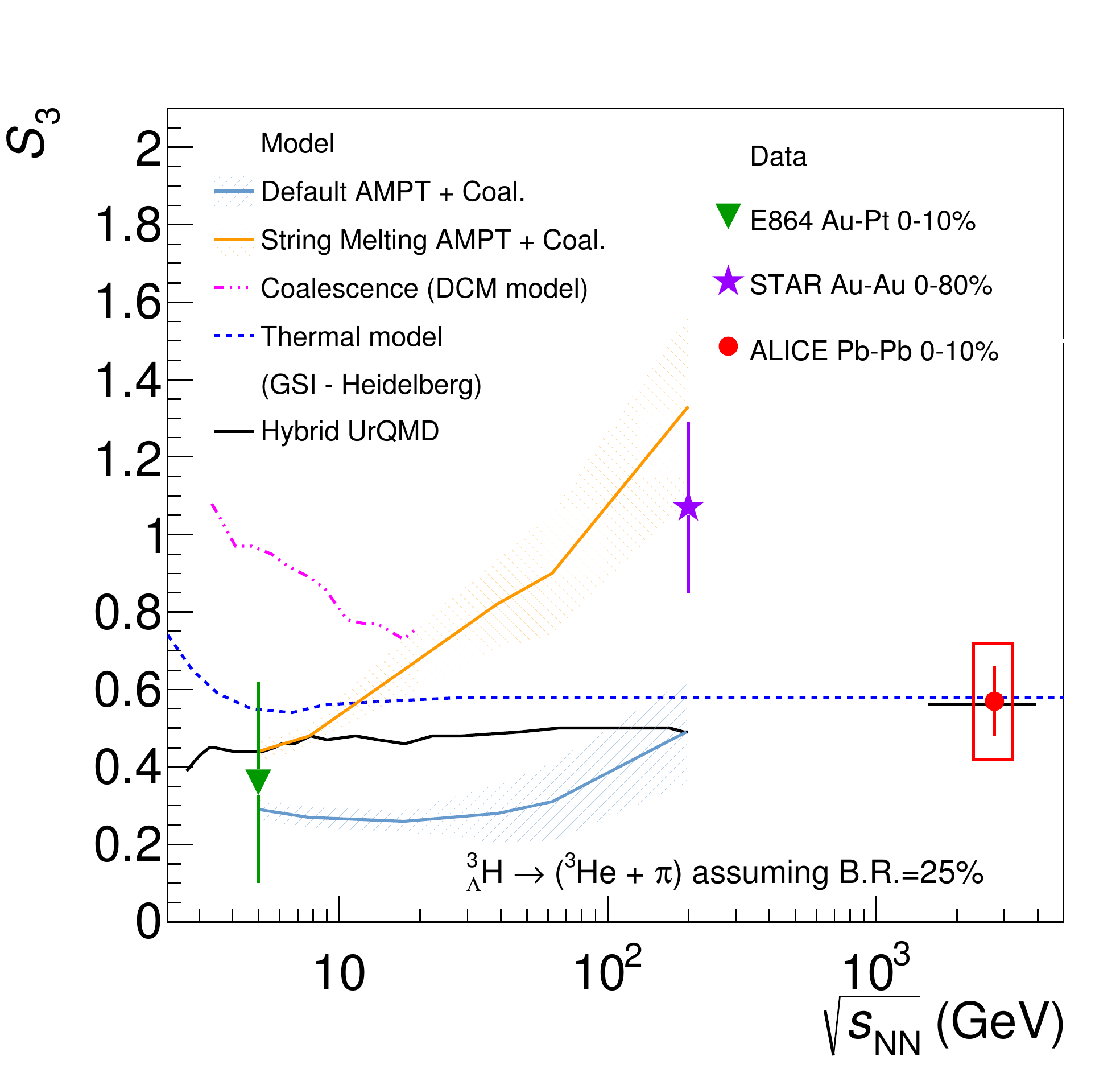}
\caption{\label{s3} Strangeness population factor $S_3$ as a function of the $\sqrt{s_\mathrm{NN}}$ compared with different model predictions. Figure from~\cite{hypertriton}. For details see text.}
\end{center}
\end{figure}

The STAR collaboration has proposed to use the strangeness population factor $S_3$  defined as the ratio of the production yield of hypertriton to that of $^3$He times the production yield ratio of p over $\Lambda$ to characterise the production process. The $S_3$ factor is thought as a quantity describing the local baryon-strangeness correlations~\cite{Koch:2005vg, Majumder:2006nq,Cheng:2008zh}. $S_3$ is shown in Figure~\ref{s3} as a function of $\sqrt{s_\mathrm{NN}}$ to compare the measurements at AGS energies (E864), with those at RHIC energy (STAR) and with the most recent measurement from the ALICE Collaboration at the LHC. The data points are compared with theory predictions from different models, namely three versions of coalescence model implementation (two different versions of AMPT and the DCM model) and two models belonging to different thermal model approaches. The data are well compatible with the thermal model and hybrid UrQMD predictions. String models and simple coalescence models do not describe the observations. The somewhat high value from the STAR collaboration is surprising given the new ALICE data.

\subsection{Hypertriton lifetime}
\label{lifetime}
Another important topic is the measurement of the lifetime of the hypertriton, especially because it goes hand in hand with the branching ratio and the binding energy. Since its observation its lifetime has been determined in several experiments~\cite{Prem:1964hyp, Keyes:1968zz, Phillips:1969uy, Bohm:1970se,Keyes:1970ck, Keyes:1974ev, Abelev:2010sci, Rappold:2013fic}. Until the first measurement in heavy-ion collisions by the STAR Collaboration in the year 2010, the results stem mainly from strangeness-exchange reactions in emulsions. At the same time calculations were done connecting the experimental information of the binding energy $E_B$, or better the separation energy of the $\Lambda$ $B_\Lambda$ being only about 130 keV~\cite{Juric:1973mh}, and the expected lifetime. \\
The theoretical predictions \cite{Dalitz:1959zz, Dalitz:1962eb, Rayet:1966fe, Ram:1971tf, Mansour:1979xw, Kolesnikov:1988uy,Congleton:1992kk, Gloeckle:1998ty, Kamada:1997rv} 
span a range of $0.7\times\tau_\Lambda$ to $0.97\times\tau_\Lambda$. The latter value is the most recent calculation by Kamada et al.~\cite{Kamada:1997rv}.

\begin{figure}[!htb]
\begin{center}
\includegraphics[width=0.9\textwidth]{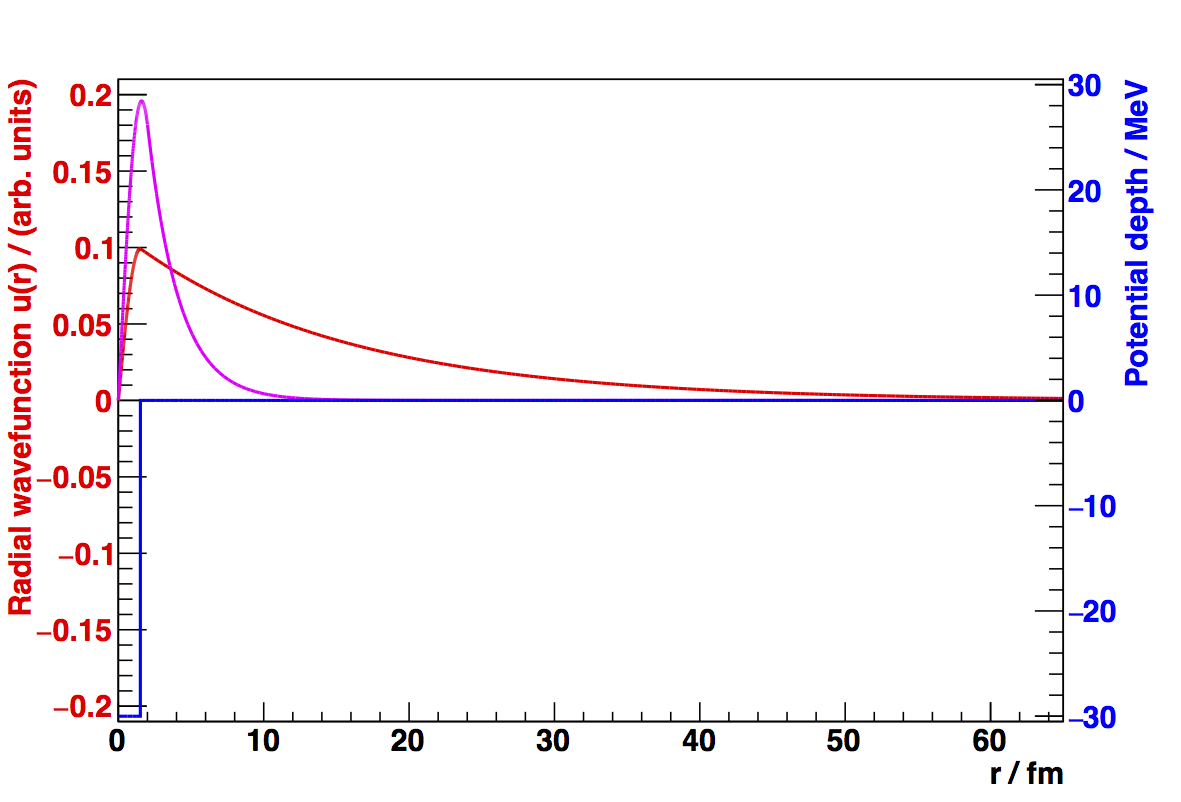}
\caption{\label{wavefunction} Wave function (red) of the hypertriton assuming a s-wave interaction for the bound state of a $\Lambda$ and a deuteron. The root mean square value of the radius of this function is $\sqrt{\langle r^2 \rangle} =  10.6$~fm. In blue the corresponding square well potential is shown. In addition, the magenta curve shows a "triton" like object using a similar calculation as for the hypertriton, namely a deuteron and an added nucleon, resulting in a much narrower object.}
\end{center}
\end{figure}

A simple model already shows that the lifetime of the $\Lambda$ in the hypertriton should be very close to the free one. If one assumes a pure s-wave interaction between the $\Lambda$ and the deuteron a similar calculation can be done as for the deuteron~\cite{segre,weisskopf,kamal}.  This pure bound state of a $\Lambda$ and a deuteron is described by a non-relativistic Schr\"odinger equation with a square well potential of depth $V_0$. The solution now depends on the range of the potential. This range is usually indicated by $R$. For a potential depth $V_0 = -30$ MeV with a bound state of a $\Lambda$ separation energy of $E_B = 130$ keV, $R$ is around 1.5~fm. This means the attractive potential is only acting in a very restricted region. The calculated wave function and the potential are depicted in Figure~\ref{wavefunction}. From the wave function one can clearly see that the probability to find the $\Lambda$ close to the potential/deuteron is very small. A simple calculation leads to a probability of about 90\% to find the $\Lambda$ outside the potential well, which shows that the $\Lambda$ in this simple model is most of the time outside of the potential region and thus its wave function should not be modified too strongly. As a consequence the lifetime of the hypertriton should not be too much different from that of the free $\Lambda$. From this simple quantum mechanical approach one can also estimate the rms radius of the object to be $\sqrt{\langle r^2 \rangle} = 10.6$~fm, so clearly larger than a lead nucleus. This value coming from the integration of the displayed wave function is very close to that coming from the approximation without the acting potential, namely $\sqrt{\langle r^2 \rangle} = \frac{1}{\sqrt{4\mu E_B}}$, where $\mu$ is the reduced mass of the $\Lambda$-deuteron system. 
A completely different view on this comes from the discussions of halo nuclei, and under certain assumptions the hypertriton can be seen as one or even as an Efimov state~\cite{Hammer:2017tjm}. 

\begin{figure}[!htb]
\begin{center}
\includegraphics[width=0.8\textwidth]{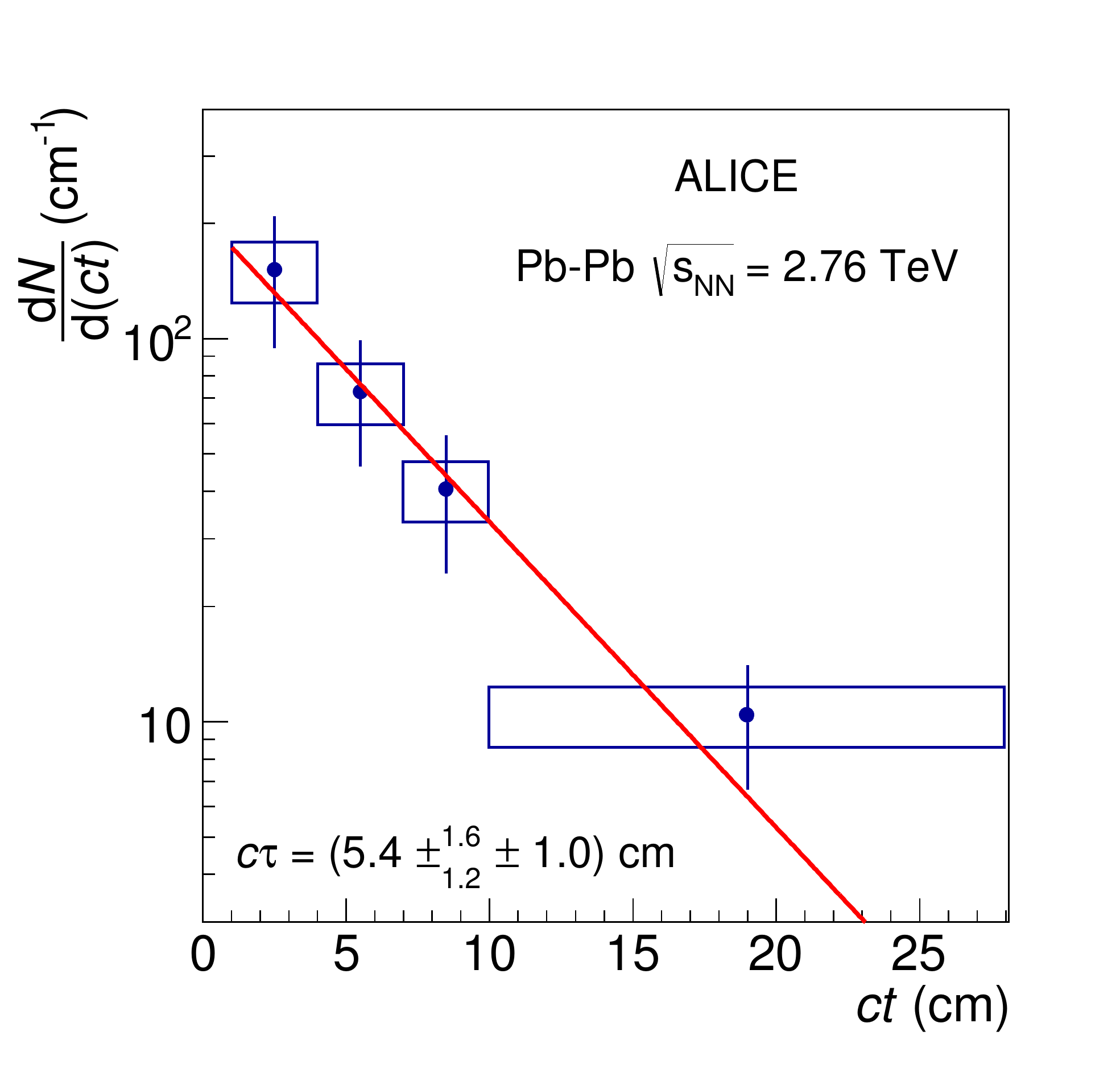}
\caption{\label{ctau_hypt} Measured d$N$/d($ct$) spectrum shown together with an exponential fit to determine the lifetime, taken from~\cite{hypertriton}.}
\end{center}
\end{figure}

The data taken by the ALICE Collaboration was split into four $c$t bins combining the signal of hypertriton and anti-hypertriton to extract the lifetime experimentally, then an exponential fit was performed to determine the lifetime. 
The d$N$/d($c$t) distribution and the exponential fit are shown in Figure~\ref{ctau_hypt}. \\
The fit results in a proper decay length of $c\tau = \left(5.4^{+1.6}_{-1.2}(\mathrm{stat.})\pm 1.0 (\mathrm{syst.})\right)$~cm.

\begin{figure}[!htb]
\begin{center}
\includegraphics[width=0.85\textwidth]{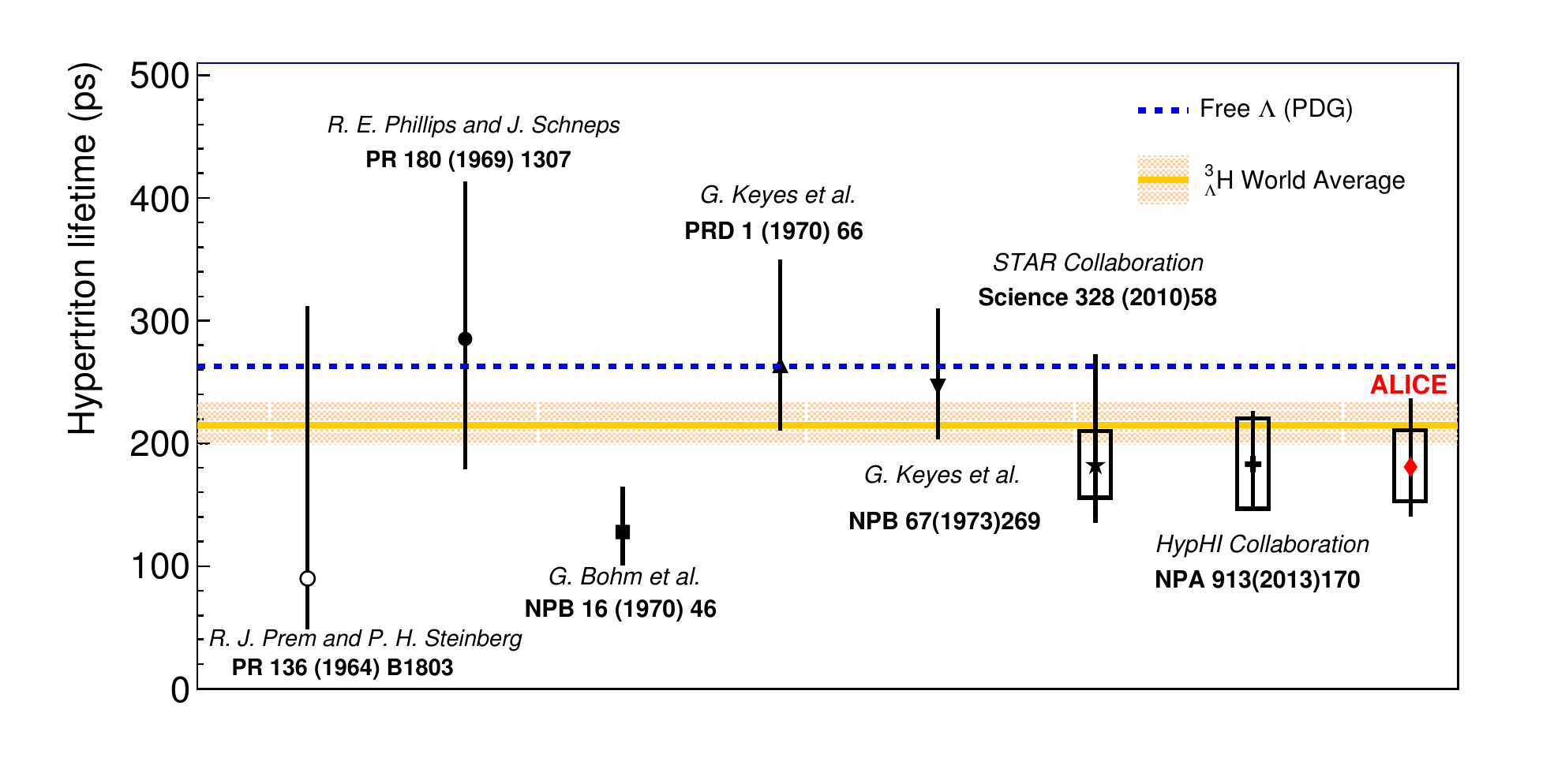}
\caption{\label{lifetime_hypt_paper} Hypertriton lifetime measured by the ALICE Collaboration (red symbol) and compared with previously published results (see text for details). The band represents the world average of the hypertriton lifetime measurements. The dashed line indicates the lifetime of the $\Lambda$ hyperon as reported by the Particle Data Group. Taken from~\cite{hypertriton}}
\end{center}
\end{figure}

A new statistical combination of the experimental lifetime measurements of the hypertriton was recently performed~\cite{Rappold:2014jqa}, leading to an average value of $\tau = \left(216^{+19}_{-18}\right)\ \mathrm{ps}$ ($0.82\times\tau_\Lambda$).\\
The results from the measurement of the ALICE Collaboration yields a lifetime of $\tau = \left(181^{+54}_{-39} (\mathrm{stat.})\pm 33 (\mathrm{syst.})\right)$ ps, as shown together with the previously published results~\cite{Prem:1964hyp, Keyes:1968zz, Phillips:1969uy, Bohm:1970se,Keyes:1970ck, Keyes:1974ev, Abelev:2010sci, Rappold:2013fic} in Figure~\ref{lifetime_hypt_paper}.
This and the previous measured values were used to calculate a new world average as described in \cite{Rappold:2014jqa}. It is evaluated to be $\tau = \left(215^{+18}_{-16}\ \mathrm{ps}\right)$ and shown as a band in   
Figure~\ref{lifetime_hypt_paper} together with the data. The published ALICE result is compatible with this average.  

\begin{figure}[!htb]
\begin{center}
\includegraphics[width=0.98\textwidth]{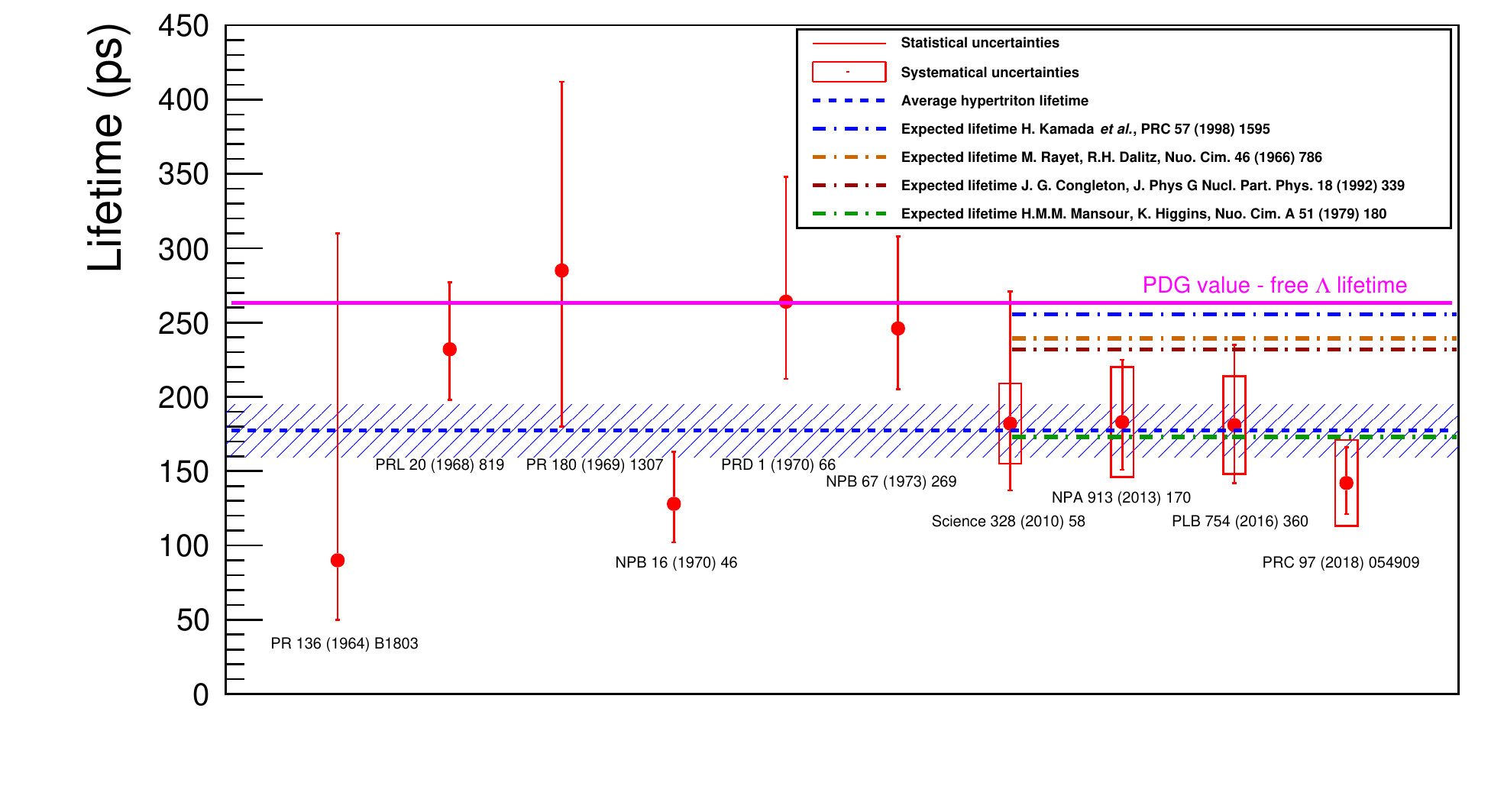}
\caption{\label{lifetime_new} An updated collection of the hypertriton lifetime measurements. The additional point corresponds to the newly released measurement of the lifetime by the STAR Collaboration~\cite{Adamczyk:2017buv}. }
\end{center}
\end{figure}

Recently, the STAR Collaboration published a new measurement and gives a rather low new value of the lifetime combining the results of the 2-body decay channel and the 3-body decay channel using the statistics gathered in the RHIC energy scan program~\cite{Adamczyk:2017buv}. To estimate a new world average value of the lifetime we added this value to the previous results and performed a new fit, shown in~\ref{lifetime_new} together with the current existing experimental results. The corresponding new world average shown there is $(177 \pm 18)$~ps together with the most reliable model calculations of the hypertriton lifetime~\cite{Rayet:1966fe,Congleton:1992kk,Kamada:1997rv}. 

In any case, the uncertainties, in particular the systematic ones, have to be taken seriously and only lead to a deviation of 2$\sigma$ of the world average compared with the free $\Lambda$ lifetime. For an indication of the expected improvement in Run2 see figure~\ref{hypt_lukas} below. From these data ALICE extracted a preliminary value of $\tau = \left(237^{+33}_{-36} (\mathrm{stat.})\pm 17 (\mathrm{syst.})\right)$ ps which is in agreement with the new world average and the free $\Lambda$ lifetime~\cite{Trogolo:2017oii}. Using this most precise preliminary value in the calculation of the world average the mean value would go up to $(188 \pm 16)$~ps.

\section{Results of searches for dibaryon states}
\label{exotica}

In addition to the previously discussed (anti-)nuclei and (anti-)hypernuclei the ALICE Collaboration has started to investigate the hyperon-nucleon and hyperon-hyperon interaction through the search of signals of possible bound states involving hyperons. This is done in invariant mass analyses of different possible decay channels of these objects.\\

\begin{figure}[!htb]
\begin{center}
\includegraphics[width=0.49\textwidth]{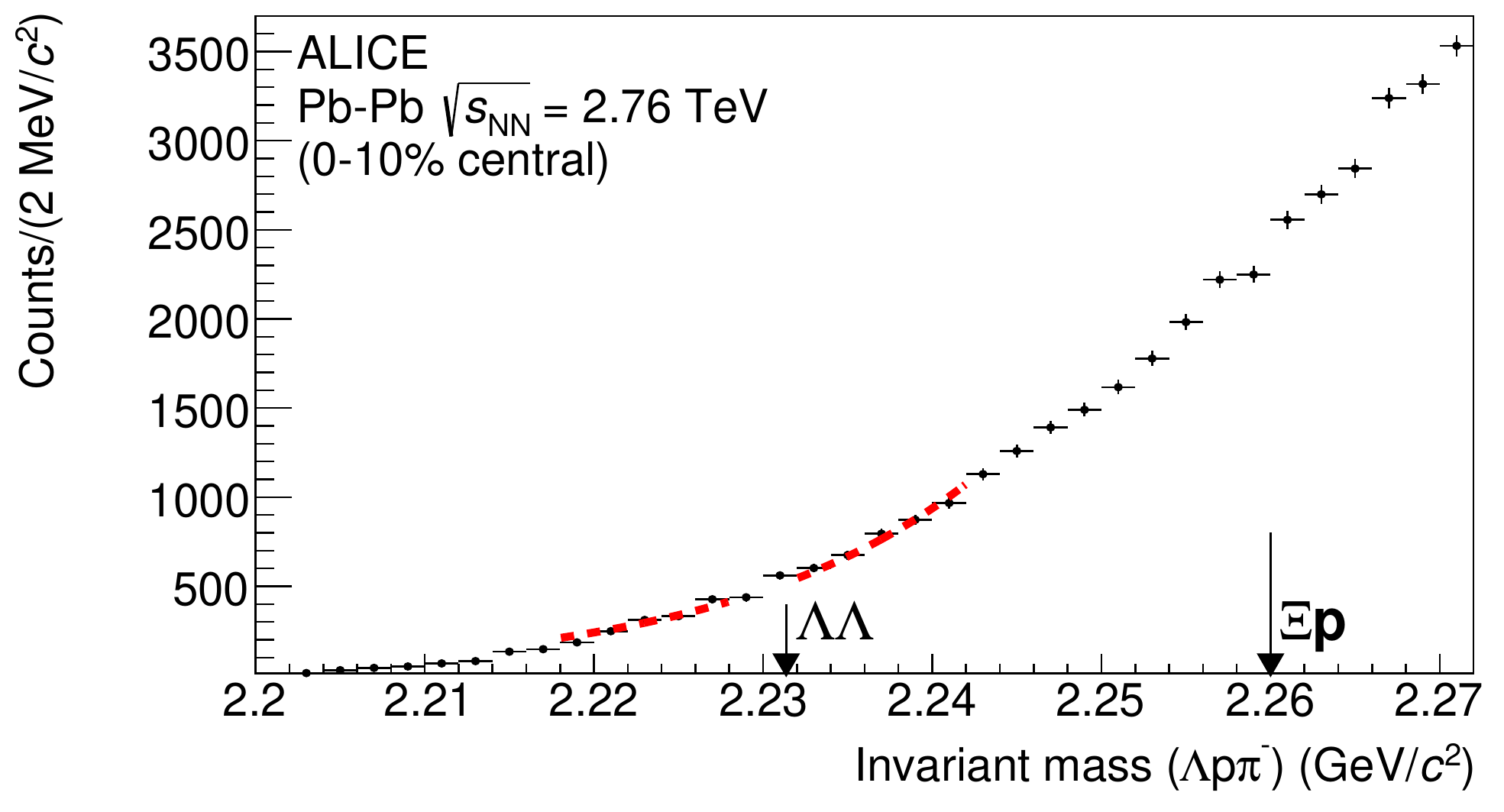}
\includegraphics[width=0.49\textwidth]{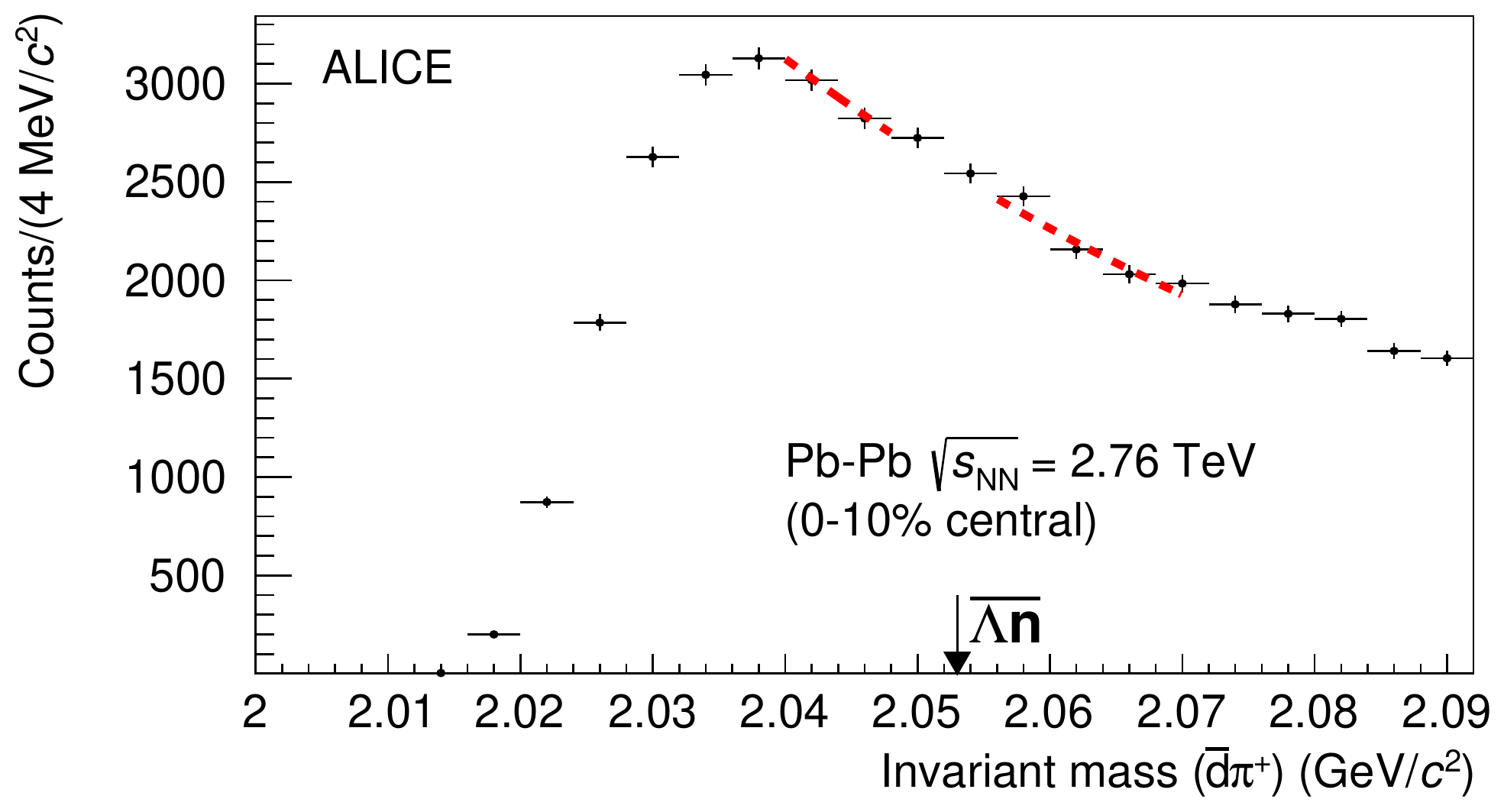}
\caption{\label{inv_mass_exotica} Invariant mass of $\Lambda+$p$+\pi^-$ (left) and $\bar{\mathrm{d}}+\pi^+$ (right) reconstructed in $19.3\times10^6$ central events. Taken from~\cite{searches}.}
\end{center}
\end{figure}

The most prominent example is the H-dibaryon, a hexaquark state consisting of $uuddss$, which has the same quantum numbers as a bound state of two $\Lambda$ hyperons. It was first predicted by R. Jaffe in a bag model calculation~\cite{jaffe}. Recent studies on the lattice lead to a potentially bound H-dibaryon, in calculations connected to the pursuit  to calculate light nuclei in QCD from first principles~\cite{beane,inoue}. These calculations are still done at unphysically high pion masses, and therefore need an chiral extrapolation to the physical point~\cite{shanahan,haidebauer}. With these further steps the H-Dibaryon becomes rather unbound, by 13$\pm$14 MeV, or could be still weakly bound by about 1 MeV. The study of double $\Lambda$ hyper-nuclei and there the well known $Nagara$ event lead also to a possible binding energy of 1 MeV for the $\Lambda\Lambda$ system~\cite{Botta:2012xi,Gal:2016boi}.\\
Tp provide new information on a possible $\Lambda \Lambda$ bound state the ALICE collaboration studied the possible H-dibaryon in the region around the $\Lambda\Lambda$ threshold. For this the analysis was done in the invariant mass of $\Lambda+$p$+\pi$ as shown in Figure~\ref{inv_mass_exotica}.\\
The other candidate which was studied by the ALICE Collaboration is a possible bound state of $\Lambda$n which would decay into a deuteron and a $\pi^-$. The HypHI Collaboration observed a signal in this channel which was first interpreted in a preliminary analysis as the possible decay of the $\Lambda$n bound state~\cite{take}, since it was lying below the threshold. The published result is then clearly above the threshold and an interpretation as a bound state is withdrawn~\cite{hyphi}. The ALICE result in the same decay channel is depicted in Figure~\ref{inv_mass_exotica}.\\

\begin{figure}[!htb]
\begin{center}
\includegraphics[width=0.7\textwidth]{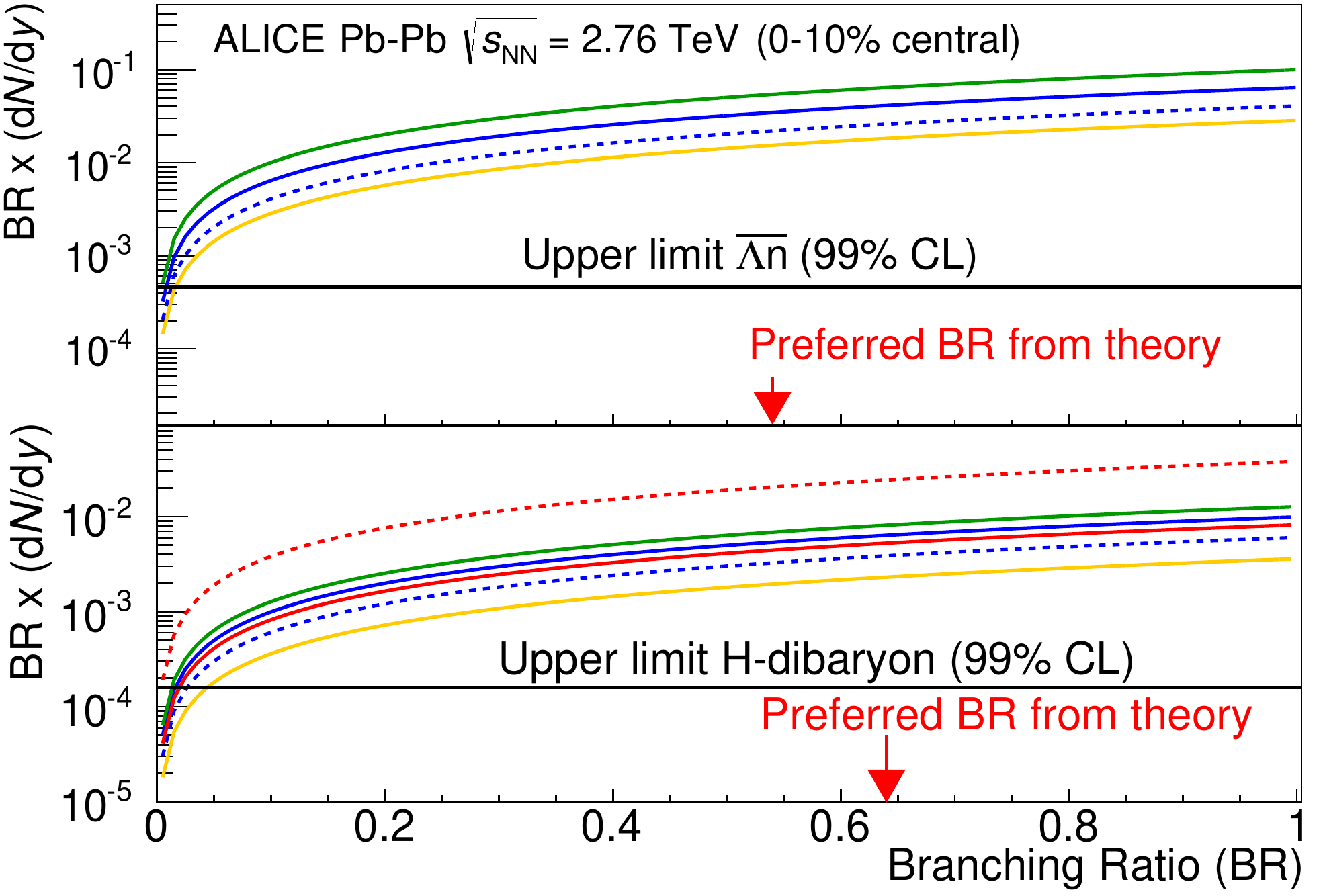}
\caption{\label{ul_br} Extracted upper limit d$N$/d$y$ as function of the branching ratio compared with different models. For details see~\cite{searches}.}
\end{center}
\end{figure}

\begin{figure}[!htb]
\begin{center}
\includegraphics[width=0.7\textwidth]{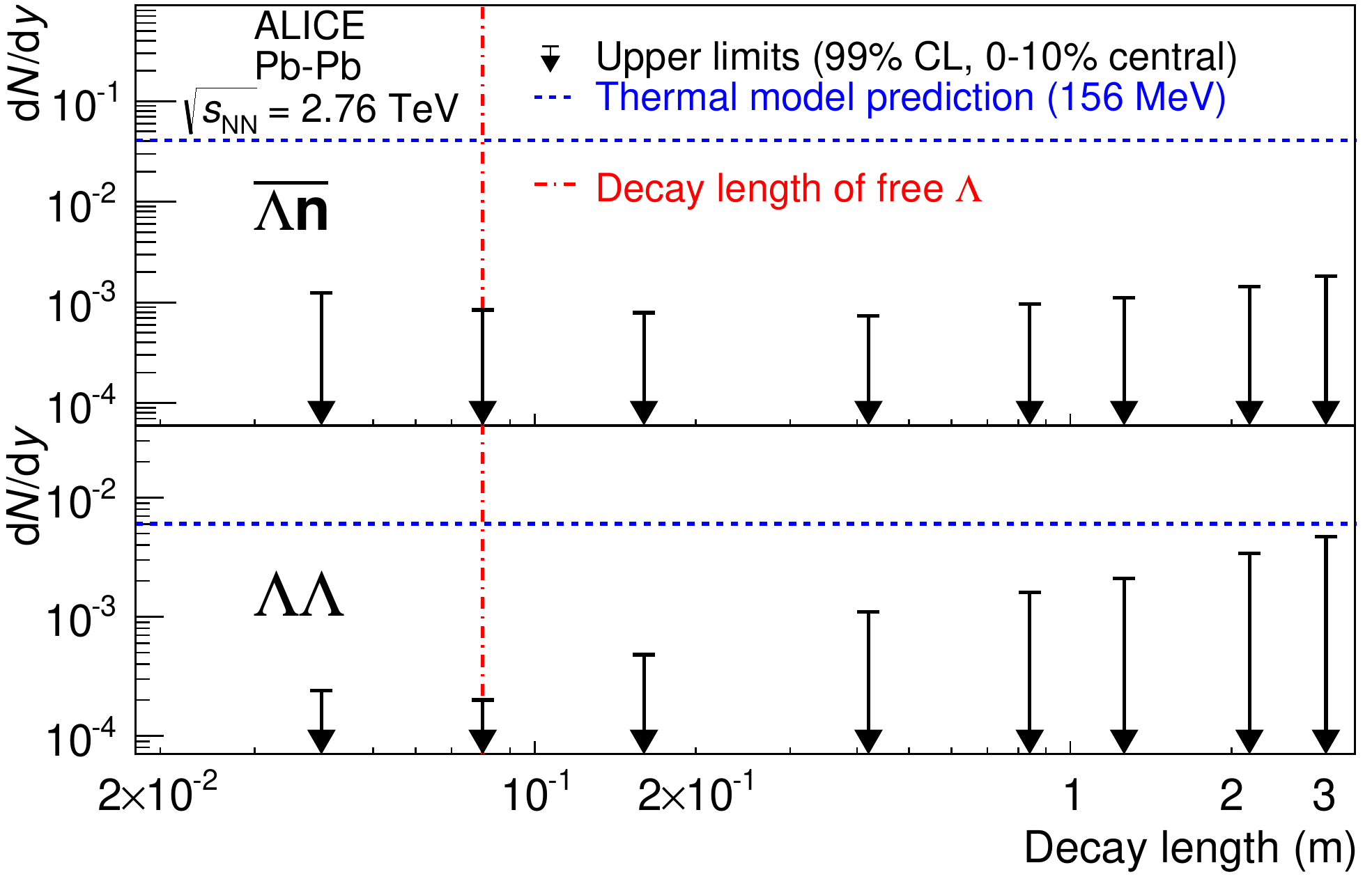}
\caption{\label{ul_lifetime} Upper limit d$N$/d$y$ as function of the decay length compared with the thermal model prediction for the hypothetical $\Lambda\Lambda$ and $\Lambda$n bound states. For details see~\cite{searches}.}
\end{center}
\end{figure}

Since both analyses did not yield a signal, upper limits were calculated being far from any  model predictions, including thermal models. Therefore, the analysis was extended in the phase space of decay length and branching ratio. The results are displayed in Figures~\ref{ul_br} and \ref{ul_lifetime}. For the study versus branching ratio the lifetime of the free $\Lambda$ hyperon was assumed and for this case only for unreasonable very small branching ratios the upper limit and the model predictions agree. For the theoretically preferred branching ratios the upper limits are factors of more than 20 away from the predicting models.\\
To analyze the upper limits in the phase space of decay length the branching ratio was set to the theoretically most favoured value. Also here the upper limits are far away from the model expectations as long as reasonable decay lengths are assumed.

Recently a theoretical study was performed~\cite{Sun:2016rev} using a more sophisticated method for the quark coalescence. In fact, these results mainly focus on the comparison with the thermal model which nicely describes the experimental results of the known states. This is done especially, because the predicting power of the (quark) coalescence is limited due to its dependence on the unknown wave function of the dibaryons and thus the additional input needed to get an estimate for the production probability, as already indicated in section~\ref{sect:hypernuclei}.   

\begin{figure}[!htb]
\begin{center}
\includegraphics[width=0.95\textwidth]{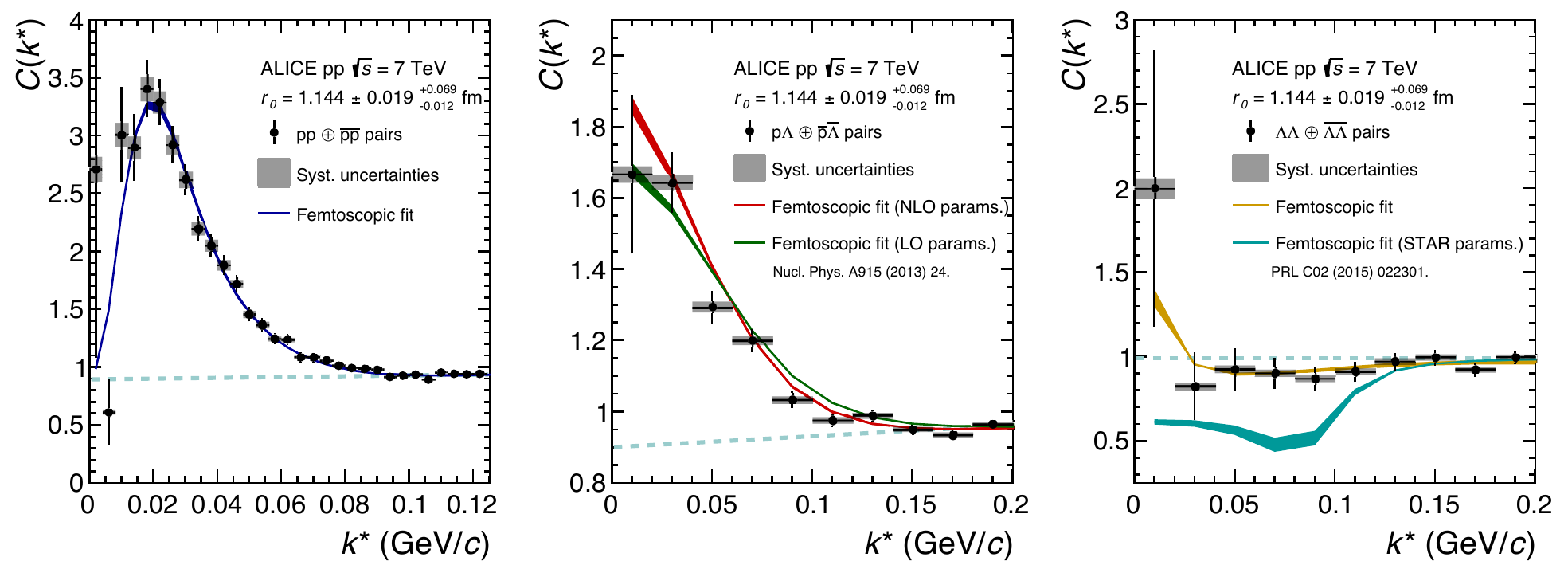}
\caption{\label{correlation_lp} The pp (\textit{left}), $\Lambda$p (\textit{centre}) and $\Lambda\Lambda$ (\textit{right}) correlation functions with a simultaneous fit with the Next-to-Leading-Order expansion (red line) for the scattering parameter of $\Lambda$p. The dashed line denotes the assumed linear baseline. After the fit is performed the Leading-Order~parameter set (green curve) is plugged in for the $\Lambda$p system and the scattering length obtained from~ for the $\Lambda\Lambda$  system (cyan curve). As shown in~\cite{Acharya:2018gyz}}
\end{center}
\end{figure}

Another way to study the interaction of baryons is through the two-particle intensity correlation function. Thus the hyperon-nucleon and  the hyperon-hyperon interaction can be extracted from the correlation functions~\cite{star_h,kenji}. This is possible by a precise modeling but is complicated because of the influence of feed-down on the correlation function and with this to the extracted interaction.\\ 
The first published results for the pp, $\Lambda$p and $\Lambda\Lambda$ correlation functions are depicted in Figure~\ref{correlation_lp}~\cite{Acharya:2018gyz}. A detailed study is described there to extract the interaction parameter from these correlation functions. The extracted scattering parameters for the $\Lambda\Lambda$ interaction from the data are allowing the existence of bound H-dibaryon if the effective range $d_0$ would be small and the scattering length $f_0$ would be negative. Nevertheless, the data and many models favour the possible existence of resonance state.   

\section{Discussion}
\label{discussion}

\begin{figure}[!htb]
\begin{center}
\includegraphics[width=0.95\textwidth]{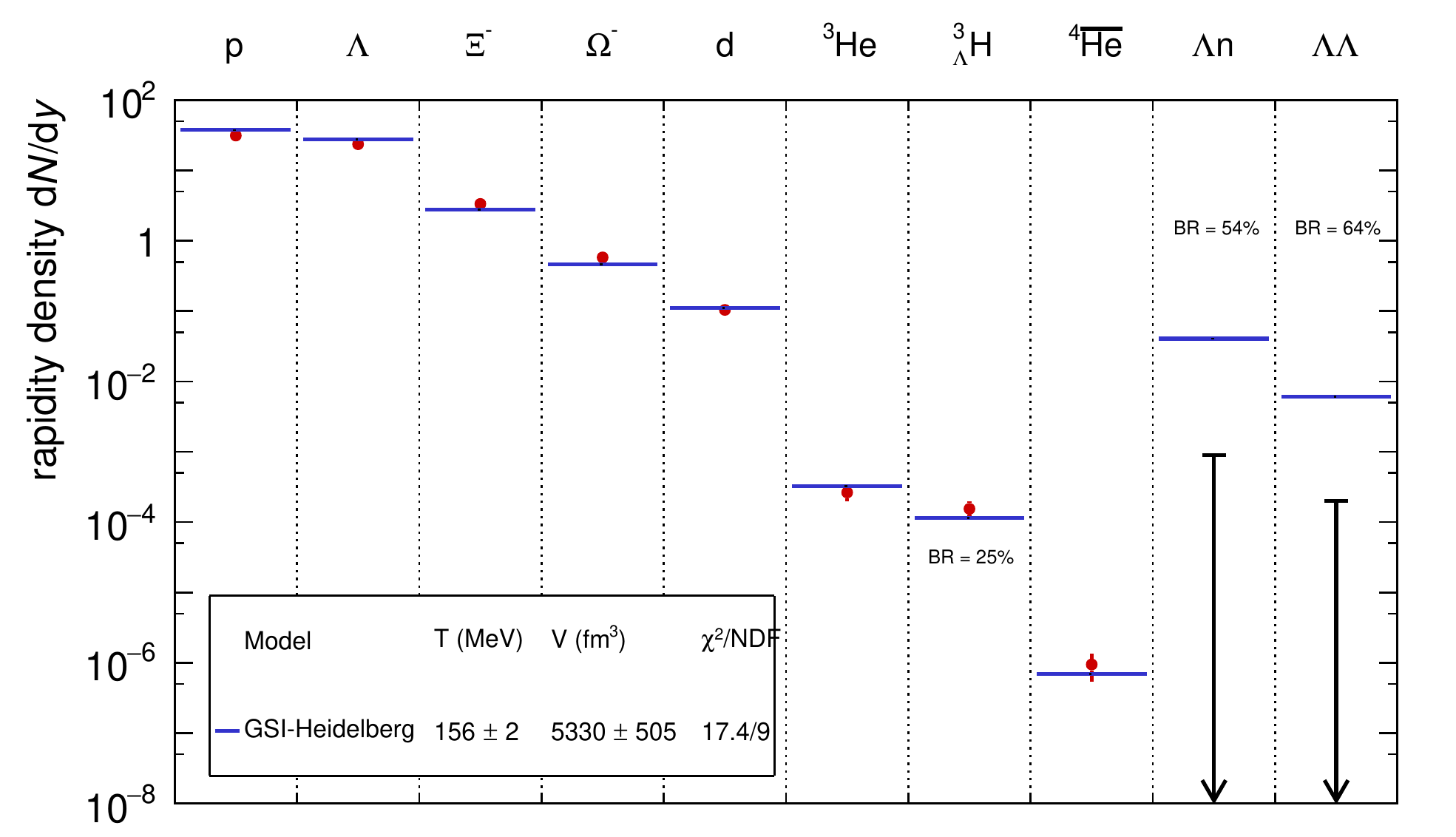}
\caption{\label{thermal_model_all} Comparison of ALICE production yields d$N$/d$y$ of baryons, light (hyper-)nuclei and exotica with thermal model predictions. The yield for hypertriton production was corrected by 25\% for the branching ratio, $\Lambda$n by 54\% and $\Lambda\Lambda$ by 64\%.}
\end{center}
\end{figure}

In the previous sections we have presented results on strange hadrons, light nuclei, the hypertriton and exotic bound states. A thermal model fit including the (hyper-)nuclei yields is displayed in Figure~\ref{thermal_model_all}. The outcome is basically the same as when the nuclei are not included in the fit, namely the temperature is 156 MeV (as also visible in Figure~\ref{thermal_fit_lhc}). This alone is very interesting because it means the yields of loosely-bound objects can be predicted by the thermal model, as also shown in Figure~\ref{thermal_model_all} for the $\Lambda\Lambda$ and $\Lambda$n bound state where upper limits have been estimated as discussed in Sec.~\ref{exotica}. The upper limits are more than a factor 20 away from the thermal model expectations.\\ 
On the other hand, the yields of loosely-bound objects as the deuteron (2.225 MeV binding energy), ${}^3_{\Lambda}$H (binding energy of 2.355 MeV) and ${}^3$He (2.57 MeV binding energy per nucleon) are nicely reproduced by the thermal model with a temperature of 156 MeV, which is 60 times above the binding energy of the (hyper-)nuclei.\\
Even more when only the separation energy of the $\Lambda$ inside the hypertriton is considered which is only 130 keV, thus the temperature is 1000 times higher than the separation energy and still the hypertriton yield is well described by the model.   

This fact can be understood when a isentropic expansion of the fireball is considered which means that the entropy per (net-)baryon is fixed at the chemical freeze-out as first pointed out in~\cite{Siemens:1979dz}. This means that since entropy is conserved the even very fragile objects survive the expansion and can be detected in the final state.

On top of the previously discussed facts, the produced nuclei exhibit a significant flow following the mass ordering as introduced in the very beginning. This means that the 'loosely-bound objects' experience a radial flow field similar to that for the other hadrons, leading to a significant increase of the transverse momenta (for the deuteron for instance the mean $p_\mathrm{T}$ is about 0.9 GeV/$c$ in pp collisions and reaches 2.2 GeV/$c$ in central Pb--Pb collisions). For the most extreme case, the hypertriton, this push is thus even significantly higher.

In fact, the nuclei even "feel" the anisotropy of the flow field, leading to significant values of  $v_2$ for the deuteron as shown above, which follows the expected trend given by the blast-wave model. 

Within the currently available limited statistics the transverse momentum spectra of hypertriton and anti-hypertriton indicate hydrodynamic flow very comparable to that measured for the nearly equal mass $^3$He nuclei. This could lend support to the assumption that such nuclei are rather formed via coalescence of nucleons and hyperons, thereby somehow naturally inheriting the flow of the baryons. However, connecting hydrodynamic flow with the coalescence picture has not been very successful so far. In particular because the existing model was not yet applied on data except on some preliminary data from the SPS.

\begin{figure}[!htb]
\begin{center}
\includegraphics[width=0.95\textwidth]{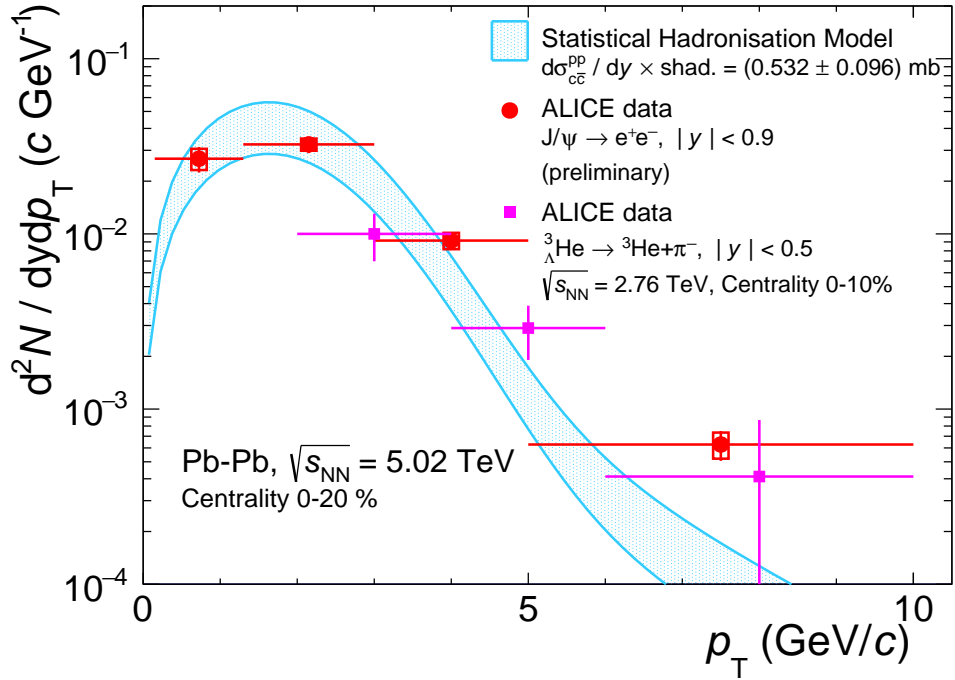}
\caption{\label{fig:shm_jpsi_hypertriton} Comparison between the scaled hypertriton and J/$\psi$ transverse momentum spectra, together with a hydrodynamic model calculation. Figure modified from~\cite{Andronic:2018vqh}. For details see text.}
\end{center}
\end{figure}

Furthermore, we show, in Fig.~\ref{fig:shm_jpsi_hypertriton}, the transverse momentum spectra of J/$\psi$ mesons decaying into dielectrons at mid-rapidity compared with a statistical hadronisation model approach~\cite{BraunMunzinger:2000px,Andronic:2017pug} for the production which is combined with a hydrodynamic model calculation to describe its spectrum in the shown band~\cite{Andronic:2018vqh}. In addition, the appropriately scaled $p_\mathrm{T}$ spectrum for the hypertriton is shown. Within uncertainties, the two transverse momentum distributions agree in shape. One should note that the masses of the two objects are rather equal, namely 2.99 GeV/$c$ for the hypertriton and 3.10 GeV/$c$ for the J/$\psi$. The interesting fact is here that the binding energy differs dramatically for these hadrons: the binding energy of the hypertriton is about 2.3 MeV, with a $\Lambda$ separation energy of a few hundreds of keV, whereas the J/$\psi$ is bound by about 600 MeV. Even more important would be a measurement of the elliptic flow or even higher flow coefficients with suitable statistics.   

Below we will contrast the thermal and coalescence description for loosely-bound states and offer an admittedly rather speculative way out of the apparent dilemma that such fragile objects seem to be produced thermally nearly the phase boundary but flow like all other hadrons.

Recently~\cite{Botvina:2017yqz} showed a nice agreement with the ALICE data in a phase-space approach, whereas one should note that the spectra are well described using different relative velocity cut-off values between the involved nucleons for the deuteron and $^3$He spectra. The predicted elliptic flow $v_2$ is nevertheless not well described. The most recent phase-space coalescence calculation was shown at the Quark Matter Conference 2018 in Venice and uses a set of values for the maximum relative momentum and the maximum distance which is fitted from the AGS data~\cite{Sombun:2018yqh}. Here the d/p ratio can be described reasonably well, using pure UrQMD for the pp data and a hybrid approach (hydro model + UrQMD) for the production of the nucleons which are forming the nuclei then by final-state coalescence.

The transverse momentum spectrum of deuterons and the $B_2$ are also well reproduced in a similar approach using the SMASH transport model~\cite{Oliinychenko:2018ugs}.

The recent work by Zhang and Ko~\cite{Zhang:2018euf} shows that the coalescence formation of the hypertriton is driven by two processes, the three-body channel binding together $\Lambda$, p and n as assumed in general for the formation of A=3 nuclei. In addition, they show the necessity of the d + $\Lambda$ coalescence with even larger weight compared to the three-body coalescence, and conclude that both are needed to get close to the measured yields.

Very recently, it was shown that this coalescence approach actually fails to describe $^3$He production if parameters based on deuteron production are used~\cite{Zhao:2018lyf}, calling the coalescence approach into question.

The general issue of these approaches in these recent publications is connected with the densities they assume at freeze-out~\cite{Rapp:2000gy,Rapp:2001bb}. In a very recent work, this concept was revived and used to calculate the production yields, yield ratios and transverse momentum spectra parameterised by blast-wave functions changing the mass of the different (anti-)nuclei rather successfully~\cite{Xu:2018jff}.

If one assumes that coalescence only takes place at or after thermal freeze-out the relevant densities should be large enough that the coalescing nucleons are sufficiently off-shell so that the energy conservation problem can be avoided. On the other hand, the density has to be low enough that the objects formed are not immediately destroyed again. If we conservatively assume that the hypertriton has a radius of larger than 5 fm, see ~\ref{impact}, then the pion or nucleon density  has to be  $< 10^{-3}/\mathrm{fm}^{3}$ for the hypertritons to survive. At such densities one would normally assume that all nucleons are on the mass shell. On the other hand, this simple estimate also would lead to the conclusion that no hypertriton can survive the phase after chemical freeze-out, calling the thermal approach into question.

A microscopic coalescence model taking into account all the above mentioned issues would be desirable for the future and predictions from it would be well appreciated but the above considerations indicate that this will not be straightforward.

An interesting and promising discussion on the difference between thermal and coalescence model predictions are given here~\cite{Bellini:2018epz}.
The authors make similar statements as above that the wave function plays an important role for the coalescence approach but not for the thermal model.

An elegant but very speculative way out of this dilemma was recently suggested~\cite{Andronic:2017pug}. There it is proposed that composite objects such as d and hypertriton are actually formed at the phase boundary as compact multi-quark states which would develop into the nuclear wave functions over a time scale of $> 20$ fm/c, i.e. corresponding to an excitation energy of approximately 10 MeV. At such time scales the density of the fireball should be well below $10^{-3}$/ fm$^3$ such that even the hypertriton can survive. In this picture also the surprising flow behaviour of the loosely-bound states, see the discussion below and in~\cite{Yang:2018ghi} would find its natural explanation as due to quark flow from the QGP. 

One way to make a precision test of this hypothesis is connected to plans for LHC Run3 and Run4 in the coming decade. There, measurements discussed above such as the comparison in Fig.~\ref{fig:shm_jpsi_hypertriton} could be performed with a hundred-fold improved statistical accuracy, see the discussion in section~\ref{expectations}.   

\section{Expectations for Run2,  Run3 and Run4 of the LHC}
\label{expectations}

\begin{figure}[!htb]
\begin{center}
\includegraphics[width=0.7\textwidth]{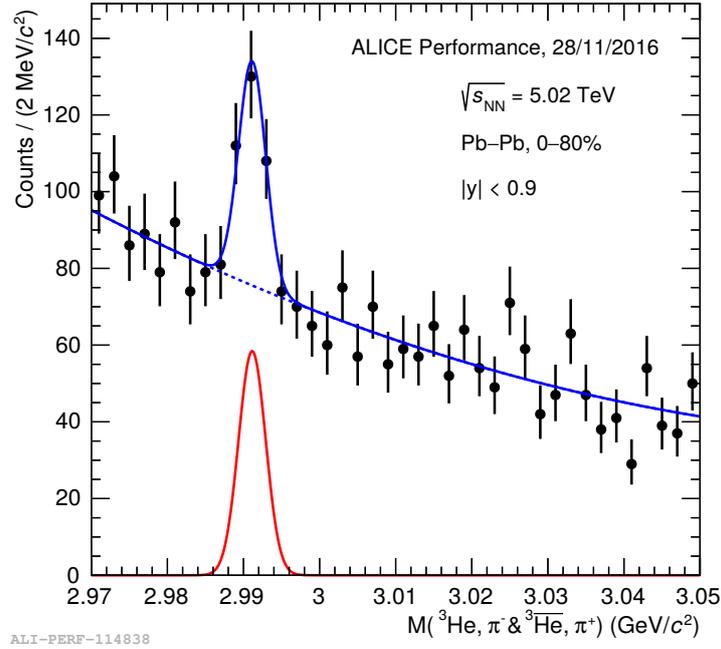}
\caption{\label{hypt_lukas} Invariant mass distribution for the $^3_\Lambda$H and $^3_{\bar{\Lambda}}\bar{\mathrm{H}}$ from a first analysis of the Pb--Pb data taken in 2015, shown first at the LHCC meeting autumn 2016~\cite{lhcc_os2016}.} 
\end{center}
\end{figure}

Higher precision results on the aforementioned topics are the goal in the next running phases of the LHC. The LHC Run2 has already started in 2015 and the aim of Run2 is to collect $2 \cdot 10^8$ Pb--Pb minimum bias events at $\sqrt{s_\mathrm{NN}} = 5$~TeV. For these statistics only a moderate reduction of the uncertainties is expected. Nevertheless, it will allow to study the observables more differentially as reported here. In addition, an online trigger on $z = 2$ particles is in place for the pp and p--Pb data taking and will provide the possibility to enhance the samples with light (hyper-)nuclei. A first example of the results on the hypertriton (using the Pb--Pb minimum bias data taken in 2015) was shown first at the LHCC open session at CERN and is displayed in Fig.~\ref{hypt_lukas}.

\begin{figure}[!htb]
\begin{center}
\includegraphics[width=0.8\textwidth]{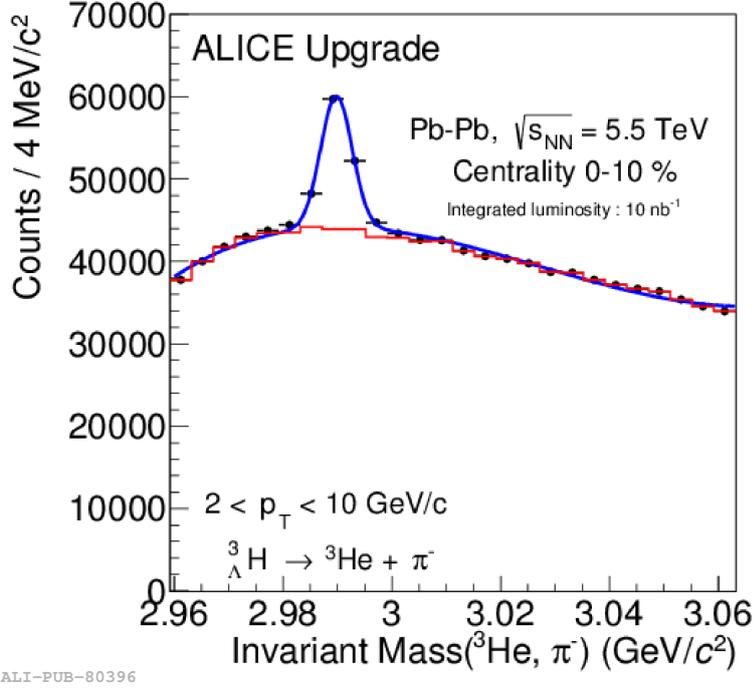}
\caption{\label{hypt_pred_run3} Expected invariant mass distribution for $^3_\Lambda$H reconstructed in Pb--Pb collisions (0-10\% centrality class) at the top LHC energy of $\sqrt{s_\mathrm{NN}} = 5.5$~TeV, corresponding to $L_{int} = 10 \mathrm{nb}^{-1}$. Figure from~\cite{its_upgrade}}.
\end{center}
\end{figure}

For Run3 a huge increase of statistics is expected in data from the upgraded ALICE experiment~\cite{Adam:2015xke,Abelevetal:2014dna,Abelevetal:2014cna} ( about $10^{10}$ central events) which will allow precise measurements in the sector of loosely-bound objects. For instance Figure~\ref{hypt_pred_run3} shows the expected signal for the hypertriton decaying in the two-body decay channel. In the peak a signal of around 44000 hypertriton candidates is expected. This will be possible because the collisions will happen at 50 kHz and the detector will be read-out continuously. For this purpose the TPC and the ITS will be upgraded~\cite{Adam:2015xke,Abelevetal:2014dna}: the multi-wire proportional chambers of the TPC will be replaced by GEMs (Gas Electron Multiplier) and the new ITS will consist of 7 concentric layers of pixel detectors, using Monolithic Active Pixel Sensors (MAPS). 

\begin{figure}[!htb]
\begin{center}
\includegraphics[width=0.8\textwidth]{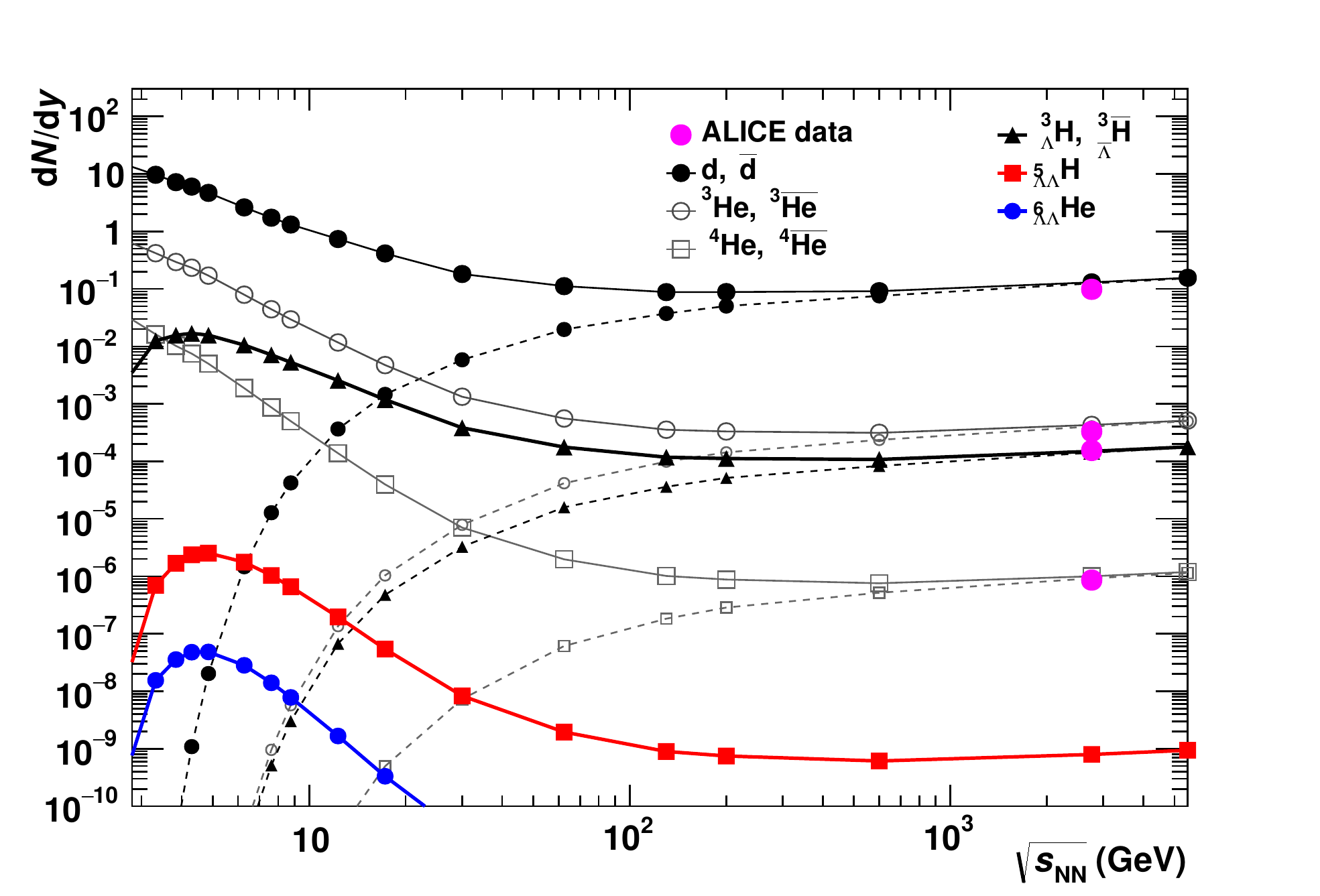}
\caption{\label{hypt_sqrts_new} Predictions of the thermal model for different loosely-bound objects as function of $\sqrt{s_{\mathrm{NN}}}$. In addition the ALICE results of Run1 are shown in magenta. Similar figure as in~\cite{thermalModel}.}
\end{center}
\end{figure}

Table~\ref{tableExotics} shows the expected yields for $10^{10}$ central events expected in Run3. This will allow to study the A=2 and A=3 objects with high precision and will allow to investigate the A=4 (hyper-)nuclei with the accuracy currently possible for the A=2, 3 states. This means the ALICE Collaboration will be able to study the mass (difference of particle and anti-particle) lifetime and the production yields of this objects, as shown in the body of this review.
Figure~\ref{hypt_sqrts_new} shows the predictions for (hyper-)nuclei as a function of $\sqrt{s_\mathrm{NN}}$ in comparison with the ALICE measured values from Run1. Clearly, the Run1 and Run2 measurements will be improved significantly in Run3, Run4 and, in addition there is a good chance that a first observation of the indicated A=5 hyper-nucleus can be made.  

\begin{table}[htp]

\caption{Expected yields of (hyper-)nuclei and exotica per 10$^{10}$ central collisions which are envisaged in Run3. The acceptance $\times$ efficiency for detection of charged tracks and weak decays (secondary vertex reconstruction) are included, as well as the branching ratios.}
	\label{tableExotics}

	\begin{center}
	\begin{tabular}{l|l}
	
Particle & Yield \\ \hline \hline & \\ 
Anti-alpha
${^{4}\overline{\rm{He}}}$ & $5.5 \cdot 10^{3}$ \\ & \\ 
Anti-hypertriton
$^{3}_{\bar{\Lambda}}{\overline{\mathrm{H}}} \; \; (\bar{\Lambda}
\bar{\mathrm{p}} \bar{\mathrm{n}})$ & $4.4 \cdot 10^{4}$ \\ &
\\ 
$^{4}_{\bar{\Lambda}}{\overline{\mathrm{H}}} \; \; (\bar{\Lambda}
\bar{\mathrm{p}} \bar{\mathrm{n}}\bar{\mathrm{n}})$ & $\, 
10^{2}$ \\ & \\ 
$^{5}_{\bar{\Lambda}}{\overline{\mathrm{H}}} \; \;
(\bar{\Lambda} \bar{\mathrm{p}}
\bar{\mathrm{n}}\bar{\mathrm{n}}\bar{\mathrm{n}})$ & $2$ \\ &
\\$^{4}_{\bar{\Lambda}\bar{\Lambda}}{\overline{\mathrm{He}}} \; \;
(\bar{\Lambda} \bar{\Lambda} \bar{\mathrm{p}}
\bar{\mathrm{n}})$ & $1$ \\ &
\\ 
H-Dibaryon $(\Lambda\Lambda)$ & $ > \,10^{6}$ \\ & \\ $\Lambda
\mathrm{n}$ & $ > \,8 \cdot 10^{6}$ \\ \\
$\Xi\Xi$ & $  \, 5 \cdot 10^{4}$ \\ \hline \hline

	\end{tabular}
	\end{center}
	
\end{table}%

In addition, a recent review~\cite{Cho:2017dcy} from the ExHIC Collaboration shows the huge variety of possible states to investigate in the next years. A particular focus will be also put on the searches for kaonic bound states, states which consist of nucleons and negatively charged kaons which bind these systems and shrink them strongly in size~\cite{Yamazaki2004167,Maeda:2013zha}. 
The existence of such kaonic bound states with relatively narrow widths are currently under strong debate because several different experiments conflicting signals in different channels~\cite{Gal:2016boi,Ajimura:2018iyx}. ALICE will make a dedicated search in Run3. In the most prominent channel ppK$^{-}$, which decays strongly, we expect a production yield of the order of $\textrm{d}N/\textrm{d}y \approx 3\times10^{-3}$ in central Pb--Pb collisions~\cite{Andronic:2005hh}. 

\section{Summary and outlook}
\label{summary}

We have presented a survey of results on the production of loosely-bound objects in hadronic and nuclear collisions at LHC energies. The measured production yields of light (anti-)(hyper-)nuclei clearly match the thermal model expectations. Including them or excluding them in a thermal model fit does not change the fit outcome significantly. The survival of the nuclei through the fireball evolution from chemical to kinetic freeze-out can be explained by an isentropic evolution of the fireball. The measured (anti-)(hyper-)nuclei follow the same radial expansion as the light hadrons, and can be explained with one common set of blast-wave parameters. 

Despite the very high collision energy and the more than 6 orders of magnitude lower binding energy of such objects, such bound states are copiously produced. In particular, their production yield is determined by their mass and does not depend on their binding energy. Surprisingly, all measured yields closely follow the prediction of the thermal model with the same temperature as obtained from an analysis of light hadron yields.

The model predictions for the investigated bound states of $\Lambda\Lambda$ and $\Lambda$n are well above the upper limits which have been set. The latter also holds in a reasonable part of the phase space of decay length and branching ratio of these states. On the other hand, the thermal model can also describe the production yield of the hypertriton reasonably well, which binding energy is only 130 keV. The investigated bound states could nevertheless still be unbound and resonances above the threshold. In this case they might be so broad that they are not observed.

The current world average of the lifetime of the hypertriton is around 2$\sigma$ away from the most recent theoretical estimate, expecting it to be only 3\% below the lifetime of the free $\Lambda$. This world average is currently dominated by the measurements in heavy-ion collisions. These measurements being already more precise as the measurements in emulsions need a significant improvement on their uncertainties, both statistical and systematic. One way to study the systematic effects in heavy-ions was started by the STAR Collaboration looking also into the three-body-decay channel. The more fundamental point is to significantly reduce the statistical uncertainties, which will be possible with the discussed upgrade of the ALICE detector.    

The current data taking period will reduce the uncertainties significantly, whereas the expected statistics of Run3 will allow the study of A=4 hyper-nuclei and lead to  much improved measurement of the lifetime of the hypertriton. In addition to this, the measurement of the production of  light (anti-)nuclei will become very precise and also the phase space for the investigation of other hypothetical exotic bound states is opening up in Run3.  

These results will be complemented by the measurements at lower energies at the upcoming facilities CBM at FAIR, BM and MPD at NICA and the heavy-ion program at J-PARC where the physics aims include plans to study hypernuclei and exotic objects. Furthermore, plans are ongoing to build a hadron accelerator machine at energies even higher than the LHC. Current planning foresees nucleus-nucleus collisions at a centre-of-mass energy of $\sqrt{s_\mathrm{NN}} = 39$\,TeV~\cite{Dainese:2016gch}.

Altogether this will increase our knowledge of the production of loosely-bound objects and will help to finally reveal their production mechanism.  

We finally note that a review of anti(hyper)-matter production in 
nuclear collisions with different focus just became available~\cite{Chen:2018tnh}.

\section*{Acknowledgements}

The authors would like to thank the BMBF for their support through the FSP202 (F\"orderkennzeichen 05P15RFCA1). This work is part of and supported by the DFG Collaborative Research Center "SFB1225/ISOQUANT". We thank the members of the ALICE Collaboration and in particular Anton Andronic, Stefania Bufalino, Boris Hippolyte, Alexander Kalweit, Nicole L\"oher, the late Helmut Oeschler, Krzysztof Redlich, J\"urgen Schukraft and Johanna Stachel for many helpful discussions.





%\begin{thebibliography}{00}


%\bibitem{}

%\bibliographystyle{utphys} 
\bibliographystyle{elsarticle-num} 
\bibliography{bib_review_merged}

%\end{thebibliography}
\end{document}